%% file: MEx2GPhysicsPaper.tex
\documentclass[epjc3,twocolumn,a4paper,fleqn]{svjour3} % epjc
\usepackage[fleqn]{amsmath}
\usepackage{graphicx}
\usepackage{epstopdf}
\usepackage{fix-cm}
\usepackage{import}
\usepackage{txfonts}
\usepackage{epsfig}
\usepackage{listings}
\usepackage{lineno}
\usepackage{mdwlist}
\usepackage{subfigure}
\usepackage{pennames}
\usepackage{hyperref}
\usepackage[british]{babel}
\usepackage{color}
\usepackage{lipsum}
\usepackage{mathtools}
\usepackage{cuted}
\usepackage{multirow}
\usepackage{comment}
\usepackage{cite}
\usepackage{empheq}
\usepackage[titletoc,title]{appendix}
\RequirePackage[normalem]{ulem} %DIF PREAMBLE
\RequirePackage{color}\definecolor{RED}{rgb}{1,0,0}\definecolor{BLUE}{rgb}{0,0,1} %DIF PREAMBLE
\RequirePackage[normalem]{ulem} %DIF PREAMBLE
\RequirePackage{color}\definecolor{RED}{rgb}{1,0,0}\definecolor{BLUE}{rgb}{0,0,1} %DIF PREAMBLE
\graphicspath{{./}{../00_introductionLFV/}{../01_detector/}{../02_run/}{../03_reconstruction/}{../04_analysis/}{../05_results/}{../06_conclusions/}}
\DeclareGraphicsExtensions{.eps,.pdf,.jpeg,.jpg,.png}

\def\vector#1{\mbox{\boldmath $#1$}}
\newcommand{\meg}{\ifmmode\muup^+ \to {\rm e}^+ \gammaup\else$\muup^+ \to \mathrm{e}^+ \gammaup$\ \fi}
\newcommand{\michel}{\ifmmode\muup^+ \to {\rm e}^+ \nuup\bar{\nuup}\else$\muup^+ \to \mathrm{e}^+ \nuup\bar{\nuup}$\ \fi}
\newcommand{\radiative}{\ifmmode\muup^+ \to {\rm e}^+ \nuup\bar{\nuup}\gammaup\else$\muup^+ \to \mathrm{e}^+ \nuup\bar{\nuup}\gammaup$\ \fi}
\newcommand{\mextwog}{\ifmmode\muup^+ \to \mathrm{e}^+\mathrm{X}, \mathrm{X} \to \gammaup\gammaup\else$\muup^+ \to \mathrm{e}^+\mathrm{X}, \mathrm{X} \to \gammaup\gammaup$\ \fi}
\newcommand{\conv}{\ifmmode\muup^- \to {\rm e}^-\else$\muup^- \to \mathrm{e}^-$\fi}
\newcommand{\mute}{\ifmmode\muup^+ \to 3{\rm e}\else $\muup^+ \to 3\mathrm{e}$\fi}
\newcommand{\aif}{\ifmmode\mathrm{e}^+ \mathrm{e}^- \to \gammaup\gammaup \else$\mathrm{e}^+ \mathrm{e}^- \to \gammaup \gammaup$\ \fi}

\newcommand{\nn}{\nonumber \\}
\newcommand{\bea}{\begin{eqnarray}}
\newcommand{\eea}{\end{eqnarray}}
\newcommand{\be}{\begin{equation}}
\newcommand{\ee}{\end{equation}}
\newcommand{\fref}[1]{Fig.~\ref{#1}}
\newcommand{\Fref}[1]{Figure~\ref{#1}}
\newcommand{\sref}[1]{Sect.~\ref{#1}}

\newcommand{\tref}[1]{Table~\ref{#1}}
\newcommand{\eref}[1]{Eq.~(\ref{#1})}
\newcommand{\erefs}[1]{Eqs.~(\ref{#1})}

\newcommand{\teg}{\ifmmode t_{\gammaup_1\mathrm{e^+}}\else$t_{\gammaup_1\mathrm{e^+}}$\fi}
\newcommand{\tgg}{\ifmmode t_{\gammaup\gammaup}\else$t_{\gammaup\gammaup}$\fi}

\newcommand*{\mathtentative}{}
\def\mathtentative#1#{\mathcoloraux{#1}}
\newcommand*{\mathcoloraux}[3]{%
  \protect\leavevmode
  \begingroup
    \color#1{#2}#3%
  \endgroup
}
\journalname{Eur. Phys. J. C} % epjc

\begin{document}

\title{Search for lepton flavour violating muon decay mediated by a new light particle in the MEG experiment}

\author{The MEG collaboration}
\include{author-meg-epjc} % epjc

\thankstext[*]{e1}{Corresponding author: mitsutaka.r.nakao@gmail.com} % epjc
%\thankstext[$\dagger$]{e2}{Deceased } % epjc
\maketitle % epjc
 
\begin{abstract}

We present the first direct search for lepton flavour violating muon decay mediated by a new light particle X, $\mextwog$. This search uses a dataset resulting from $7.5\times 10^{14}$ stopped muons collected by the MEG experiment at the Paul Scherrer Institut in the period 2009--2013.
No significant excess is found in the mass region 20--45~MeV/c$^2$ for lifetimes below 40 ps, and we set the most stringent branching ratio upper limits in the mass region of 20--40~MeV/c$^2$, down to 
%in particular upper limits of
$\mathcal{O}(10^{-11})$ at 90\%\ confidence level.
%in the mass region 20--30~MeV/c$^2$

\end{abstract}

 % epjc

\tableofcontents 

%\linenumbers

% --------------------------- 0_introduction -------------------
\section{Introduction}
\label{sec:introduction}
%\subsection{Physics motivation}
%Recent efforts in elementary particle physics research have been made to explore new physics beyond the Standard Model (SM).
%However, there is no clear evidence of new physics to date but for some anomalies.
%In this paper, we try to tackle this situation by combining two different directions:
%charged lepton flavour violation and light new physics.

The search for charged lepton flavour violating (CLFV) processes is one of the key tools to probe for physics beyond the Standard Model (SM) of elementary particles and interactions.
The observation of neutrino oscillations\cite{SK:1998,SNO:2002,PDG2018} showed that lepton flavour is not conserved in nature.
As a consequence, charged lepton flavour is violated, even though the rate is unobservably small ($<\!10^{-50}$) in an extension of the SM accounting 
for measured neutrino mass differences and mixing angles~\cite{Petcov:1977,Cheng:1980}.
In the context of new physics, in the framework of grand unified theories for example, CLFV processes can occur at an
experimentally observable rate~\cite{barbieri1994}. %, while they are practically forbidden in the SM.
%, while the SM allows CLFV processes with only extremely small branching ratios ($<\!10^{-50}$) even when accounting 
% for measured neutrino mass differences and mixing angles. 
Therefore, such processes are free from SM physics backgrounds and a positive signal 
would constitute unambiguous evidence for physics beyond the SM.
This motivates the effort to search for evidence of new physics through CLFV processes \cite{mori_2014,calibbi_2018}.

The MEG experiment at the Paul Scherrer Institut (PSI) in Switzerland searched for one such CLFV process, \meg decay, with the highest sensitivity in the world.
No evidence of the decay was found yet, leading to an upper limit on the branching ratio $\mathcal{B}(\meg) < 4.2\times10^{-13}$ at 90\% confidence level (C.L.)~\cite{baldini_2016}.
%The collaboration is preparing the upgraded experiment MEG II to achieve a sensitivity below $6\times10^{-14}$ after three-year-data taking\,\cite{baldini_2018}.
%
Models that allow \meg decay at an observable rate usually assume that CLFV couplings are introduced through an exchange of new particles much heavier than the muon.
%Negative results by CLFV searches suggest another possibility: new physics exists at a lighter scale but with very weak coupling to SM particles.
Negative results by CLFV searches leave open another possibility: new physics exists at a lighter scale but with very weak coupling to SM particles.

If a new particle X (with mass $m_\mathrm{X}$ and lifetime $\tau_\mathrm{X}$) lighter than the muon exists, the CLFV two-body decay $\muup \to \mathrm{eX}$ may be a good probe for such new physics.
The experimental signature depends on how the new particle X decays.
In this paper, we report a search for \mextwog (MEx2G) decay using the full dataset collected in the MEG experiment.
Here, we assume that X is an on-shell scalar or pseudo-scalar particle.
%A possible search for such a new light particle in the MEG experiment is the search for the \mextwog\,(MEx2G) decay, where X is a scalar or pseudo-scalar particle generated via a CLFV coupling and the on-shell X decays back into SM particles, $\gammaup$.
%In this paper, we searched for this decay mode using the full dataset taken in the MEG experiment.
%In this search, we assume X is long-lived and the total decay width is narrow.
Axion-like particles~\cite{Peccei1977,Weinberg1978,Wilczek1978,Cornella2019}, Majoron~\cite{Chikashige1981,Gelmini1981}, familon~\cite{Reiss1982,Wilczek1982,Berezhiani1990,Jaeckel2014}, flavon~\cite{Tsumura2010,Bauer2016}, flaxion~\cite{Ema2017,calibbi_2017}, hierarchion~\cite{Davidi2018}, and strongly interacting massive particles~\cite{Hochberg2014,Hochberg2015} are candidates for X.

%\subsection{Experimental searches}
A dedicated search for the MEx2G decay has never been done, although some constraints on the X particle parameter space can be deduced by experimental results from both related muon decay modes and non-muon experiments; these are discussed below. 

%\paragraph*{Constraints from other modes}\,
Current upper limits on the inclusive decay $\muup^+\to \mathrm{e^+ X}$ are given at $\mathcal{O}(10^{-5})$ for $m_\mathrm{X}$ in the range 13--80~MeV/c$^2$~\cite{Bayes2015}.\footnote{In these searches, only e$^+$ is looked at.}
However, the current limits do not impose any constraints on the MEx2G decay in the target region of this search.
They are complementary, relevant for cases where X is either stable or decays invisibly. 
%
%In the future, the Mu3e experiment has a possibility to push the upper limit down to $5\times10^{-9}$\,\cite{Perrevoort2018}, which is comparable to the current upper limit of the MEx2G decay in the higher mass region.
%
For X resulting from muon decays, the only kinematically allowed visible decay channels are $\mathrm{X \to e^+e^-}$ and $\mathrm{X} \to \gammaup\gammaup$. The former can occur at tree level while the latter can occur via a fermion loop.
The current upper limit on $\muup^+\to \mathrm{e^+ X}, \mathrm{X}\to \mathrm{e^+e^-}$ at a level of $\mathcal{O}(10^{-12})$ \cite{Eichler1986} give stringent constraints on the MEx2G decay if we assume that X is more likely to decay into an e$^+$e$^-$ pair. %, it becomes more stringent than the target sensitivity of this search.
However, there is a possibility for X to be electrophobic, as pointed out in~\cite{Anderson2003,Liu2016}, and searches for both decay modes can hint at the model behind these decay modes.

%\paragraph*{MEG2012}\,
%The MEx2G direct search requires experimental features in common with the \meg\, search, and hence, the \meg\, search experiments, like MEG, are suitable for this search. 
%The MEG experiment utilizes the world's highest muon beam and high-performance gamma and positron detectors. Therefore, it is a unique experiment enabling the best search of this decay to date. The MEx2G direct search was actually performed, for the first time, by the MEG collaboration using the datasets taken in 2009 and 2010.
%No significant excess was found and upper limits on the branching ratio for the mass range of 10--45 MeV and lifetime equal or less than 10 ns were set.
%The upper limits for a mass range below $\sim$30 MeV have been significantly updated.
%The results are summarized in a Ph.D. thesis published in 2012\,\cite{Natori2012}.

\begin{figure}[t!]
    \centering
     \includegraphics[width=\columnwidth]{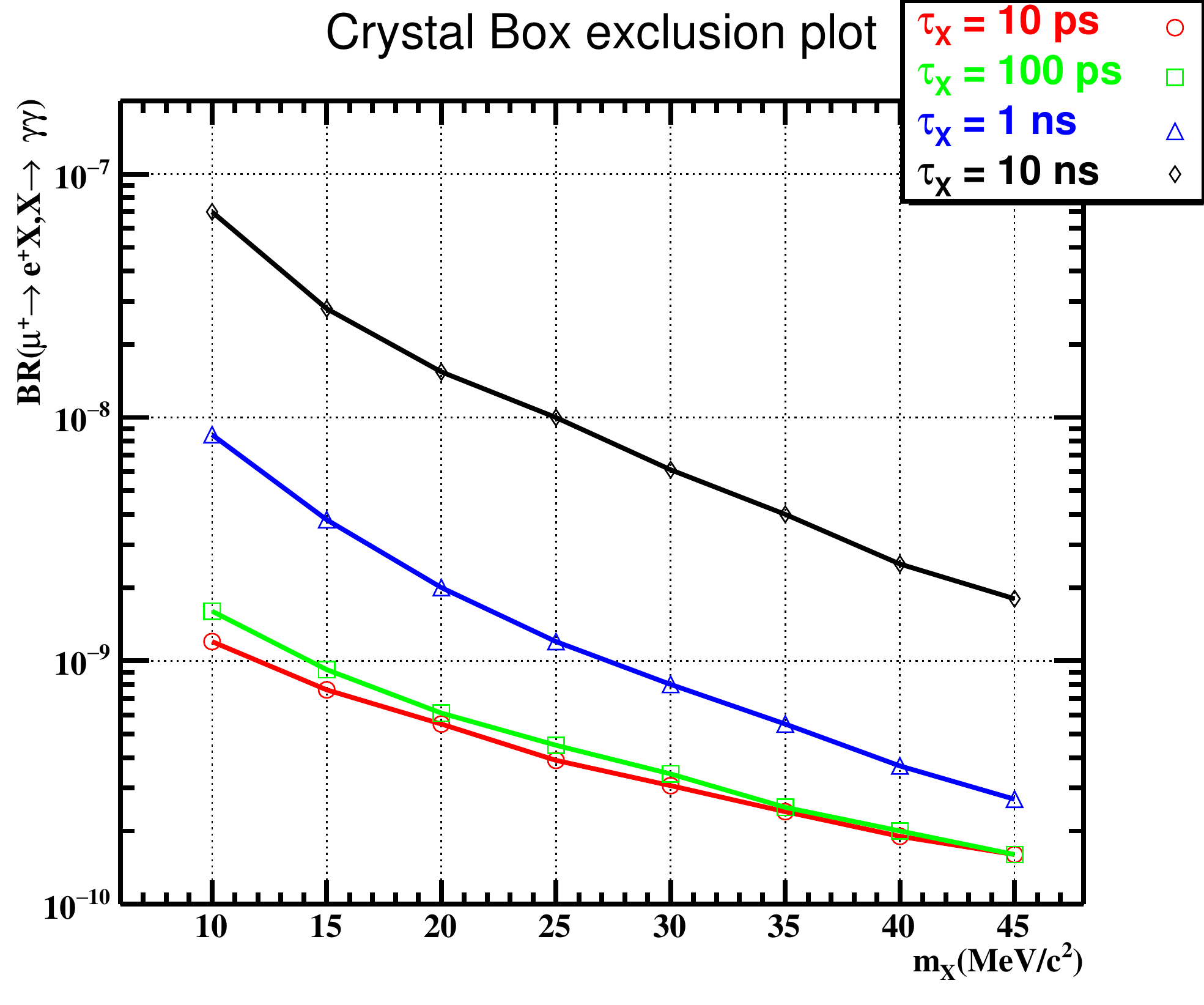}
    \caption[Upper limits estimated from the Crystal Box experiment and previous MEx2G search]{Upper limits on MEx2G decay estimated by converting the upper limits on  $\muup^+\rightarrow\mathrm{e^+}\gammaup\gammaup$ from the Crystal Box experiment as a function of $m_\mathrm{X}$. Lines with different markers and colours correspond to different $\tau_\mathrm{X}$. %In this plot (from \,\cite{Natori2012}), X is denoted by $\phi$.}
    }
    \label{fig:crystal_box_limit}
\end{figure}

%\paragraph*{The Crystal Box experiment}\,
The current upper limit on the decay $\muup^+\rightarrow\mathrm{e^+}\gammaup\gammaup$, $\mathcal{B}(\muup^+\rightarrow\mathrm{e^+}\gammaup\gammaup)< 7.2\times10^{-11}$ (90\% C.L.) from the Crystal Box experiment\,\cite{bolton_1988_prd} can be converted into an equivalent MEx2G upper limit by taking into account the difference in detector efficiencies\,\cite{Natori2012};
the converted limits are shown in \fref{fig:crystal_box_limit}.

\begin{figure}[tb]
    \centering
    \includegraphics[width=\columnwidth]{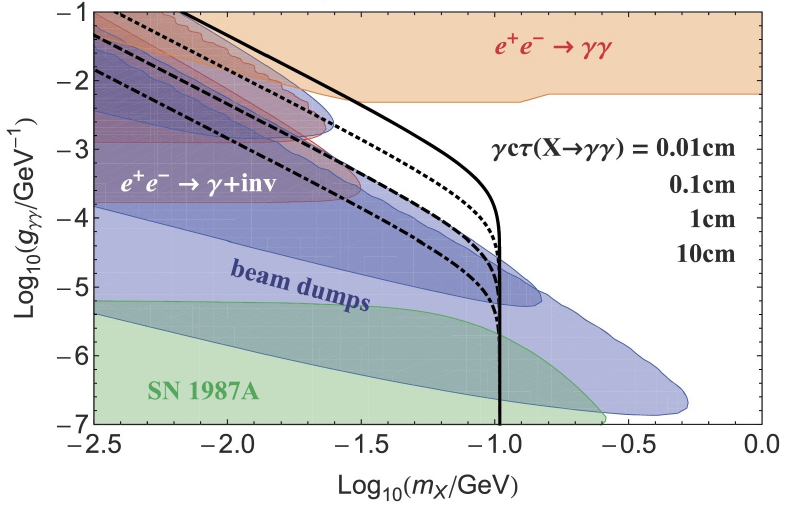}
    \caption[Excluded parameter region from collider, beam dumps, and supernova]{Excluded parameter regions for a scalar X with mass $m_{\rm{X}}$ and coupling $g_{\gamma\gamma}$ to 2$\gammaup$s from collider, beam dumps, and supernova \cite{Dolan2017,Dobrich2016,Jaeckel2016} (from \cite{Heeck2018}). In black we show contours of the boosted decay length $\gamma\mathrm{c}\tau_\mathrm{X}$ of $\mathrm{X}\rightarrow\gammaup\gammaup$, assuming X to be produced from an at-rest muon decay $\muup^+\to \mathrm{e^+ X}$. The solid black line corresponds to $\gamma\mathrm{c}\tau_\mathrm{X}=0.01$ cm, the dotted one to 0.1 cm, the dashed one to 1 cm and the dot-dashed line to 10 cm.}
    \label{fig:displaced}
\end{figure}

%\paragraph*{Constraints from other experiments}\,
Axion-like particle searches from collider and beam dump experiments and from supernova observations also constrain the branching ratio $\mathrm{X}\rightarrow\gammaup\gammaup$ if the axion-like particles are generated from coupling to photons \cite{Heeck2018}.
\Fref{fig:displaced} summarises the parameter regions excluded by these experiments.
%A part of parameter space for the MEx2G decay is excluded by lepton collider experiments and beam dump experiments.
A region with decay length c$\tau_{\rm{X}}\gamma<$ 1\,cm and $m_{\rm{X}}>20$\,MeV/c$^2$ still has room for the MEx2G decay.

\begin{figure}[tb]
	\centering
	\includegraphics[width=\columnwidth]{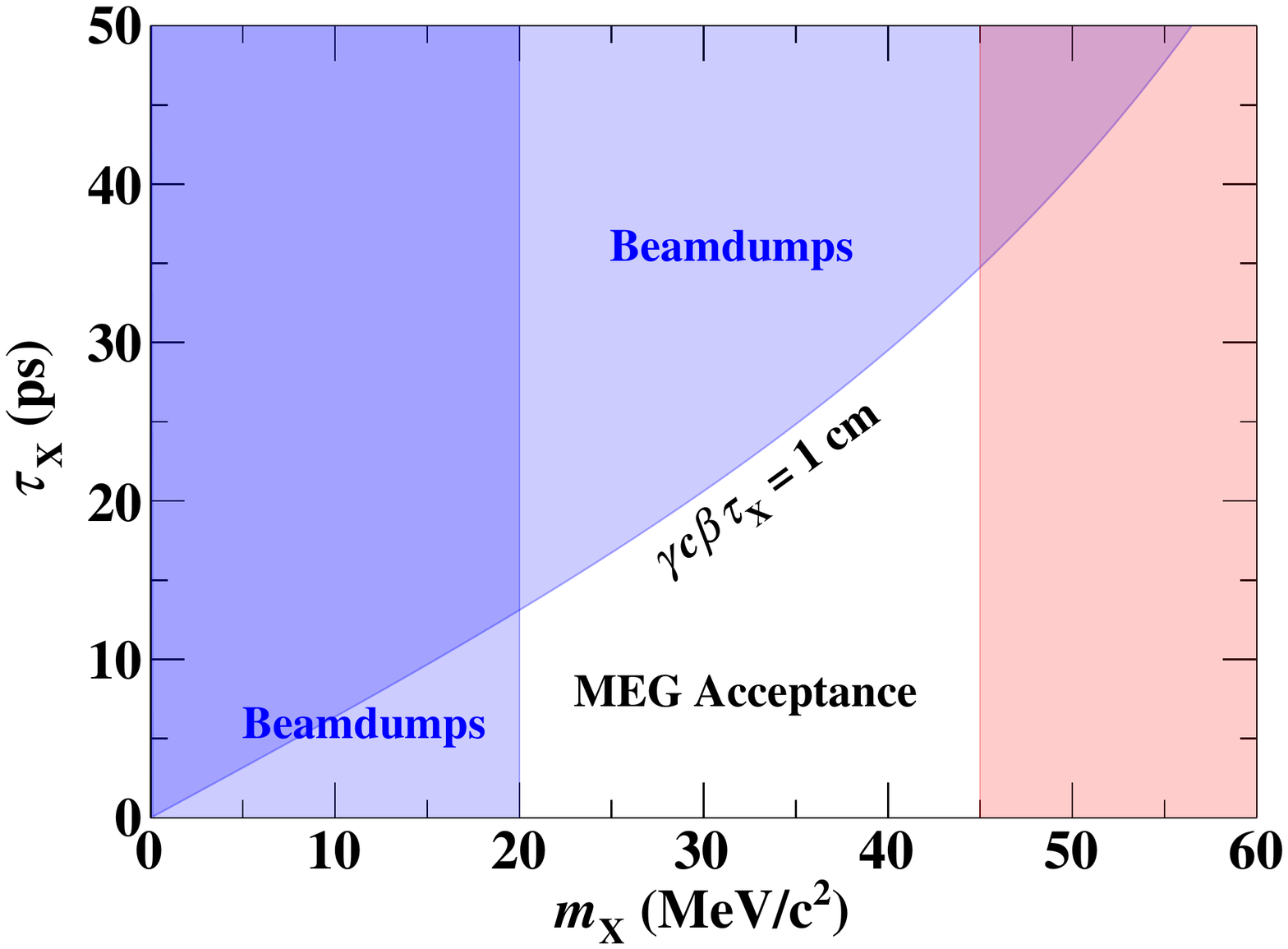}
	\caption[Allowed parameter space]{Allowed X particle parameter space (white). The blue region has already been excluded~\cite{Dobrich2016} and the red shaded region on the right  ($m_\mathrm{X}\gtrsim45$~MeV/c$^2$) is inaccessible to MEG.}
	\label{fig:parameter_space}
\end{figure}

%\paragraph*{Target parameter space in this analysis}\,
Based on limits discussed above, we define the target parameter space of this search in the  $\tau_\mathrm{X}$--$m_{\mathrm{X}}$ plane as shown in \fref{fig:parameter_space}.
%The decay length is converted into the lifetime of X($\tau_\mathrm{X}$) by using \eref{eq:decay_length} and the subsequent assumption, which gives:
%\bea
%l(=c\beta\gammaup\tau_\mathrm{X}) < 1\,\mathrm{cm}
%&\Leftrightarrow& \tau_\mathrm{X} < \frac{P_{\mathrm{X}}}{cm_{\mathrm{X}}},
%\eea
%where $P_{\mathrm{X}}$ is the momentum of X.
%The blue region has already been excluded and the red region ($m_\mathrm{X}\gtrsim45$~MeV/c$^2$) cannot be searched using the MEG experiment setup, as shown in \sref{sec:detector}.
%Therefore we focus on the following parameter space:
%\bea
%m_{\mathrm{X}} = (20, 25, 30, 35, 40, 45) \,\mathrm{MeV}
%\eea
%\bea
%\tau_\mathrm{X} = (5, 20, 40) \,\mathrm{ps}
%\eea

%\subsection{Overview of the search}

%\subsection{Overview of this paper}
%After a brief introduction to the detector and 
%to the data acquisition system (Sect.~\ref{sec:detector}), the 
%reconstruction algorithms are presented in detail (Sect.~\ref{sec:reconstruction}), followed by
%an in-depth discussion of the analysis of the full MEG dataset 
%and of the results (Sect.~\ref{sec:analysis}).
%Finally, in the conclusions, some prospects for future improvements are outlined (Sect.~\ref{sec:conclusions}).

% --------------------------- 1_detector -----------------------
\begin{figure*}[tb]
\centering
  \includegraphics[width=\textwidth,angle=0] {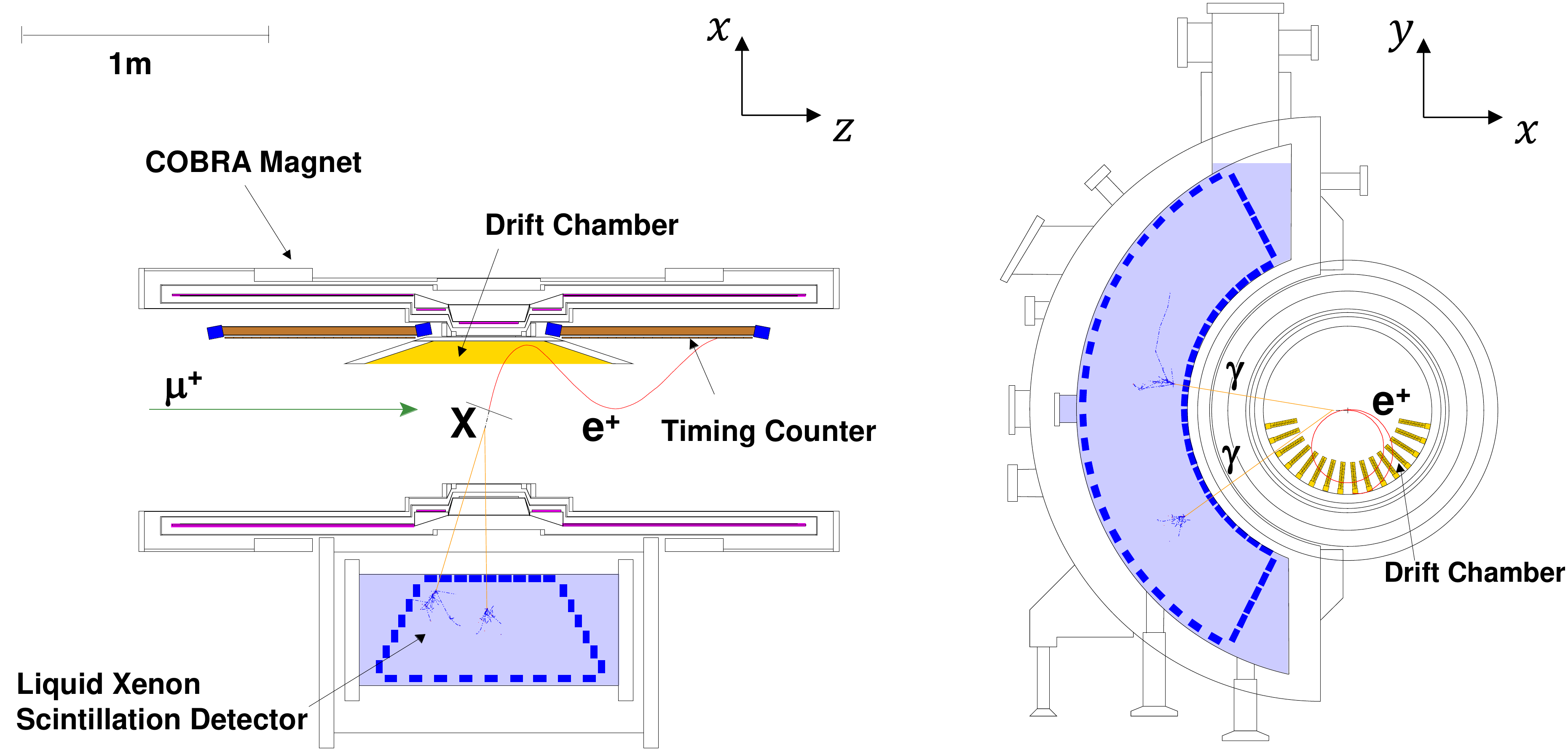}
 \caption{The figure shows a schematic view of the MEG detector with a simulated MEx2G event emitted from the target.
 The top view is shown on the left, the view from downstream on the right.
}
 \label{introduction:megdet}
\end{figure*}

\section{Detector}
\label{sec:detector}

The MEG detector is briefly presented in the following, emphasising aspects relevant to this search;
a detailed description is available in \cite{megdet}. 
%Briefly, it consists of the $\muup^+$ beam, a thin stopping target, a thin-walled,  superconducting magnet,
%a drift chamber array (DCH), scintillating timing counters (TC), and a liquid xenon calorimeter (LXe detector). 

In this paper we adopt a Cartesian coordinate system $(x,y,z)$ shown in Fig.~\ref{introduction:megdet} with the origin
 at the centre of the magnet. When necessary, we also refer to the cylindrical coordinate system ${\it (r,\phi,z)}$ as well as the polar angle $\theta$.
%The {\it z}-axis is parallel to the magnet axis and directed along the $\muup^+$ beam.
%The axis defining $\phi=90^\circ$ (the {\it y}-axis of the corresponding Cartesian coordinate system)
%is directed upwards and, as a consequence, the {\it x}-axis 
%is directed opposite to the centre of the LXe detector.
%Positrons move along trajectories with decreasing $\phi$-coordinate.
%When required, the polar angle $\theta$ with respect to the {\it z}-axis is also used.
%The region with ${\it z}<0$ is referred to as upstream, that with ${\it z}>0$ as downstream.

%\subsection{Muon beam and stopping target}
Multiple intense $\muup^+$ beams are available at the $\piup$E5 channel in the 2.2-mA PSI proton accelerator complex.  %in combination with the MEG beam line.
We use a beam of {\it surface muons}, produced by $\piup^+$ decaying near the surface of a production target.
The beam intensity is tuned to a $\muup^+$ 
stopping rate of $3\times 10^7$, limited by the rate 
capabilities of the tracking system and the rate of accidental backgrounds in the \meg search.
The muons at the production target are fully polarised ($P_{\muup^+}=-1$), and they reach a stopping target with a residual polarisation 
$P_{\muup^+} = -0.86 \pm 0.02 ~ {\rm (stat)} ~ { }^{+ 0.05}_{-0.06} ~ {\rm (syst)}$ \cite{Baldini:2015lwl}.

The positive muons are stopped and decay in a thin target placed at the centre of the spectrometer at a slant angle of $\approx$\,$20^\circ$ from the $\muup^+$ beam direction. %, where they decay at rest. 
The target is composed of a 205~$\muup$m thick layer of polyethylene and 
polyester (density $0.895$~g/cm$^3$).
 
%\subsection{Positron spectrometer}
Positrons from the muon decays are detected with a magnetic spectrometer, called the COBRA (standing for COnstant Bending RAdius) spectrometer, consisting of a thin-walled superconducting magnet, a drift chamber array (DCH), and two scintillating timing counter (TC) arrays.

%\paragraph*{COBRA}\,
The magnet\,\cite{Ootani2004} is made of a superconducting coil with three different radii.
It generates a gradient magnetic field of 1.27 T at the centre and 0.49 T at each end.
%e$^+$s emitted from the target follow helical trajectories under the magnetic field.
The diameter of an emitted e$^+$ trajectory depends on the absolute momentum, independent of the polar angle due to the gradient field.
This allows us to select e$^+$s within a specific momentum range by placing the TC detectors in a specific radial range;
e$^+$s whose momenta are larger than $\sim$45~MeV/c fall into the acceptance of the TC.
Furthermore, the gradient field prevents e$^+$s emitted nearly perpendicular to the $\muup^+$ beam direction from looping many times in the spectrometer.
This results in a suppression of hit rates in the DCH.
The thickness of the central part of the magnet is 0.2~radiation~length to maximise transparency to $\gammaup$;
85\% of the signal $\gammaup$s penetrate the magnet without interaction and reach the photon detector.
%Since the performance of Photomultiplier tubes (PMT) deteriorates under a magnetic field, a compensation coil for COBRA is placed outside the detector, resulting in a reduction of the leak magnetic field around the LXe detector down to 50 Gauss.

%\paragraph*{Drift Chamber}\,
\label{sec:intro_dch}
Positrons are tracked in the DCH~\cite{Hildebrandt2010}. 
It is composed of 16 independent modules.
Each module has a trapezoidal shape with base lengths of 104 cm (at smaller radius, close to the stopping target) and 40 cm (at larger radius).
%The directions of base lengths are parallel with the beam direction.
These modules are installed in the bottom hemisphere in the magnet at 10.5$^{\circ}$ intervals.
The DCH covers the azimuthal region between 191.25$^{\circ}$ and 348.75$^{\circ}$ and the radial region between 19.3~cm and 27.9~cm.
%It has a two-layered structure and wires in each layer are stretched in the axial direction (beam direction).
%The distance between adjacent cells is 9 mm and these cells are staggered among layers.
%Thanks to these moduled structures and the gradient magnetic field, DCH is operational under a high rate environment.
%The positron hit rate at the most inner part is suppressed down to 10 kHz (while the original muon beam rate is 30 MHz).
It is composed of low mass materials and helium-based gas ($\mathrm{He}:\mathrm{C_2H_6} = 1:1$) to suppress Coulomb multiple scattering;
$2.0\times10^{-3}$ radiation length path is achieved for the e$^+$ from \meg decay at energy of $E_\mathrm{e^+}=52.83$~MeV ($= m_\muup \mathrm{c}^2/2$, where $m_\muup$ is the mass of muon).
%As mentioned above, wires are stretched along the beam direction and position resolution along the beam axis is not good.
%To get a better e$^+$ resolution along the beam axis, vernier pads are placed in each module.

%\paragraph*{Timing Counter}\,
The TC\,\cite{DeGerone2011,DeGerone2012} is designed to measure precisely the e$^+$ hit time.
%It consists of two parts: scintillation bars and scintillation fibers.
Fifteen scintillator bars are placed at each end of the COBRA.
They are made of $4\times4\times80$~cm$^3$ plastic scintillators with fine-mesh PMTs attached to both ends of the bars.
%Since the detector is placed in the high magnetic field, fine-mesh 2'' PMTs are used.
%The scintillation fibers readout by avalanche photodiodes were originally developed to get independent position information and improve time resolution, but they did not work in the experiment.
%TC is covered with a N$_2$ bag to prevent chamber gas (He) from entering TC PMTs.

\begin{figure}[t!]
    \centering
    \includegraphics[width=\columnwidth]{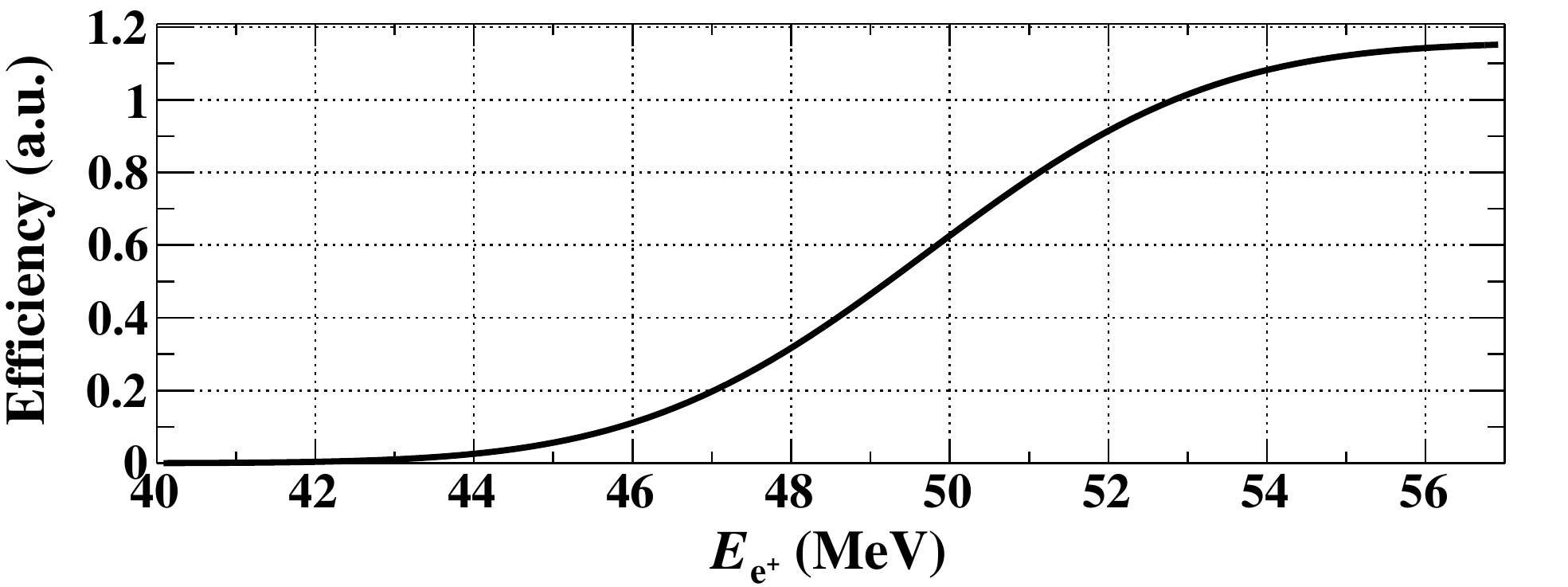}
    \caption{COBRA spectrometer relative efficiency as a function of $E_\mathrm{e^+}$ normalised to $\epsilon_\mathrm{e^+}(52.83$ MeV) = 1.}
    \label{fig:EeEfficiency}
\end{figure}

The efficiency of the spectrometer significantly depends on $E_\mathrm{e^+}$ as shown in \fref{fig:EeEfficiency}.
The e$^+$ energy from the MEx2G decay is lower than that from \meg depending on $m_{\mathrm{X}}$, and the efficiency is correspondingly lower.
The large $m_{\mathrm{X}}$ search range is limited by this effect as shown in Fig.~\ref{fig:parameter_space}.

%\subsection{Liquid xenon detector}
The photon detector is a homogeneous liquid-xenon (LXe) detector relying on scintillation light\footnote{In the high rate MEG environment, only scintillation light with its fast signal, is detected.} for energy, position, and timing measurement \cite{Sawada2010258,baldini_2005_nim}. %requires excellent position, time and energy resolutions to minimise the number of accidental coincidences between photons and positrons from different muon decays, which comprise the dominant background process.
%
%It 
%The photon direction is not directly measured in the LXe detector, rather it is inferred by the direction of a line between the photon interaction vertex in the LXe detector and 
%the intercept of the positron trajectory at the stopping target. 
%
%Liquid xenon, with its high density and short radiation length, is an efficient detection medium for photons; optimal resolution is achieved, at least at low energies, if both the ionisation and scintillation signals are detected. 
%
As shown in Fig.~\ref{introduction:megdet},
it has a C-shaped structure fitting the outer radius of the magnet.
The fiducial volume is $\approx$ 800~$\ell$, covering 11\% of the solid angle viewed from the centre of the stopping target in the radial range of $67.85<r<105.9$~cm, corresponding to $\approx 14$~radiation length. 
It is able to detect a 52.83-MeV $\gammaup$ with high efficiency and to contain the electromagnetic shower induced by it. 
 The scintillation light is detected by 846 2-inch PMTs submerged directly in the liquid xenon.
They are placed on all six faces of the detector, with different PMT coverage on different faces.
On the inner face, which is the densest part, the PMTs align at intervals of 6.2~cm.
%The detector's depth is 38.5~cm, corresponding to $\approx 14$~radiation length.

%\subsection{Trigger and DAQ}
One of the distinctive features of the MEG experiment is that it digitises and records all waveforms from the detectors using the Domino Ring Sampler v4 (DRS4) chip\,\cite{Ritt2010}. 
%It enables us to apply complicated algorithms to the acquired data and it makes it easy to reanalyze the data when we change algorithms or parameters.
The sampling speeds are set to 1.6~GSPS for TC and LXe photon detector  and 0.8~GSPS for DCH.
This lower value for DCH is selected to match the drift velocity and the required precision.

The DAQ event rate was kept below 10 Hz in order to acquire the full waveform data ($\approx$\,1~MB/event).
It was accomplished using a highly efficient online trigger system \cite{trigger2013, Galli:2014uga}.

Several types of trigger logic were implemented and activated during the physics data-taking each with its own prescaling factor.
%In the MEG experiment, the \meg dedicated trigger and triggers were implemented.
However, a dedicated trigger for the MEx2G events was neither foreseen nor implemented.
Thus, we rely on the \meg triggered data in this search.

\begin{figure}[tb]
    \centering
    \includegraphics[width=\columnwidth]{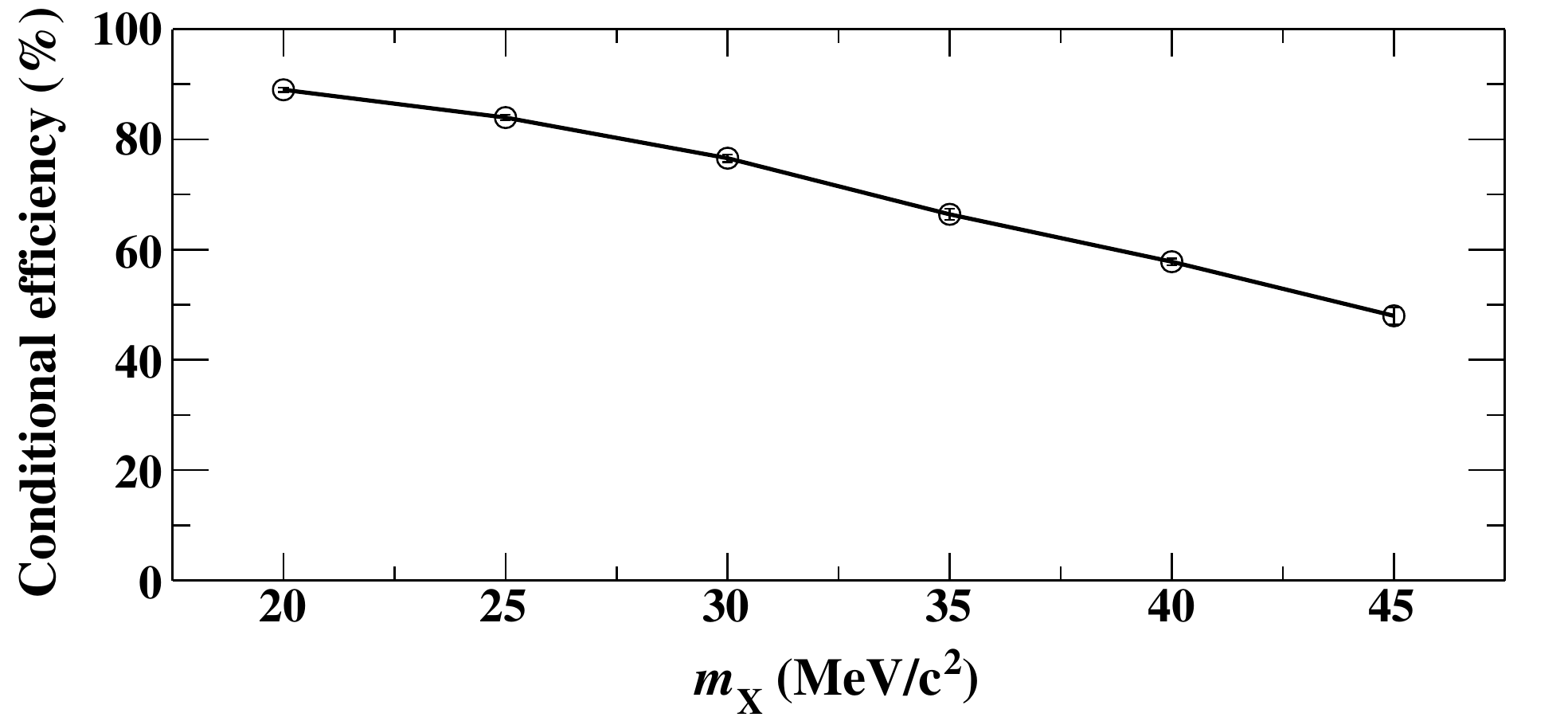}
    \caption{Trigger direction match efficiency for the MEx2G decay conditional to $\mathrm{e^+}$ and $2 \gammaup$ detection as a function of $m_\mathrm{X}$ evaluated with a Monte Carlo simulation (\sref{sec:simulation}).}
    \label{fig:DMEfficiency}
\end{figure}

The main \meg trigger, with a prescaling of 1, used the following observables: 
$\gammaup$ energy, time difference between e$^+$ and $\gammaup$, and relative direction of e$^+$ and $\gammaup$. The DC was not used in the trigger due to the slow drift velocity.
%A total charge of PMTs in the LXe detector is used as $\gammaup$ energy trigger to select high energy $\gammaup$ events.
%e$^+$ time is calculated from TC.
%The time difference trigger is used to select coincident events.
The condition on the relative direction is designed to select back-to-back events.
To calculate the relative direction, the PMT that detects the largest amount of scintillation light is used for the $\gammaup$, %assuming emission point locates at the centre of the target.
while the hit position at the TC is used for the e$^+$. %since the DCH information is too slow to be usable in the trigger. %making use of the fact that the TC bar IDs and the $z$ hit position correlate with the e$^+$ emission angle.
This direction match requirement results in inefficient selection of the MEx2G signal because, unlike the \meg decay, the MEx2G decay has 2$\gammaup$s with a finite opening angle, resulting in events often failing to satisfy the direction trigger.
The selection inefficiency for MEx2G events is 10--50\% depending on $m_{\mathrm{X}}$ as shown in \fref{fig:DMEfficiency}.

%DAQ systems needs to perform the complete read-out of all detector wavefroms while maintaining the system efficiency;the systems should have long live time and high online efficiency as much as possible.
%DAQ efficiency defined as the product of these two items was estimated to be  75\% (2009--2010) and 97\% (2011--2013).
%A multiple buffer scheme enabled us to improve the DAQ efficiency.

Finally, the detector has been calibrated and monitored over all data-taking period with various methods \cite{calibration_cw, papa_2007}, ensuring
that the detector performances have been under control over the duration of the experiment.

% --------------------------- 2_reconstruction -------------------------------

\section{Search strategy}
\label{sec:strategy}
%\textcolor{red}{To be updated. The signal kinematics and backgrounds should be summarized.
%The principle of background rejection with which variables  should be discussed.
%All the variables that will be reconstructed should be summarized.
%}

The MEx2G signal results from the sequential decays of $\muup^+\to\mathrm{e^+}\mathrm{X}$ followed by $\mathrm{X}\to\gammaup\gammaup$.
The first part is a two-body decay of a muon at rest, signalled by a mono-energetic e$^+$. The energy $E_\mathrm{e^+}$ is determined by $m_\mathrm{X}$: $E_\mathrm{e^+}(m_\mathrm{X}=0) = 52.83$~MeV and is a decreasing function of $m_\mathrm{X}$.
The sum of energies of the two $\gammaup$s is also mono-energetic and an increasing function of $m_\mathrm{X}$. %, and it is larger than 52.83~MeV.
The momenta of the two $\gammaup$s are Lorentz-boosted along the direction of X, which increases the acceptance in the LXe photon detector compared to the three-body decay $\muup^+\to\mathrm{e^+}\gammaup\gammaup$.
The final-state three particles is expected to have an invariant mass of 105.7~MeV$/\mathrm{c}^2 (=m_\muup)$ and the total momentum vector equal to 0.

A physics background that generates time-coincident $\mathrm{e^+}\gammaup\gammaup$ in the final state is $\radiative\gammaup$. This mode has not yet been measured but exists in the SM. The branching ratio is calculated to be $\sim \mathcal{O}(10^{-14})$ for the MEG detector configuration without any cut on $E_\mathrm{e^+}$ ~\cite{Pruna:2017upz,Banerjerr:2020}. Therefore, its contribution is certainly negligible in this search where we apply cuts on $E_\mathrm{e^+}$.

The dominant background is the accidental pileup of multiple $\muup^+$s decays. There are three types of accidental background events:
\begin{description}
    \item [\textbf{Type 1:}] The e$^+$ and one of the $\gammaup$s originate from one $\muup^+$, and the other $\gammaup$ from a different one.
    \item [\textbf{Type 2:}] The two $\gammaup$s share the same origin, and the e$^+$ is accidental.
    \item [\textbf{Type 3:}] All the particles are accidental.
\end{description}
The main source of a time-coincident $\mathrm{e^+}\gammaup$ pair in type 1 is the radiative muon decay \radiative \cite{megrmd}. 
The sources of time-coincident $\gammaup\gammaup$ pairs in type 2 are \aif (e$^+$ from $\muup^+$ decay and e$^-$ from  material along the e$^+$ trajectory), \radiative with an additional $\gammaup$, e.g. by a bremsstrahlung from the e$^+$\footnote{In the case of type 2, the e$^{+}$ can have low energy and be undetected.}, or a cosmic-ray induced shower.
%The rates of these accidental backgrounds can be estimated from the data in off-timing regions.

\begin{figure}[t!]
    \centering
    \includegraphics[width=\columnwidth]{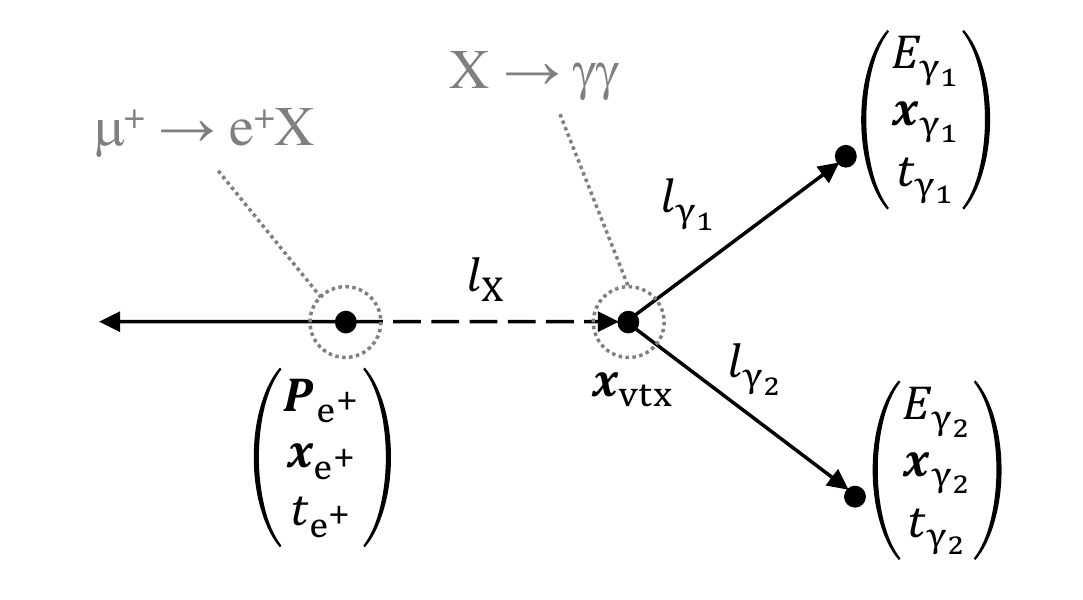}
    \caption{Decay kinematics and kinematic variables.}
    \label{fig:tdiff_rec}
\end{figure}

\Fref{fig:tdiff_rec} shows the decay kinematics and the kinematic variables.
The muon decay vertex and the momentum of the e$^+$ are obtained by reconstructing the e$^+$ trajectory using the hits in DCH and TC and the intersection of the trajectory with the plane of the muon beam stopping target (\sref{sec:positron_rec}).
The interaction positions and times of the two $\gammaup$s within the LXe photon detector and their energies are individually reconstructed using the PMT charge and time information of the LXe photon detector (\sref{sec:gamma_rec}).

Given the muon decay vertex, the two $\gammaup$s' energies and positions, and $m_\mathrm{X}$, the X decay vertex $\vector{x}_\mathrm{vtx}$ can be computed. Therefore, we reconstruct $\vector{x}_\mathrm{vtx}$ by scanning the assumed value of $m_\mathrm{X}$ (\sref{sec:vertex_rec}). 
If the final-state three particles do not originate at a single muon decay vertex, these variables will be inconsistent with originating from a single point.
After reconstructing $\vector{x}_\mathrm{vtx}$, the relative time and angles (momenta) between X and e$^+$ are tested for consistency with a muon decay (\sref{sec:momentum} and \ref{sec:relative_timing}).

The MEx2G decay search analysis is performed within 
the mass range 20~MeV/c$^2<m_\mathrm{X}<45~$MeV/c$^2$ at 1~MeV/c$^2$ step.
This step is chosen small enough not to miss signals in the gaps.
Therefore, adjacent mass bins are not statistically independent.
The analysis was performed assuming lifetimes $\tau_\mathrm{X}= 5, 20$, and $40$~ps; the value affects only the signal efficiency.

We estimate the accidental background by using the data in which the particles are not time coincident.
To reduce the possibility of experimental bias, a blind analysis is adopted; the blind region is defined in the plane of the relative times of the three particles (\sref{sec:dataset}). 

The signal efficiency is evaluated on the basis of a Monte Carlo simulation (\sref{sec:simulation}). Its tuning and validation are performed using pseudo-2$\gammaup$ data as described in \sref{sec:pseudo2gamma}.

%A blind analysis is used to reduce the experimenter's bias.
%We determine conditions of event selection and estimate the number of background events without looking at the signal region.
%The blind region is defined to be $|\teg| < 1\,\mathrm{ns} \land |\tgg| < 1\,\mathrm{ns}$\footnote{See \sref{sec:relative_timing} for the definitions.}, which is large enough to blind the signal. 
%The region is shown in blue shadow in \fref{fig:timesideband2}.

%As shown in \fref{fig:timesideband2}, we define time sideband regions A, B, and C and these regions are used as control samples to estimate backgrounds and check the analysis method. 
%See \sref{sec:BG_estimation} and \sref{sec:cut_condition} for the other definitions in \fref{fig:timesideband2}.
%\begin{figure}[t!]
%    \centering
%    \includegraphics[width=\columnwidth]{../3_analysis/timesideband2.pdf}
%    \caption[Definitions of blind region, signal region, and time sideband regions]{Definitions of blind region, signal region, and time sideband regions.}
%    \label{fig:timesideband2}
%\end{figure}

\section {Simulation}
\label{sec:simulation}

%\subsection{Monte Carlo simulation}
%%
The technical details of the program of Monte Carlo (MC) simulation are presented in \cite{softwaretns}
and an overview of the physics and detector simulation is available in \cite{megdet}.
In the following we report a brief summary.

%\subsection{Physics event simulation}

The first step of the simulation is the generation of the physics events.
That is realised with custom written code for a large number of relevant physics channels.
%The most relevant channels required by the \meg analysis are described in \cite{Adam2013}.    
%Additional decay channels have been added for this analysis.
The MEx2G decay is simulated starting from a muon at rest in the target; the decay products are 
generated in accordance with the decay kinematics for the given $m_\mathrm{X}$ and $\tau_\mathrm{X}$.

%\subsection{Detector simulation}

The muon beam transport, interaction in the target, and propagation of the decay
products in the detector are simulated with a MC program 
based on GEANT3.21 \cite{GEANT3} that describes the detector response.
%The detector package stops short of simulating the read-out, which is performed separately.
Between the detector simulation and the reconstruction program, an intermediate program processes the MC information, adding readout simulation and allowing event mixing 
to study the detector performance under combinatorial background events.
Particularly, the $\muup^+$ beam, randomly distributed in time at a decay rate of $3\times 10^{7}~\muup^+\mathrm{s^{-1}}$,
%and decaying within the SM framework
is mixed with the MEx2G decay to study the e$^+$ spectrometer performance.
The detectors' operating condition, such as the active layers of DCH and the applied high-voltages, are implemented with the known time dependence.

In order to simulate the accidental activity in the LXe photon detector, data collected with a random-time trigger are used.
%The MC event is overlaid with the random trigger data on the photoelectron basis.
A MC event and a random-trigger event are overlaid by summing the numbers of photo-electrons detected by each PMT.

\subsection{Pseudo two $\gammaup$ data}
\label{sec:pseudo2gamma}
To study the performance of the 2$\gammaup$ reconstruction, we built pseudo 2$\gammaup$ events using calibration data.
The following $\gammaup$-ray lines are obtained in calibration runs:
\begin{itemize}
\item 54.9~MeV and 82.9~MeV from $\piup^-\mathrm{p}\to\piup^0\mathrm{n}\to\gammaup\gammaup\mathrm{n}$ reaction,
\item 17.6~MeV and 14.6~MeV from $^7\mathrm{Li}(\mathrm{p},\gammaup)^8\mathrm{Be}$ reaction,
\item 11.7~MeV from $^{11}\mathrm{B}(\mathrm{p},2\gammaup)^{12}\mathrm{C}$ reaction.
\end{itemize}
The selection criteria for those calibration events are detailed in \cite{calibration_cw} and \cite{Adam2013}.
We take two events from the above calibration data and overlay them, summing the number of photo-electrons PMT by PMT. 
These pseudo 2$\gammaup$ events are generated using both data and MC events.

\section{Event reconstruction}
\label{sec:reconstruction}

%\begin{figure*}[t]
%	\centering
%	\includegraphics[width=0.8\textwidth]{../2_reconstruction/analysis_flow.pdf}
%	\caption[Overview of the reconstruction]{Overview of the event reconstruction. All the reconstructions start from waveform analysis. 
%	%Firstly, the reconstructions of each detector are performed separately, and their results are combined at the later step. 
%	The blue regions indicate detector-wise event reconstruction while the other regions indicate inter-detector reconstruction.
%	Detailed reconstruction methods will be described in the corresponding sections.
%	}
%	\label{fig:analysis_flow}
%\end{figure*}

We describe here the reconstruction methods and their performance, focusing on high-level objects; descriptions of the manipulation of low-level objects, including waveform analysis and calibration procedures, are available in \cite{baldini_2016,megdet}.
The e$^+$ reconstruction (\sref{sec:positron_rec}) is identical to that used in the \meg decay analysis in \cite{baldini_2016}.
The 2$\gammaup$ reconstruction was developed originally for this analysis (\sref{sec:gamma_rec}).
After reconstructing the e$^+$ and two $\gammaup$s, the reconstructed variables are combined to reconstruct the X decay vertex (\sref{sec:combined_rec}).
%The flow of reconstruction is summarized in \fref{fig:analysis_flow}.

\subsection{Positron reconstruction}
\label{sec:positron_rec}

Positron trajectories in the DCH are reconstructed using the Kalman filter technique~\cite{Billoir1984,Fruhwirth1987} based on the GEANE software~\cite{Fontana2008}.
This technique takes the effect of materials into account.
After the first track fitting in DCH, the track is propagated to the TC region to test matching with TC hits.
The matched TC hits are connected to the track and then the track is refined using the TC hit time.
Finally, the fitted track is propagated back to the stopping target, and the point of intersection with the target defines the muon decay vertex position ($\vector{x}_\mathrm{e^+}$) and momentum vector that defines the e$^+$ emission angles ($\theta_\mathrm{e^+}, \phi_\mathrm{e^+}$). The e$^+$ emission time ($t_\mathrm{e^+}$) is reconstructed from the TC hit time minus the e$^+$ flight time.

%\subsubsection{TC reconstruction}
%The signals from TC PMTs are processed using Double Thresholds Discriminators (DTD) to minimise the time-walk effect.
%The DTD outputs a NIM pulse at the timing when the input signal crosses the lower threshold if the signal is higher than the higher threshold.
%A TC hit is reconstructed if both PMTs in a bar have signals higher than the higher threshold of the DTD.
%The output NIM pulses are fitted with a template waveform to extract the timing information ($t_{\mathrm{IN}}$ and $t_{\mathrm{OUT}}$), where IN (OUT) corresponds to the PMT close to (far from) the target.
%Then TC hit time is given by
%\bea
%t_{\mathrm{TC}}=\frac{t_{\mathrm{IN}}+t_{\mathrm{OUT}}}{2}-\frac{b_{\mathrm{IN}}+b_{\mathrm{OUT}}}{2}-\frac{w_{\mathrm{IN}}+w_{\mathrm{OUT}}}{2}-\frac{L}{2 v_\mathrm{eff}},
%\nonumber
%\eea
%where $b$ is the time offset and $W$ is a correction value of the time-walk effect.
%$v_\mathrm{eff}$ is the effective velocity of the scintillation light inside the scintillator bar and $L$ is the length of the bar.
%The hit position along the bar is given by
%\bea
%z_{\mathrm{TC}}=\frac{v_\mathrm{eff}}{2}\left\{\left(t_{\mathrm{IN}}-t_{\mathrm{OUT}}\right)-\left(b_{\mathrm{IN}}-b_{\mathrm{OUT}}\right)-\left(w_{\mathrm{IN}}-w_{\mathrm{OUT}}\right)\right\}.
%\nonumber
%\eea

%%%% track selection %%%%
Positron tracks satisfying the following criteria are selected:
the number of hits in DCH is more than six, the reduced chi-square of the track fitting is less than 12, the track is matched with a TC hit, and the track is successfully propagated back to the fiducial volume of the target.
If multiple tracks in an event pass the criteria, only one track is selected and passed to the following analysis, based on the covariance matrix of the track fitting as well as the number of hits and the reduced chi-square.

%%%%% performance %%%%
\begin{figure}[tb!]
    \centering
    \includegraphics[width=\columnwidth]{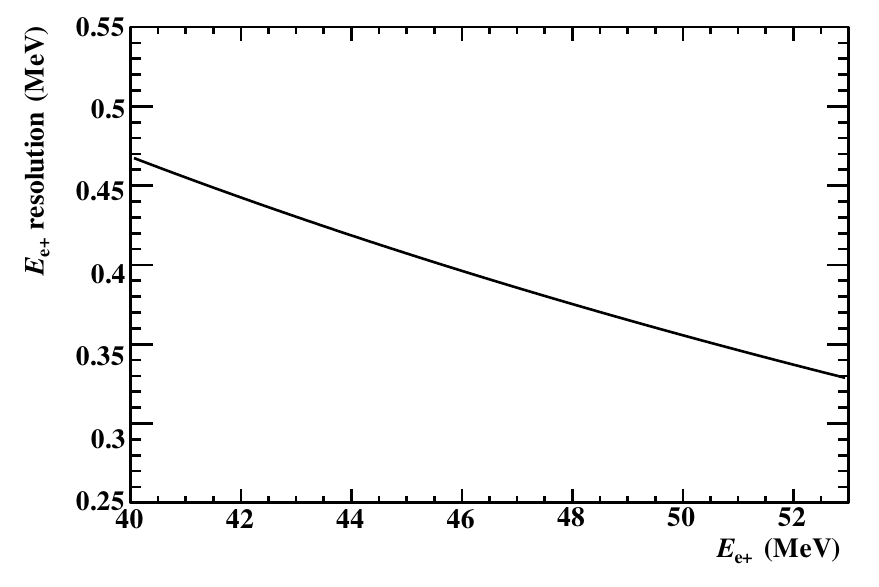}
    \caption{$E_{\mathrm{e^+}}$ resolution as a function of $E_{\mathrm{e^+}}$.}
    \label{fig:positronEnergyResolution}
\end{figure}
The resolutions are evaluated based on the MC, tuned to data using double-turn events;
tracks traversing DCH twice (two turns) are selected and reconstructed independently by using hits belonging to each turn.
The difference in the reconstruction results by the two turns indicates the resolution. The MC results are smeared so that the 
double-turn results become the same as those with the data.
\Fref{fig:positronEnergyResolution} shows the $E_\mathrm{e^+}$ resolution as a function of $E_\mathrm{e^+}$.  
The angular resolutions also show a similar $E_\mathrm{e^+}$ dependence. The $\phi_\mathrm{e^+}$- and $\theta_\mathrm{e^+}$-resolutions for $m_\mathrm{X}=20\, (45)$~MeV/c$^2$ are 
$\sigma_{\phi_\mathrm{e^+}}\sim 12\,(15)$~mrad and $\sigma_{\theta_{\mathrm{e^+}}}\sim 10\,(11)$~mrad, respectively.
The time resolution is
$\sigma_{t_\mathrm{e^+}}\sim 100\,(130)$~ps.

\subsection{Photon reconstruction}
\label{sec:gamma_rec}

Coordinates $(u, v, w)$ are used in the LXe photon detector local coordinate system rather than the global coordinates $(x, y, z)$: $u$ coincides with $z$, 
%$v$ is directed along the negative $\phi$-direction at the radius of the inner face ($r_{\mathrm{in}}=67.85\,$cm) (bottom to top along the inner face), 
$v = r_\mathrm{in}(\pi - \phi)$ where $r_{\mathrm{in}}=67.85$~cm is the radius of the inner face,
and $w=r-r_{\mathrm{in}}$ is the depth measured from the inner face.
%
%The reconstruction starts from waveform analysis.
%The charge and time of each PMT are obtained from the waveform of each PMT.
%A digital constant fraction method is used to determine the time of the signal.
%The crossing time is calculated by interpolating adjacent two points.
%The charge is calculated by integrating the waveform.
%
%Next, the charge is converted into the number of photoelectrons by using a measured PMT gain.
%The number of photoelectrons of $i$-th PMT ($N_{\mathrm{pe}, i}$) is given by
%\bea
%N_{\mathrm{pe}, i} = \frac{Q_i}{e\cdot G_i},
%\eea
%where $Q_i$ is the measured charge and $G_i$ is the gain of the PMT.
%Then, the number of photoelectrons is converted into the number of photons by using a measured PMT quantum efficiency (QE):
%\bea
%N_{\mathrm{pho}, i} = \frac{N_{\mathrm{pe}, i}}{QE_i}.
%\eea
\begin{figure*}[tb]
    \centering
    \includegraphics[width=\linewidth]{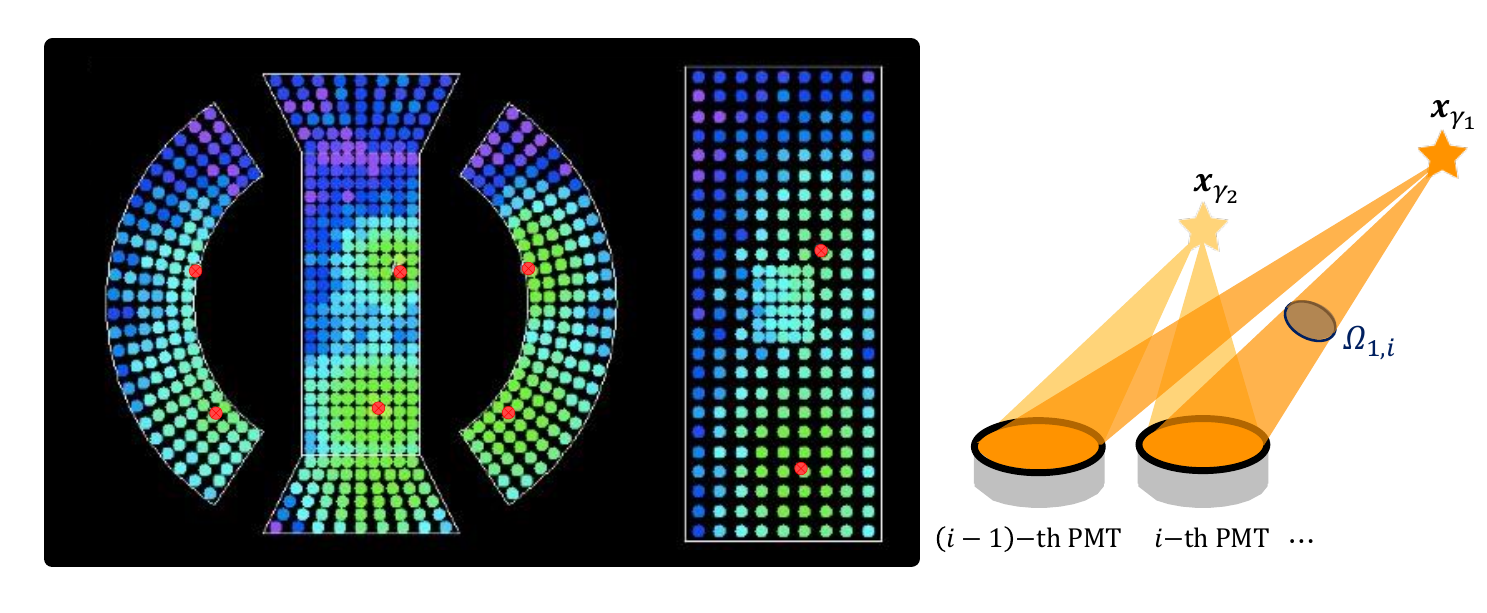}
    \caption[Position and energy reconstruction for 2$\gammaup$s]{Event display of the LXe photon detector for a 2$\gammaup$ event (in a development view). The red points show the interaction positions of the two $\gammaup$s projected to each face. Each circular marker denotes a PMT. The colour indicates the measured light yield, which is the sum of photons from the two showers induced by the two $\gammaup$s as depicted in the right figure.}
    \label{fig:2g_rec_position_energy}
\end{figure*}

\subsubsection{Multiple photon search}
\label{sec:peak_search}
A peak search is performed based on the light distributions on the LXe photon detector inner and outer faces by using {\it TSpectrum2}\,\cite{TSpectrum2,Morhac:2000}.
%Peaks in the two-dimensional histogram of the number of photons are searched for.
The threshold of the peak light yield is set to 200 photons.
Events that have more than one peak are identified as multiple-$\gammaup$ events.

\subsubsection{Position and energy}
\label{sec:2gamma_position_energy}
Hereafter, only the multiple-$\gammaup$ events are analysed.
%When the number of found peaks is larger than two, the following the fittings are iterated and the largest two peaks are selected.
When more than two $\gammaup$s are found, we select the two with the largest energy by performing the position-energy fitting described in this subsection on different combinations of two $\gammaup$s.
 
\Fref{fig:2g_rec_position_energy} shows a typical event display of a 2$\gammaup$ event.
Each PMT detects photons from the two $\gammaup$s.
The key point of the 2$\gammaup$ reconstruction is how to divide the number of photons detected in each PMT into a contribution from each $\gammaup$.

\paragraph*{Calculation of initial values}\,
First, the positions of the detected peaks in ($u, v$) are used as the initial estimate with $w=1.5$~cm.
Given the interaction point of each $\gammaup$ within the LXe photon detector, the contribution from each $\gammaup$ to each PMT can be calculated as follows.
Assuming the ratio of the energy of $\gammaup_1$ to that of $\gammaup_2$ to be $E_{\gammaup_1}\colon E_{\gammaup_2}=R_1:(1-R_1)$ ($0<R_1<1$, at first $R_1$ is set to 0.5), the fractions of the number of photons from $\gammaup_1$ is calculated as 
\paragraph*{}
\vspace{-2\baselineskip}
\bea
%\[
%\begin{empheq}[left={
R_{1,i}=%\empheqlbrace}]{align}
    \frac{R_1\Omega_{1,i}}{R_1\Omega_{1,i} + (1 - R_1)\Omega_{2,i}},  %\quad w\leq 2.0~\mathrm{cm} \label{eq:solidr} \\
%    \frac{RD_{1,i}}{RD_{1,i} + (1 - R)D_{2,i}} & \quad w\geq 2.0~\mathrm{cm} \label{eq:disr}
%\end{empheq}
%R'_{1,i} = \left\{ 
%\begin{eqnarray}
%\frac{R\Omega_{1,i}}{R\Omega_{1,i} + (1 - R)\Omega_{2,i}} & w\leq & 2.0~\mathrm{cm} \label{eq:solidr} \\
%\frac{RD_{1,i}}{RD_{1,i} + (1 - R)D_{2,i}} & w\geq 2.0~\mathrm{cm} & \label{eq:disr}
%\end{eqnarray}
%\right.
%\frac{R\Omega_{1,i}}{R\Omega_{1,i} + (1 - R)\Omega_{2,i}}, \qquad\qquad %w\leq 2.0~\mathrm{cm}
\label{eq:solidr}
\eea
%or
%\bea
%R'_{1,i }= \frac{RD_{1,i}}{RD_{1,i} + (1 - R)D_{2,i}}, \qquad \qquad w\geq 2.0~\mathrm{cm}
%\label{eq:disr}
%\eea
%\]
where $\Omega_{1,i}$ is the solid angle subtended by the $i$-th PMT from the $\gammaup_1$ interaction point%, and $D_{1,i}$ is the distance between them.
%\Eref{eq:solidr} is used for shallow events and \eref{eq:disr} is used for deeper events fow which the solid angle is less sensitive.
.
%By using the ratio $R_{1,i}$ and the number of photons at each PMT $N_{\mathrm{pho},i}$, the initial value of the total number of photons generated by $\gammaup_{1(2)}$, $M_{\mathrm{pho}, 1(2)}$, is calculated.
The total number of photons generated by $\gammaup_{1(2)}$, $M_{\mathrm{pho}, 1(2)}$, is calculated from the  ratio $R_{1,i}$ and the number of photons at each PMT $N_{\mathrm{pho},i}$ as
\be
M_{\mathrm{pho}, 1(2)} = \sum_i^{n_\mathrm{{PMT}}^\mathrm{all}}\left( R_{1,i} \times N_{\mathrm{pho},i} \right).
\label{eq:mpho}
\ee
Then, $R_1$ is updated to $R_1 = M_{\mathrm{pho}, 1} / (M_{\mathrm{pho}, 1} + M_{\mathrm{pho}, 2})$ and 
calculations (\ref{eq:solidr}) and (\ref{eq:mpho}) are repeated with the updated $R_1$. This procedure is iterated four times.

\paragraph*{Position pre-fitting}\,
Inner PMTs that detect more than 10~photons are selected to perform a position pre-fitting. %is performed to reproduce light distributions of the LXe detector.
The following quantity is minimised during the fitting:
\bea
\chi^2_{2\gammaup}=\sum_i^{n_\mathrm{{PMT}}^\mathrm{selected}}\frac{\left(N_{\mathrm{pho}, i}-M_{\mathrm{pho},1}\Omega_i(\vector{x}_{\gammaup_1})-M_{\mathrm{pho}, 2}\Omega_i(\vector{x}_{\gammaup_2})\right)^2}{\sigma^2_{\mathrm{pho}, i}(N_{\mathrm{pho}, i})},
\label{eq:prefit}
\eea
where $\sigma^2_{\mathrm{pho}, i}(N_{\mathrm{pho}, i}) = N_{\mathrm{pho}, i}/\epsilon_{\mathrm{PMT},i}$ with $\epsilon_{\mathrm{PMT},i}$ being the product of quantum and collection efficiencies of the PMT.  % = N_{\mathrm{pe},i}$
This fitting is performed\footnote{This fitting is performed by a grid search in $\vector{x}_{\gammaup_{1(2)}}=(u, v, w)_{\gammaup_{1(2)}}$ space for good stability, while subsequent fittings are performed with MINUIT \cite{MINUIT} for better precision.} 
separately for each $\gammaup$: first, the light distribution is fitted with $\{\vector{x}_{\gammaup_1}, M_{\mathrm{pho},1}\}$ as free parameters,
while the other parameters are fixed; next, the light distribution is fitted with $\{\vector{x}_{\gammaup_2}, M_{\mathrm{pho},2}\}$ as free parameters, while the other parameters are fixed.
%The minimisation is performed by a grid-search in $\vector{x}_{\gammaup_{1(2)}}$ space. %while the other parameters are fixed to the initial values.

%Next, to find a better parameter set for each $\gammaup$, the following $\chi^2$ is minimised:
%\bea
%\chi^2=\sum_i^{\mathrm{nPMT}}\frac{\left[N_{\mathrm{pho}, i}-M_{\mathrm{pho},1}\times\Omega_i(\vector{x}_{\gammaup_1})-M_{\mathrm{pho}, 2}\times\Omega_i(\vector{x}_{\gammaup_2})\right]^2}{\sigma_{\mathrm{pho}, i}(N_{\mathrm{pho}, i})^2}.
%\label{eq:fit}
%\eea

\paragraph*{Energy pre-fitting}\,
To improve the energy estimation,  $M_{\mathrm{pho},1(2)}$ are fitted while the other parameters are fixed.
The same $\chi^2_{2\gammaup}$ (\eref{eq:prefit}) is used but only with PMTs that detect more than 200 photo-electrons.
%This time, the minimisation is performed using MINUIT \cite{MINUIT} instead of the grid search.

%If there are 3 or more peaks, the two largest peaks are selected and 
The $\gammaup$ with the larger $M_\mathrm{pho}$ is defined as $\gammaup_1$ and the second largest one is defined as $\gammaup_2$ in the later analysis.

\paragraph*{Position and energy fitting}\,
At the final step, all the parameters are fitted simultaneously to eliminate the dependence of the fitted positions on the value of $M_{\mathrm{pho},1(2)}$ initially assumed.
The best-fit value of $M_{\mathrm{pho},1(2)}$ is used to update $R_1$ and calculations (\ref{eq:solidr}) and (\ref{eq:mpho}) are repeated again to 
obtain the final value of $M_{\mathrm{pho},1(2)}$.
Finally, it is converted into $E_{\gammaup_{1(2)}}$:
\bea
E_{\gammaup_{1(2)}}&=&U(\vector{x}_{\gammaup_{1(2)}})\times H(T) \times S \times M_\mathrm{pho, 1(2)},
\eea
where $U(\vector{x}_{\gammaup_{1(2)}})$ is a uniformity correction factor, $H(T)$ is a time variation correction factor with $T$ being the calendar time when the event was collected, and $S$ is a factor to convert the number of photons to energy.
The functions  $U(\vector{x}_{\gammaup_{1(2)}})$  and $H(T)$  are mainly derived from the 17.6-MeV line from $^7\mathrm{Li}(\mathrm{p},\gammaup)^8\mathrm{Be}$ reaction, which was measured twice per week.
The factor $S$ is calibrated using the 54.9-MeV line from $\piup^0$ decay, taken once per year.

\begin{figure}[tb]
    \centering
    \includegraphics[width=\columnwidth]{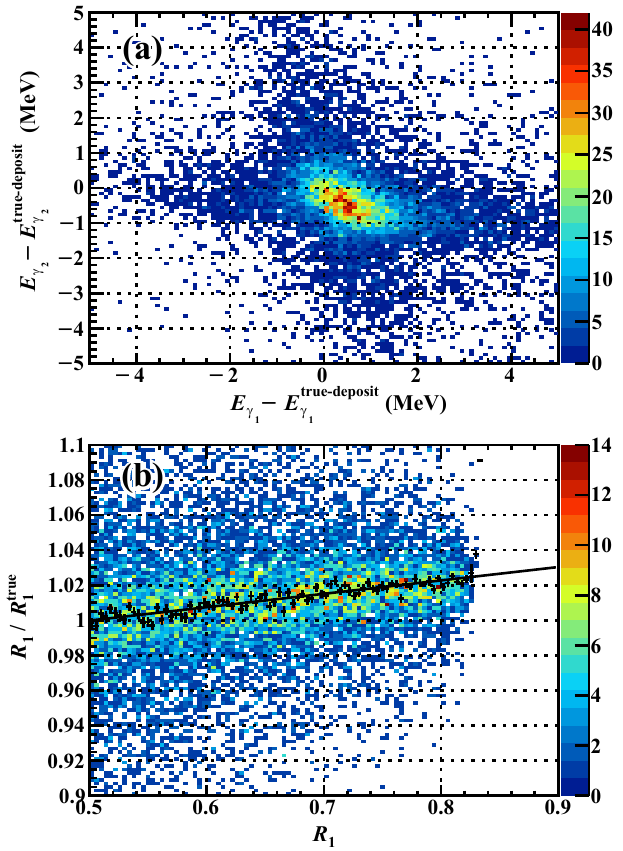}
    \caption{(a) Scatter plot of the energy reconstruction errors (MC). $E^\mathrm{true-deposit}_{\gammaup_{1(2)}}$ is the MC true value of the energy deposited in the LXe. (b) Dependence of the reconstructed energy ratio bias as a function of the reconstructed energy ratio (MC).}
    \label{fig:ratio_cor}
\end{figure}

\paragraph*{Energy-ratio correction}\,
Both the MC data and the pseudo-2$\gammaup$ data show an anti-correlation between the errors\footnote{Error is defined as the difference between the reconstructed energy and the true energy deposit for MC data and between the reconstructed one with 2$\gammaup$ and that with single $\gammaup$ for pseudo-2$\gammaup$ data.} 
in $E_{\gammaup_{1}}$ and $E_{\gammaup_{2}}$ as shown in \fref{fig:ratio_cor}a, while their sum is not biased. %This is specific to the two $\gammaup$ fitting.
Defining %$R_{1}^{\mathrm{rec}}$ as the  $R_1$ for reconstructed energies and
$R_{1}^{\mathrm{true}}$ as the $R_1$ for true energies for MC data and that for energies reconstructed without the overlay for real data,
the reconstruction bias in both the MC data and the pseudo-2$\gammaup$ data is apparent by the linear dependence of $R_{1}/R_{1}^{\mathrm{true}}$
on $R_{1}$ as shown in \fref{fig:ratio_cor}b.
This bias is removed by applying a correction to the reconstructed energies; the correction coefficients are evaluated from the pseudo-2$\gammaup$ data with different combinations of calibration data.

\paragraph*{Position correction}\,
Oblique incidence of $\gammaup$s to the inner face results in a bias of the fitted positions. This bias was checked and corrected for using the MC simulation.
No bias is observed in the $v$ direction while a significant bias is observed in the $u$ direction. This is because the $\gammaup$s from the MEx2G decay enter the LXe photon detector almost perpendicularly in the $x$-$y$ view but enter with angles in the $z$-$r$ view. Since the $u$ bias arises from the direction and the size of the shower, it depends on the $u$ coordinate and the energy. 
Therefore, the correction function is prepared as a function of $u_{\gammaup_{1(2)}}$ and $E_{\gammaup_{1(2)}}$.

\paragraph*{Selection criteria}\, 
To guarantee the quality of the reconstruction, the following criteria are imposed on the reconstruction results: the fits for both $\gammaup$s converge; the two $\gammaup$ positions are both within the detector fiducial volume defined as $|u|<25$~cm $\land$ $|v|<71$~cm; the distance between the two $\gammaup$s on the inner face is $d_{uv}> 20$~cm; $E_{\gammaup_{1(2)}} > 10$~MeV; and $E_{\gammaup_1} + E_{\gammaup_2} > 40$~MeV. 

\paragraph*{Probability density function for $E_{\gammaup}$}\,
%\paragraph*{Resolutions}\,
%\begin{figure}[tb!]
%    \centering
%    \includegraphics[width=\columnwidth]{../2_reconstruction/SignalMC_energy2.pdf}
%    \caption{$E_{\gammaup_{1(2)}}$ resolution as a function of the energy deposit in LXe. The bands show systematic uncertainties stemming from the MC smearing parameters.}
%    \label{fig:gammaEnergyResolution}
%\end{figure}
%
\begin{figure}[tb]
    \centering
    \includegraphics[width=\columnwidth]{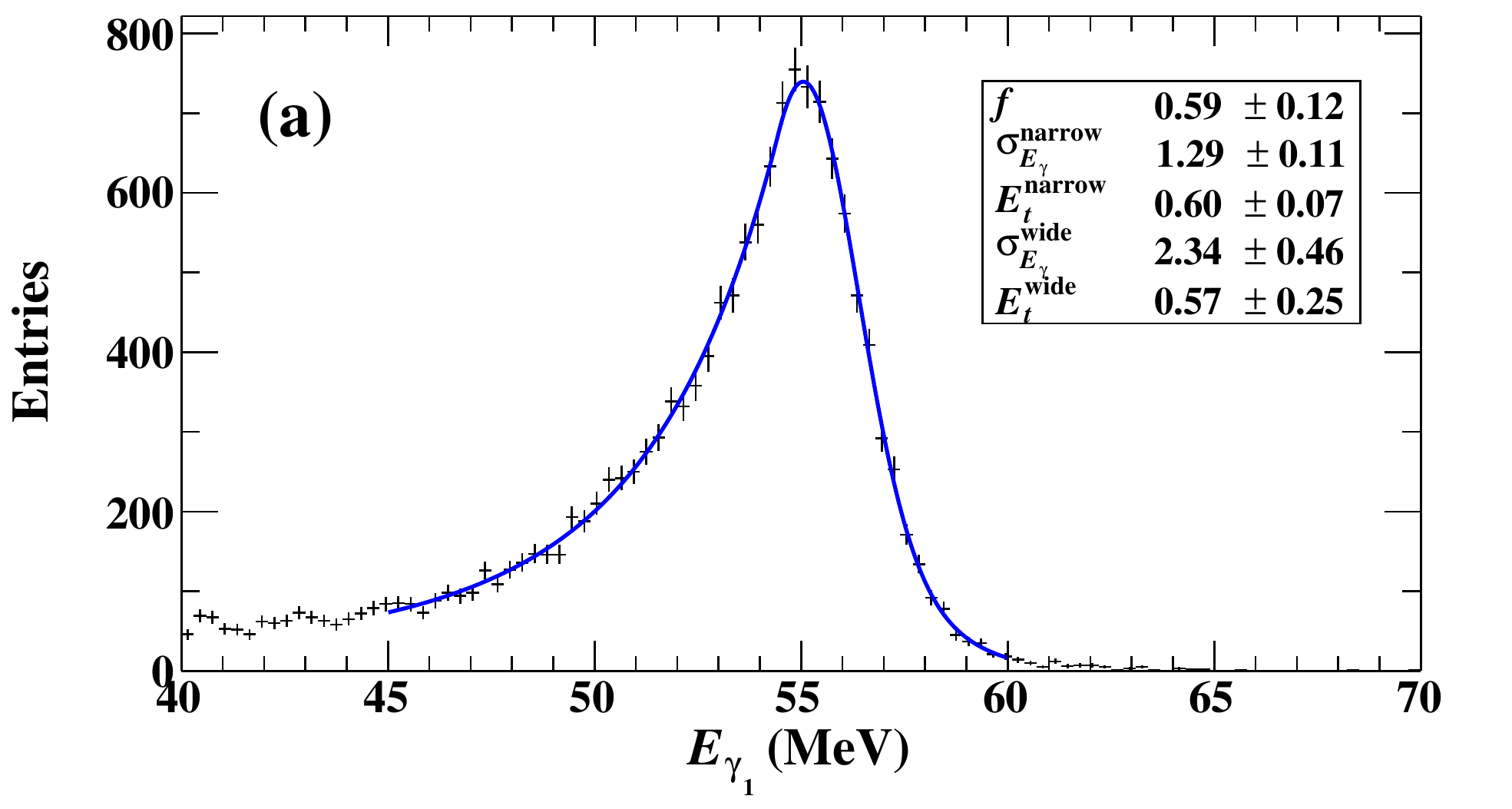}
    \includegraphics[width=\columnwidth]{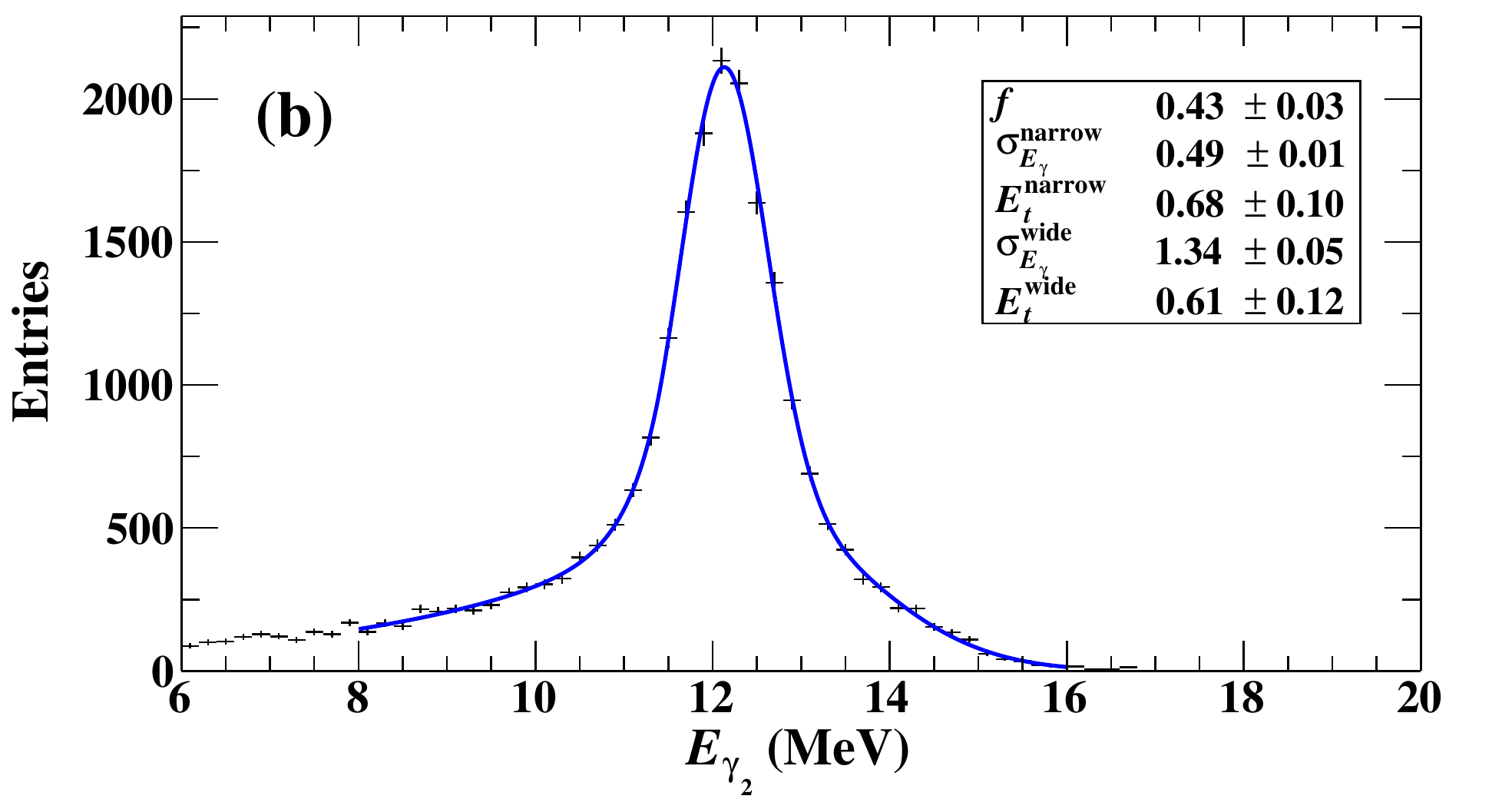}
    \caption{Energy response to MC $2\gammaup$ events with $E^\mathrm{true}_{\gammaup_{1}}=55$~MeV and  $E^\mathrm{true}_{\gammaup_{2}}=12$~MeV. The blue curves are the PDFs fit to the distributions. See text for the formula of the PDFs.}
    \label{fig:gammaEnergyResolution}
\end{figure}
The %$\gammaup$ energy line shape 
probability density function (PDF) for $E_{\gammaup_{1(2)}}$
is evaluated by means of the MC simulation.
To tune the MC, the pseudo-2$\gammaup$ data of MC and data are used.
%It is asymmetric with a lower tail and modelled by the sum of two (core and tail) components of the following function:
%\bea
%P(E_\gammaup)=
%\left\{
%\begin{array}{ll}
%A \exp \left\{-\frac{\left(E_\gammaup-E_0\right)^2}{2\sigma^2_{E_\gammaup}}
%\right\}
%& E_\gammaup>E_0+E_1 \\
%A \exp \left\{\frac{E_1}{\sigma^2_{E_{\gammaup}}} \left(\frac{E_1}{2}-(E_\gammaup-E_0)\right)\right\} & E_\gammaup \leq E_0+E_1
%\end{array}
%\right.,
%\label{eq:egammaresponse}
%\eea
%where $A$ is a scale parameter, $E_\gammaup$ is a generic $\gammaup$ energy, $E_0$ is the peak position, $E_1$ is the transition parameter, and $\sigma_{E_{\gammaup}}$ is the standard deviation of the Gaussian component describing the width on the high-energy side.
%\Fref{fig:gammaEnergyResolution} shows the $E_{\gammaup_{1(2)}}$ resolution evaluated with $\sigma_{E_{\gammaup}}$ for the two  components.
%
It is asymmetric with a lower tail and modelled as follows:
\begin{align}
P(E_\gammaup \mid E_\gammaup^\mathrm{true})
 &=    f\cdot F(E_\gammaup ;  E_\gammaup^\mathrm{true}, E_{t}^\mathrm{narrow}, \sigma_{E_{\gammaup}}^\mathrm{narrow}) \nn
 &+ (1-f)\cdot F(E_\gammaup ;  E_\gammaup^\mathrm{true}, E_{t}^\mathrm{wide}, \sigma_{E_{\gammaup}}^\mathrm{wide}), 
\label{eq:egammaPDF}
\end{align}
where
\begin{align}
&F(E_\gammaup ; E_\gammaup^\mathrm{true}, E_{t}, \sigma_{E_{\gammaup}}) \nn
&=
\begin{cases}
A \exp \left(-\frac{\left(E_\gammaup- E_\gammaup^\mathrm{true}\right)^2}{2\sigma^2_{E_\gammaup}}
\right)
& E_\gammaup> E_\gammaup^\mathrm{true} - E_{t} \\
A \exp \left(\frac{E_{t}}{\sigma^2_{E_{\gammaup}}} \left(\frac{E_{t}}{2} + (E_\gammaup- E_\gammaup^\mathrm{true})\right)\right)
& E_\gammaup \leq  E_\gammaup^\mathrm{true} - E_{t}
\end{cases}
,
\label{eq:egammaresponse}
\end{align}
$E_\gammaup$ is a reconstructed $\gammaup$ energy, $E_\gammaup^\mathrm{true}$ is the true value, $f$ is the fraction of the narrow component, $A$ is a normalisation parameter, 
$E_{t}$ is the transition parameter between the Gaussian and exponential components,
and $\sigma_{E_{\gammaup}}$ is the standard deviation of the Gaussian component describing the width on the high-energy side.
The parameters $E_{t}$ and $\sigma_{E_{\gammaup}}$ are correlated with each other, different for the narrow and wide components, and are %functions of 
dependent on $E_\gammaup^\mathrm{true}$.
%\Fref{fig:gammaEnergyResolution} shows 
%%the $E_{\gammaup_{1(2)}}$ resolution evaluated with 
%$\sigma_{E_{\gammaup}}$ for the two  components.
\Fref{fig:gammaEnergyResolution} shows an example of the PDFs for $2\gammaup$ events with  $E^\mathrm{true}_{\gammaup_{1}}=55$~MeV and  $E^\mathrm{true}_{\gammaup_{2}}=12$~MeV.

\paragraph*{Probability density functions for $\gammaup$ position}\,
The PDFs of $\gammaup$ position are almost 
independent of $E_{\gammaup_{1(2)}}$ and hence $(m_\mathrm{X}, \tau_\mathrm{X})$.
They are represented by double Gaussians with
fractions of tail components of $\sim\!20$\%. 
The standard deviations of the core components are %$\sigma_u \sim 5.4$~mm, $\sigma_v \sim 4.7$~mm, and $\sigma_w\sim 6.5$~mm.
$\vector{\sigma}^\mathrm{core}_{\vector{x}_{\gammaup_{1(2)}}}=(5.4, 4.7, 6.5)$~mm in $(u,v,w)$ coordinates, those of the tail components are $\vector{\sigma}^\mathrm{tail}_{\vector{x}_{\gammaup_{1(2)}}}=(29, 19, 45)$~mm.

\subsubsection{Time}
\label{sec:2gamma_time}
The interaction time of $\gammaup_1(\gammaup_2)$ can be reconstructed using the pulse time measured by each PMT ($t_{\mathrm{PMT}, i}$)
by correcting for a delay time ($t_{\mathrm{delay}, \gammaup_{1(2)},i}$) including the propagation time of the light between the interaction point 
and the PMT and the time-walk effect, and
 a time offset due to the readout electronics ($t_{\mathrm{offset}, i}$):
\bea
t_{\gammaup_{1(2)},i}=t_{\mathrm{PMT}, i}-t_{\mathrm{delay}, \gammaup_{1(2)},i}-t_{\mathrm{offset}, i}.
\eea
The single PMT time resolution $\sigma_{t, i}$ is approximately proportional to $1/\sqrt{N_{\mathrm{pe}, \gammaup_{1(2)},i}}$ with 
$\sigma_{t, i}(N_{\mathrm{pe}, \gammaup_{1(2)},i}=500)\approx 500~\mathrm{ps}$, where $N_{\mathrm{pe}, \gammaup_{1(2)},i}$ is the number of photo-electrons from $\gammaup_1(\gammaup_2)$.

These individual PMT measurements are combined to obtain the best estimate of the interaction time of $\gammaup_1(\gammaup_2)$ ($t_{\gammaup_{1(2)}}$). 
The following $\chi^2$ is minimised:
\bea
\chi^2_{\mathrm{time}}=\sum_i^{n_\mathrm{PMT}^\mathrm{selected}}\frac{\left(t_{\gammaup_{1(2)},i}-t_{\gammaup_{1(2)}}\right)^2}{\sigma^2_{t, i}(N_{\mathrm{pe}, \gammaup_{1(2)},i})}.
\label{chi2_time}
\eea
%where $t_{\gammaup_{1(2)}}$ is the interaction time of $\gammaup_{1(2)}$ to be fitted in this minimisation.
We use PMTs whose light yield from $\gammaup_1(\gammaup_2)$ is 5 times higher than that from $\gammaup_2(\gammaup_1)$ excluding PMTs whose light yield is less than 100 photons or which give large $\chi^2$ contribution in the fitting.

%\subsubsection{Performance}
The $E_\gammaup$-dependent time resolution for single $\gammaup$ event is evaluated with the calibration runs and corrected for 2$\gammaup$ events using the MC: 
\bea
\sigma_{t_{\gammaup_{1(2)}}} = \sqrt{338^2/E_{\gammaup_{1(2)}}\mathrm{(MeV)} + 45^2} \  ~\mathrm{(ps)}.
 \eea

%\paragraph*{Efficiency}\,

%\subsubsection{Performance}
%\paragraph*{Resolution}\,
%\paragraph*{Efficiency}\,
 
\subsection{Combined reconstruction}
\label{sec:combined_rec}
In this section, we present the reconstruction method for the $\mathrm{X}\rightarrow\gammaup\gammaup$ vertex
%(not $\muup^+\rightarrow\mathrm{e^+X}$ vertex) is presented.
assuming a value for $m_{\mathrm{X}}$ in the reconstruction.
We scan $m_{\mathrm{X}}$ in 20--45 MeV/c$^2$ at 1 MeV/c$^2$ intervals; each assumed mass results in a different reconstructed $\mathrm{X}\rightarrow\gammaup\gammaup$ vertex position.

\subsubsection{X decay vertex}
\label{sec:vertex_rec}
A maximum likelihood fit is used in the reconstruction, with the following
observables:
\bea
X = (E_{\gammaup_1}, E_{\gammaup_2}, \vector{x}_{\gammaup_1}, \vector{x}_{\gammaup_2}, \vector{x}_\mathrm{e^+}, \theta_\mathrm{e^+}, \phi_\mathrm{e^+}).
\eea
The fit parameters are the following:
\bea
\Theta = (\cos \theta_{\mathrm{rest}}, \phi_{\mathrm{rest}}, \vector{x}_{\mathrm{vtx}}),
\eea
where $\theta_\mathrm{rest}$ is the $\gammaup$ emission angle in the  X rest frame, % (in this frame, 2$\gammaup$s are emitted in the opposite directions)
$\phi_\mathrm{rest}$ is the angle of the photons in the X rest frame with respect to the X momentum direction in the MEG coordinate system,
%at X-rest frame;
and $\vector{x}_{\mathrm{vtx}}$ is the X decay vertex position.
The function $L(\Theta)$ is defined as 
follows:
%the product of the likelihood with the decay length constraint:
\bea
L(\Theta)
%&=&P(E_{\gammaup_1}, E_{\gammaup_2}, \vector{x}_{\gammaup_1}, \vector{x}_{\gammaup_2}, \theta_\mathrm{e^+}}, \phi_\mathrm{e^+}}, \mid \cos \theta_{\mathrm{rest}}, \phi_{\mathrm{rest}}, \vector{x}_{\mathrm{vtx}}, \vector{x_\mathrm{e^+}}}, m_\mathrm{X}) \\
&=&P(E_{\gammaup_1}\mid \cos \theta_{\mathrm{rest}}, m_\mathrm{X}) \nn
&\times& P(E_{\gammaup_2}\mid \cos \theta_{\mathrm{rest}}, m_\mathrm{X}) \nn
&\times& P(\vector{x}_{\gammaup_1}\mid \cos \theta_{\mathrm{rest}}, \phi_{\mathrm{rest}}, \vector{x}_{\mathrm{vtx}},\vector{x}_\mathrm{e^+}, m_\mathrm{X})\nn
&\times& P(\vector{x}_{\gammaup_2}\mid \cos \theta_{\mathrm{rest}}, \phi_{\mathrm{rest}}, \vector{x}_{\mathrm{vtx}},\vector{x}_\mathrm{e^+}, m_\mathrm{X})\nn
&\times& P(\theta_\mathrm{e^+} \mid \vector{x}_\mathrm{vtx}, \vector{x}_\mathrm{e^+})\nn
&\times& P(\phi_\mathrm{e^+} \mid \vector{x}_\mathrm{vtx}, \vector{x}_\mathrm{e^+})\nn
&\times& P(l_\mathrm{X} \mid \vector{x}_\mathrm{vtx}, \vector{x}_\mathrm{e^+}, \tau_\mathrm{X}, m_\mathrm{X}),
\label{eq:likelihood}
\eea
%$E_{\gammaup_1}$ and $E_{\gammaup_2}$ are assumed to be independent thanks to the ratio correction in the 2$\gammaup$ reconstruction.
where $l_\mathrm{X}$ is the X decay length.
The term $P(\vector{x}_\mathrm{e^+}\mid \vector{x}_\mathrm{e^+}^{\mathrm{true}})$ is omitted by approximating $\vector{x}_\mathrm{e^+}^{\mathrm{true}}$ by $\vector{x}_\mathrm{e^+}$ to reduce the fitting parameters.
%PDFs used for the maximum likelihood fitting are assumed to be single Gaussian functions only for positron angles.

%\paragraph*{Gamma energy PDF}\,
The energy dependence of the $E_{\gammaup_{1(2)}}$ PDF \eref{eq:egammaPDF} is 
modelled with a morphing technique~\cite{Read1999} using two quasi-monoenergetic calibration lines:
the 11.7~MeV line from the  nuclear reaction of $^{11}\mathrm{B}(\mathrm{p}, 2\gammaup)^{12}\mathrm{C}$
and the 54.9~MeV line from $\piup^0$ decay. 
%The distribution between them are formed using a morphing technique\,\cite{Read1999}, shown in \fref{fig:morphing3}.

The PDFs of the $\gammaup$ position are approximated as double Gaussians to fit better tails in the PDF.

The positron angles are compared with those of the flipped direction of the X momentum ($-(\vector{x}_{\mathrm{vtx}} - \vector{x}_\mathrm{e^+})$)
with PDFs approximated as single Gaussians. 

%Descriptions of other PDFs are the following.

%\paragraph*{Decay length PDF}\,
%The likelihood function up to here has small information on the direction along X.
%The PDF of $l_\mathrm{X}$ is multiplied with the likelihood function to constrain the radial position of the X decay vertex.
The decay length is defined as $l_\mathrm{X} = |\vector{x}_{\mathrm{vtx}} - \vector{x}_\mathrm{e^+}|$.\footnote{
From the fifth and sixth terms in \eref{eq:likelihood}, the reason for using the absolute value of $l_\mathrm{X}$ rather than the signed value with the sign of $-(\vector{x}_{\mathrm{vtx}} - \vector{x}_\mathrm{e^+})\cdot\vector{P}_\mathrm{e^+}$ becomes apparent.
If the signed value of the decay length were negative, the X angle would be flipped by $\pi$ and the penalty would be huge preventing the fit to succeed.}
Under the approximation $\vector{\sigma}_{\vector{x}_{\mathrm{e^+}}}\to 0$, the PDF is
\bea
P(l_\mathrm{X} \mid \vector{x}_{\mathrm{e^+}}, \vector{x}_{\mathrm{vtx}}, \tau_\mathrm{X}, m_\mathrm{X}) = \frac{1}{\gamma \beta c \tau_\mathrm{X}} \cdot \exp{\left(-\frac{l_\mathrm{X}}{\gamma \beta c \tau_\mathrm{X}}\right)},
\eea
which is defined and normalised for $l_\mathrm{X}\geq0$. 
The approximation is justified because the transverse component of $\vector{\sigma}_{\vector{x}_\mathrm{e^+}}$ is $\sim\,$1--2 mm \cite{Adam2013} while the longitudinal component is largely driven by the target thickness ($\sim\,$0.2 mm), which is to be compared with $\gamma \beta c \tau_\mathrm{X}$ ranging between $\sim\,$6--30 mm.

%The following negative log likelihood is added:
%\bea
%- \log(p(l, \mathrm{m_\mathrm{X}}, \tau_\mathrm{X})) = \log(\gammaup(\mathrm{m_\mathrm{X}}) \beta(\mathrm{m_\mathrm{X}}) c \tau_\mathrm{X}) + \frac{l}{\gammaup(\mathrm{m_\mathrm{X}}) \beta(\mathrm{m_\mathrm{X}}) c \tau_\mathrm{X}}
%\eea
We fix $\tau_\mathrm{X}=20$\,ps since the vertex reconstruction performance is almost independent of $\tau_\mathrm{X}$ in the assumed range.
This likelihood term effectively penalizes non-zero decay lengths using a scale that is fixed to the average expected decay length of 20 ps.

%%%% resolution %%%%
The $\vector{x}_{\mathrm{vtx}}$ resolution of the maximum-likelihood fit  is evaluated via the MC to be $\vector{\sigma}_{\vector{x}_{\mathrm{vtx}}} = (8, 12)$~mm in the transverse and longitudinal directions.

%\paragraph*{$\chi^2$ of the vertex fit}\,
We define an expression to quantify the goodness of the vertex fit as
%\begin{strip}
%\bea
%\chi^{2} 
%=\sum_{\gammaup=\gammaup 1, \gammaup %2}\left(\frac{E_{\gammaup}-E_{\gammaup}^{\mathrm{vtxfit}}}{\sigma_{E_{\gammaup}}}\right)^{2}
%+ \sum_{\gammaup=\gammaup 1, \gammaup 2}\left(\frac{\vector{x}_{\gammaup} - %\vector{x}^{\mathrm{vtxfit}}_{\gammaup}}{\sigma_{\vector{x}}}\right)^2
%+\left(\frac{\theta_\mathrm{X}-\theta^{\mathrm{vtxfit}}}{\sigma_{\theta}}\right)^{2}
%+\left(\frac{\phi_\mathrm{X}-\phi^{\mathrm{vtxfit}}}{\sigma_{\phi}}\right)^{2}
%+\left(\frac{l_\mathrm{best}}{\gammaup \beta c \tau_\mathrm{X}}\right)^2,
%\label{eq:vertexchi2}
%\eea
%\end{strip}
\bea
\chi^{2}_\mathrm{vtx}
&=&\sum_{\gammaup=\gammaup_1, \gammaup_2}\left(\frac{E_{\gammaup}-E_{\gammaup}^{\mathrm{best}}}{\sigma_{E_{\gammaup}}}\right)^{2}
+ \sum_{\gammaup=\gammaup_1, \gammaup_2}\left(\frac{\vector{x}_{\gammaup} - \vector{x}^{\mathrm{best}}_{\gammaup}}{\vector{\sigma}_{\vector{x}_\gammaup}}\right)^2 \nn
&&+\left(\frac{\theta_\mathrm{X}-\theta_\mathrm{X}^{\mathrm{best}}}{\sigma_{\theta_\mathrm{X}}}\right)^{2}
+\left(\frac{\phi_\mathrm{X}-\phi_\mathrm{X}^{\mathrm{best}}}{\sigma_{\phi_\mathrm{X}}}\right)^{2}
+\left(\frac{l_\mathrm{X}^\mathrm{best}}{\gamma \beta c \tau_\mathrm{X}}\right)^2.
\label{eq:vertexchi2}
\eea
The variables with the superscript ``best'' indicate the best-fitted parameters in the maximum likelihood fit and the variables with no superscript indicate the measured ones. 
Here, $(\theta_\mathrm{X},\phi_\mathrm{X})=(\pi-\theta_\mathrm{e^+}, \pi+\phi_\mathrm{e^+})$ is the direction opposite to $(\theta_\mathrm{e^+}, \phi_\mathrm{e^+})$.
 
The $\sigma$ of each variable is the corresponding resolution when the distribution is approximated as a single Gaussian.
This expression is not expected to follow a %Pearson 
$\chi^2$ distribution because the PDFs of the variables are not in general Gaussian. The last term is quadratic by analogy with the other terms and its expression has been found to be effective in separating signal from background.
The rationale for using \eref{eq:vertexchi2} is to provide a powerful discriminator between signal and background as shown later in
Fig.~\ref{fig:MEx2GCutOptimizer_20_20_distribution}f.

\begin{figure}[tb!]
    \centering
    \includegraphics[width=\columnwidth]{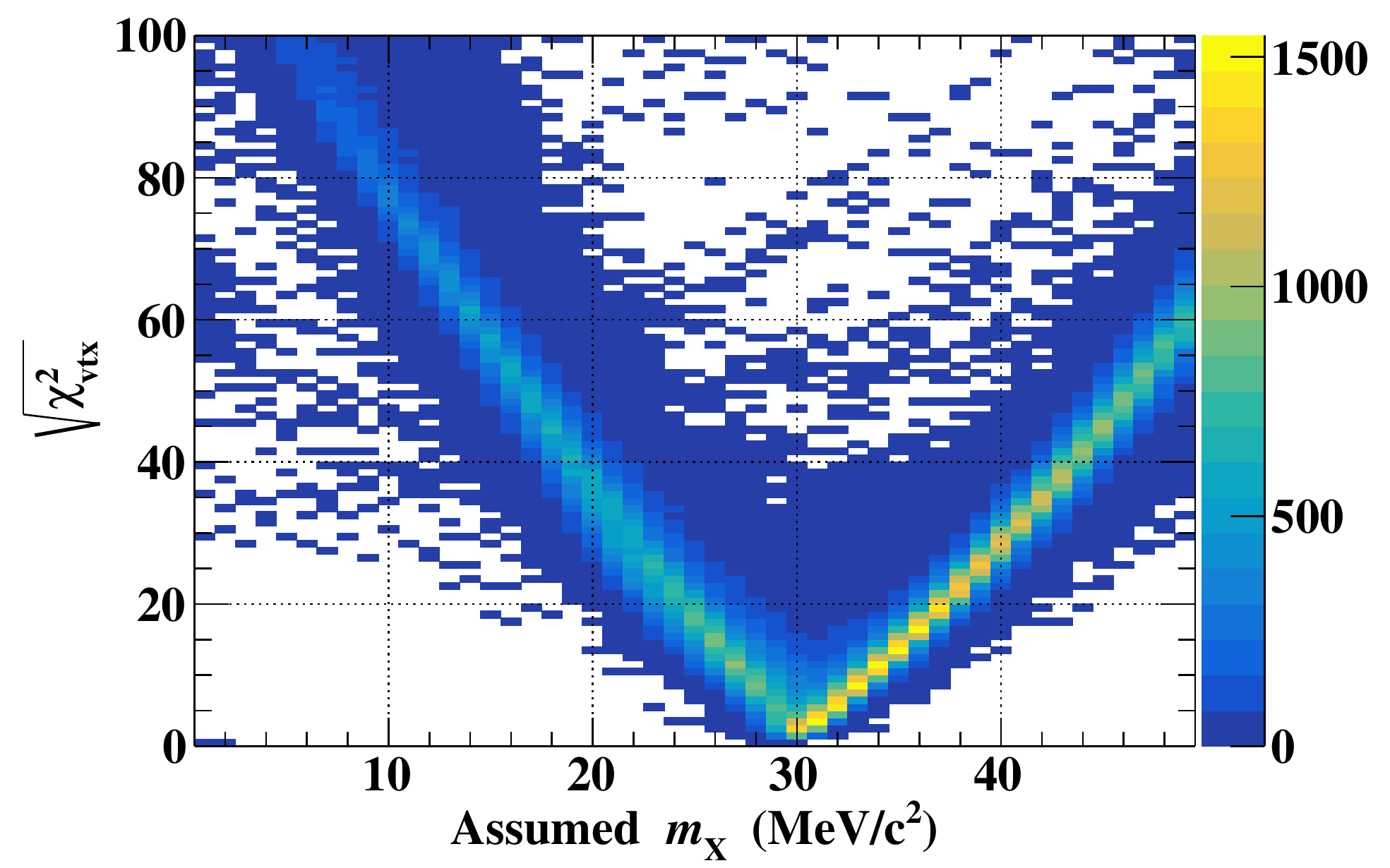}
    \caption{$\sqrt{\chi^2_\mathrm{vtx}}$ distribution for the MC signal events at $(m_\mathrm{X},\tau_\mathrm{X}) = (30~\mathrm{MeV/c^2}, 20~\mathrm{ps})$ as a function of $m_\mathrm{X}$ assumed in the reconstruction.}
    \label{fig:vertex_chi}
\end{figure}
\Fref{fig:vertex_chi} shows the dependence of $\chi^2_\mathrm{vtx}$\footnote{For better display, we take the square root of $\chi^2_\mathrm{vtx}$ here.} on the assumed value of $m_\mathrm{X}$ for the MC signal events, providing another rationale for \eref{eq:vertexchi2}.
When the assumed value is the same as the true value ($m_\mathrm{X}=30$~MeV/c$^2$ in this case), the resultant $\chi^2_\mathrm{vtx}$ becomes minimum on average.
The effective $m_\mathrm{X}$ resolution is $\sim 2.5$~MeV/c$^2$.

\subsubsection{Momentum}
\label{sec:momentum}
Given the vertex position, the momentum of each $\gammaup$ can be calculated.
The sum of the final-state three particles momenta,
\bea
\vector{P}_{\mathrm{sum}}\equiv\vector{P}_{\mathrm{e^+}}+\vector{P}_{\gammaup_1}+\vector{P}_{\gammaup_2},
\eea
should be 0 for the MEx2G events.

\subsubsection{Relative time}
\label{sec:relative_timing}
%As we will discuss in \sref{sec:analysis}, blinding and background estimation are performed on the $\tgg$--$t_{\gammaup_1\mathrm{e^+}}$ surface, where $\tgg$ is the time difference between 2$\gammaup$s at the X vertex and $t_{\gammaup_1\mathrm{e^+}}$ is the time difference between $\gammaup_1$ and positron at the muon vertex.
%These time differences are calculated from the reconstructed time: $t_{\gammaup_1}, t_{\gammaup_2},t_{\mathrm{e^+}}$.
%$t_{\gammaup_1}$ and $t_{\gammaup_2}$ are reconstructed time inside the LXe detector and $t_{\mathrm{e^+}}$ is reconstructed time at the target.
%
%\fref{fig:tdiff_rec} illustrates schematics of the MEx2G decay.
The time difference between the 2$\gammaup$s at the X vertex is calculated as
%$r_{1(2)}$ is distance between $\gammaup_{1(2)}$ and the vertex position of $\mathrm{X}\rightarrow \gammaup\gammaup$.
%$l$ is distance between the vertex position of $\mathrm{X}\rightarrow \gammaup\gammaup$ and that of $\muup^+\rightarrow \mathrm{e^+}^+\mathrm{X}$.
%$\tgg$ should be 0 at the vertex position of $\mathrm{X}\rightarrow \gammaup\gammaup$.
%Thus $\tgg$ is calculated as
\bea
t_{\gammaup\gammaup} = \left(t_{\gammaup_1}-\frac{l_{\gammaup_1}}{c}\right) - \left(t_{\gammaup_2}-\frac{l_{\gammaup_2}}{c}\right),
\label{eq:tgg}
\eea
where $l_{\gammaup_{1(2)}}$ is the distance between the $\gammaup_{1(2)}$ interaction point in the LXe photon detector and the X vertex position, 
$l_{\gammaup_{1(2)}}=|\vector{x}_{\gammaup_{1(2)}} - \vector{x}_{\mathrm{vtx}}|$. The relative position of the vertices is such that this definition is identical to the signed distance defined according to the X direction and therefore the distribution is centred at 0 for MEx2G events.
%$t_{\gammaup_1e}$ should be 0 at the vertex position of $\muup^+\rightarrow \mathrm{e^+}^+\mathrm{X}$.
%Thus $t_{\gammaup_1e}$ is calculated as

The time difference between $\gammaup_1$ and e$^+$ at the muon vertex is calculated as
\bea
\teg = \left(t_{\gammaup_1}-\frac{l_{\gammaup_1}}{c}-\frac{l_\mathrm{X}}{\beta c}\right) -t_{\mathrm{e^+}}.
\label{eq:teg}
\eea
With the unsigned definition of $l_\mathrm{X}$ the distribution is slightly offset with respect to 0 for MEx2G events as visible in Fig.~\ref{fig:MEx2GCutOptimizer_20_20_distribution}d.

% --------------------------- 3_analysis -------------------------------
%\section{Analysis}
%\label{sec:analysis}

\begin{figure}[t!]
	\centering
	\includegraphics[width=\columnwidth]{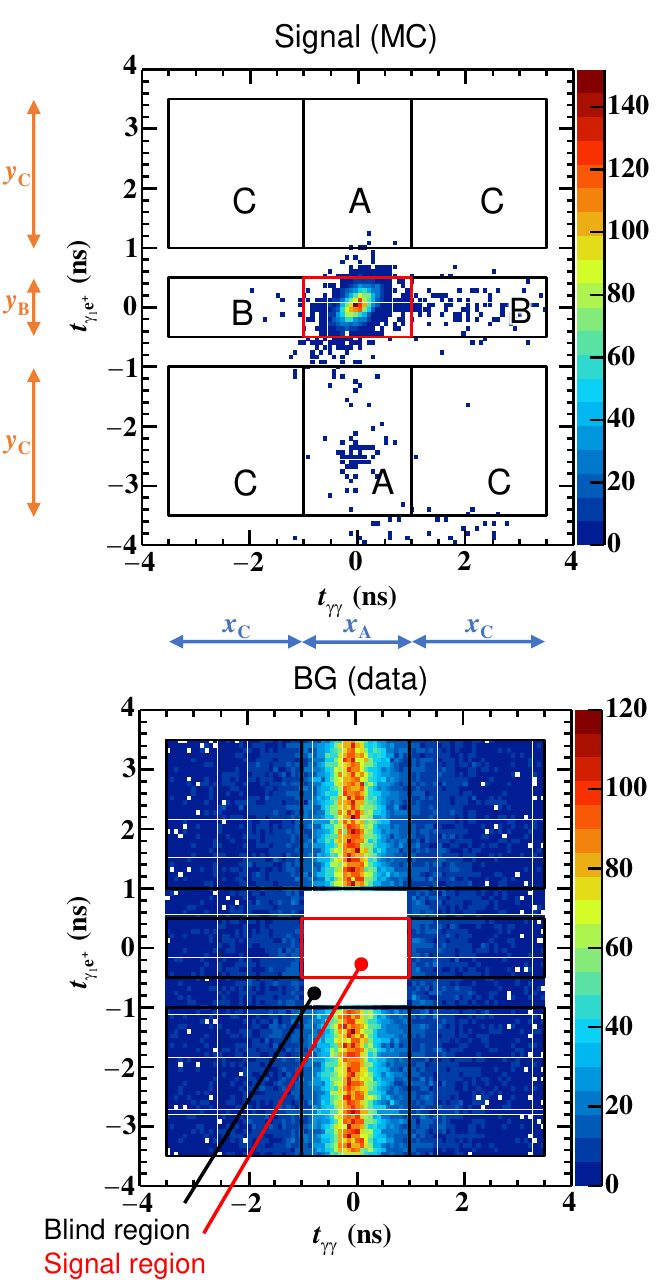}
	\caption{Top: Signal (MC) and bottom: background (sideband data) event distributions in the $\teg$--$\tgg$ plane before the signal selection criteria are applied. (20\,MeV/c$^2$, 20\,ps) case is shown as an example. The time sideband regions (A, B, C) and the signal region (the red box) are also shown.}
	\label{fig:time2D}
\end{figure}

\section{Dataset and event selection}
\label{sec:dataset}
%All the detectors of the MEG experiment were ready in 2007 and performed engineering run.
%After fixing several problems found in the engineering run, physics data taking had started in 2008.
%From 2009 to 2013, data were accumulated continually.
%as shown in \fref{fig:statistics} 
%except for the annual beam shutdown period.
%The data in 2008 was not used for the final analysis of the MEG analysis because of a discharge problem of the drift chamber, which resulted in a bad positron performance.
%It is not used for the MEx2G analysis, either.
We use the full MEG dataset, collected in 2009--2013, as was used in the \meg search reported in \cite{baldini_2016}.
As described in \sref{sec:detector}, the \meg trigger data are used in this analysis.
In total, $7.5\times10^{14}\,\muup^+$s were stopped on the target.
%The MEx2G analysis based on the first $1.8\times10^{14}\,\muup^+$s (2009 and 2010) were presented in \cite{Natori2012}.
%In the analysis presented in this paper, we use the full dataset accumulated in five years.

%During the \meg data taking, a dedicated trigger for \mextwog events was not prepared.
%We use \meg triggered events in the MEx2G search analysis.
%One of the \meg trigger conditions requires back-to-back e$^+\gammaup$ events, but it has more chance to lose the \mextwog signal for the larger mass of X.
%This inefficiency needs to be taken into account.

A pre-selection was applied at the first stage of the \meg decay analysis, requiring that at least one positron track is reconstructed and the time difference between signals in the LXe photon detector and TC is in the range  $-6.9<t_{\mathrm{LXe}-\mathrm{TC}}<4.4~\mathrm{ns}$. 
At this stage, aiming to select  the \meg decays, the time of the LXe photon detector is reconstructed with PMTs around the largest peak found in the peak search (\sref{sec:peak_search}).
This retained $\sim$16\% of the dataset, on which the full event reconstruction for the \meg decay analysis was performed.
%Some reconstructions are common between the \meg and the \mextwog decay analysis.
Before processing the MEx2G dedicated reconstruction, we applied an additional event selection using the \meg reconstruction results.
It was based on the existence of multiple ($\geq\!2$) $\gammaup$s %the reconstruction quality of the e$^+$ track, 
and the total energy of the $\gammaup$s\footnote{At this stage, the sum of the multiple $\gammaup$s' energy is reconstructed without being separated into each $\gammaup$.} and e$^+$ ($E_\mathrm{total}$) being $|E_\mathrm{total} - m_\muup|<0.2m_\muup$. This selection  reduces the dataset by an additional factor of $\sim$ 300.
%at least one high quality e$^+$ track,
%$|m_{\muup} - E_{\mathrm{Total}}| < m_{\muup} \times 20\%$,
%and $ E _ { \gammaup} > 40\,\mathrm{MeV}$.
We applied the MEx2G dedicated reconstruction (\sref{sec:2gamma_position_energy}, \ref{sec:2gamma_time}, and \ref{sec:combined_rec}) to this selected dataset.

%A blind analysis is used to reduce the experimenter's bias.
%We determine conditions of event selection and estimate the number of background events without looking at the signal region.
A blind region was defined containing the events satisfying the cuts $|\teg| < 1\,\mathrm{ns} \land |\tgg| < 1\,\mathrm{ns}$. This blind region is large enough to hide the signal. 
Those events were sent in a separated data-stream and were not used in the definition of the analysis strategy including cuts; background events in the signal region were estimated without using events in the blind region. After the analysis strategy was defined, the blind region was opened and events in this region were added to perform the last step of analysis.

%The region is shown in blue shadow in \fref{fig:timesideband2}.
The accidental background can be estimated from the off-time sideband regions defined in \fref{fig:time2D}.
%The signal region (in red) and sideband regions are defined on the $\teg$--$\tgg$ surface;
There are three such regions: A, B, and C; each containing a different combination of the types of background as defined in Sect.~\ref{sec:strategy}.
The outer boundary of the time sidebands, $|\teg| < 3.5\,\mathrm{ns} \land |\tgg| <3.5\,\mathrm{ns}$, are determined so that the background distribution is not deformed by the time-coincidence trigger condition.
The widths $x_\mathrm{A}$ and $y_\mathrm{B}$ in \fref{fig:time2D} are the same as the outer boundary of the signal region defined depending on $m_{X}$ by the signal selection criteria described below.

\begin{figure*}[t!]
	\centering
	\includegraphics[width=\textwidth]{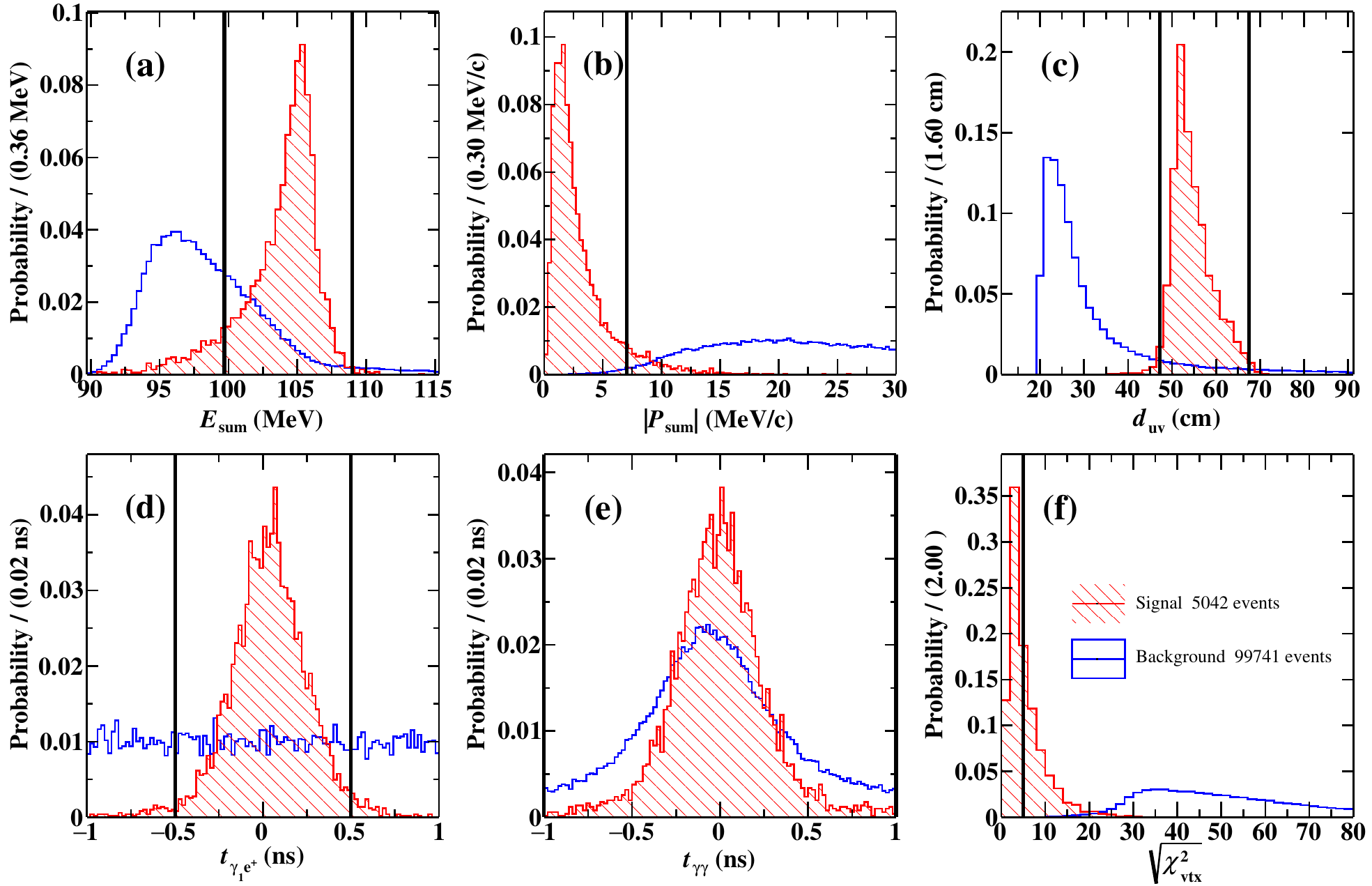}
	\caption{Distributions of variables used in the event selection for $(m_\mathrm{X}, \tau_\mathrm{X})$ = (20\,MeV/c$^2$, 20\,ps) case. The hatched histograms show the distribution of MC signal events while the blank histograms that of background events; each histogram is normalised to 1. The vertical lines show the optimised thresholds.
	(a) The peak value of the signal distribution is at $m_\muup$ with FWHM$_{E_\mathrm{sum}}$ = 2.7~MeV.
	(c) Cut-off at 20 cm in the background distribution comes from one of the 2$\gammaup$ reconstruction conditions.
	(e) The threshold lines are not visible because they are set to $\pm1$\,ns. For a detailed definition of the variables see Sect.~\ref{sec:dataset}.}
	\label{fig:MEx2GCutOptimizer_20_20_distribution}
\end{figure*}

\begin{comment}
\begin{figure}[t!]
    \centering
    \includegraphics[width=\columnwidth]{statistics.pdf}
    \caption{Number of stopped muons on target\,\cite{baldini_2016}}
    \label{fig:statistics}
\end{figure}
\end{comment}

%\section{Event selection}
%\label{sec:cut_condition}
%\subsubsection{Variables used for the event selection}
\label{sec:cut_conditions}
The following seven variables are used for the signal selection:
\begin{enumerate}
    \item \textbf{{$E_\mathrm{e^+}$}}: the e$^+$ energy.
    \item \textbf{{$E_\mathrm{sum}$}}: the total energy of the three particles.
   % It should be equal to the mass of muon from the energy conservation.
    \item $|\vector{P}_\mathrm{sum}|$: the magnitude of the sum of the three particles' momenta.
    %The magnitude should be 0 from the momentum conservation.
    \item $d_{uv}$: the distance between the 2$\gammaup$ positions on the LXe photon detector inner face.
    \item {\teg}: the time difference between $\gammaup_1$ and e$^+$  calculated in \eref{eq:teg}.
    \item {\tgg}: the time difference between 2$\gammaup$s calculated in \eref{eq:tgg}.
    \item {{$\chi^2_\mathrm{vtx}$}}: the goodness of vertex fitting calculated in \eref{eq:vertexchi2}.
\end{enumerate}

First, we fix the $E_\mathrm{e^+}$ selection to require $|E_\mathrm{e^+}-E_\mathrm{e^+}^{m_\mathrm{X}}|<1$~MeV, where $E_\mathrm{e^+}^{m_\mathrm{X}}$ is the e$^+$ energy for the MEx2G decay with $m_\mathrm{X}$.
This selection is also used in the Michel normalisation described in \sref{sec:pa_norm}.
%$|P_{\mathrm{e^+}^{+}}-P_{\mathrm{X}}|<1\,\mathrm{MeV}$

%\subsubsection{Threshold optimisation}

Next, we optimise the cut thresholds for the other variables to maximise the experimental outcomes. % without assuming the signal number.
Distributions of these variables for the signal and background at a parameter set (20\,MeV/c$^2$, 20\,ps) are shown in \fref{fig:MEx2GCutOptimizer_20_20_distribution}.
All other selection criteria, such as trigger and reconstruction conditions as well as the $E_\mathrm{e^+}$ selection, are applied.
%The optimised thresholds are shown in the back lines.
The time sideband events are used for the background distribution, while MC samples are used for the signal distribution.

%For the cut optimisation,
Punzi's expression \cite{Punzi2003} is used as a figure of merit % of the cut optimisation.
%It does not require any assumption on the signal counts. % whether 0 or not.
%The sensitivity $\sigma_{\mathrm{Punzi}}$ is defined as
\bea
%\sigma^{-1}_{\mathrm{Punzi}}
F_{\mathrm{Punzi}}
=\frac{\epsilon_{\mathrm{selection}}}{b^{2}+2 a \sqrt{N_{\mathrm{BG}}}+b \sqrt{b^{2}+4 a \sqrt{N_{\mathrm{BG}}}+4 N_{\mathrm{BG}}}},
\label{eq:Punzi}
\eea
where $a$ and $b$ are the significance and the power of a test, respectively,
$\epsilon_{\mathrm{selection}}$ is the selection efficiency for the signal, and $N_{\mathrm{BG}}$ is the expected number of background events.
The values of $a$ and $b$ should be defined before the analysis, and we set $a = 3, b = 1.28\,(=90\%)$, where $b$ is set to the value appropriate to the confidence level being used to set the upper limit when a non-significant result is obtained.
%The task is to find the optimal cuts to maximize $\sigma^{-1}_{\mathrm{Punzi}}$.
%In general, $\epsilon_{\mathrm{signal}}$ should be large and $N_{\mathrm{BG}}$ should be small to maximize $\sigma^{-1}_{\mathrm{Punzi}}$.
%The cut thresholds are optimised at 5-MeV intervals in $m_\mathrm{X}$.

The optimisation process is divided into two steps.
In the first step, we optimise the cut thresholds of variables 2--6, independently for each variable in order to maintain high statistics in the sidebands.
Because the absolute value of $N_{\mathrm{BG}}$ does not make sense in this independent optimisation process, we approximate $N_{\mathrm{BG}}$ to $\epsilon_{\mathrm{BG}}$, a selection efficiency for the background events calculated using the time sideband samples selected up to this point.
%As an approximation, before applying each cut we normalise $N_\mathrm{BG} = 1$ and then maximise 
%$F_{\mathrm{Punzi}}$;
%$\sigma^{-1}_{\mathrm{Punzi}}$;
Because of this approximation, the first step leads to suboptimal criteria.

In the second step, after all other selection criteria are applied, 
the threshold for $\chi^2_\mathrm{vtx}$ is optimised to give the highest 
$F_{\mathrm{Punzi}}$.
%$\sigma^{-1}_{\mathrm{Punzi}}$.
In this step, to estimate $N_\mathrm{BG}$ from the low statistics in the sideband regions, we use a kernel-density-estimation method \cite{Crabner2001} to model the continuous event distribution.

The cut thresholds are optimised at 5 $\mathrm{MeV/c^2}$ intervals in $m_\mathrm{X}$, while the same thresholds are used for different $\tau_\mathrm{X}$ for each $m_\mathrm{X}$.
The optimised thresholds for $m_\mathrm{X}=20~\mathrm{MeV/c^2}$ are shown as black lines in \fref{fig:MEx2GCutOptimizer_20_20_distribution}.
These cuts result in $\epsilon_\mathrm{selection} = 67\%$ ($m_\mathrm{X} =45$ MeV/c$^2$) -- 51\% ($m_\mathrm{X} =20$ MeV/c$^2$).

\begin{comment}
\begin{figure*}[!t]
\normalsize
\begin{equation}
\label{eqn_dbl_x}
x = 5 + 7 + 9 + 11 + 13 + 15 + 17 + 19 + 21+ 23 + 25
+ 27 + 29 + 31
\end{equation}
\begin{equation}
\label{eqn_dbl_y}
y = 4 + 6 + 8 + 10 + 12 + 14 + 16 + 18 + 20+ 22 + 24
+ 26 + 28 + 30
\end{equation}
\hrulefill
\end{figure*}
\end{comment}

\section{Single event sensitivity}
\label{sec:pa_norm}
%code   : Taka/physics_analysis/MEx2GNorm.cpp
The single event sensitivity of the MEx2G decay $s$ is defined as follows:
\bea
\mathcal{B}_{\mathrm{MEx2G}} = s\times N_\mathrm{MEx2G},
\label{eq:bMichel}
\eea
where $N_\mathrm{MEx2G}$ is the expected number of signal events in the signal region.
We calculate it using Michel decay ($\michel$) events taken at the same time with the \meg trigger.
This Michel normalisation is beneficial for the following reasons.
First, systematic uncertainties coming from the muon beam are cancelled because
beam instability is included in both Michel triggered and the \meg triggered events.
Moreover, we do not need to know the $\muup^+$ stopping rate nor the live DAQ time.
Second, most of the systematic uncertainties coming from e$^+$ detection are also cancelled. 
The absolute value of e$^+$ efficiency is not needed. 

The number of Michel events is given by
\bea
N_\mathrm{Michel}= N_{\muup^+}
\cdot \frac{\mathcal{B}_{\mathrm{Michel}} 
\cdot f_{\mathrm{Michel}}}{p_{\mathrm{Michel}}\cdot p_{\mathrm{correction}}}
\cdot A_{\mathrm{Michel}} \cdot \epsilon_{\mathrm{Michel}},
\label{eq:nMichel}
\eea
where
\begin{description}
%    \item []$N_\mathrm{Michel}$: the number of observed Michel events.
    \item [$N_{\muup^+}$:] the number of stopped $\muup^+$s;
    \item [$\mathcal{B}_{\mathrm{Michel}}$:] branching ratio of the Michel decay ($\approx1$);
    \item [$f_{\mathrm{Michel}}$:] branching fraction of the selected energy region (7\%--10\% depending on $m_{\mathrm{X}}$);
    \item [$p_{\mathrm{Michel}}$:] prescaling factor of the Michel trigger ($=10^7$);
    \item [$p_{\mathrm{correction}}$:] correction factor of $p_{\mathrm{Michel}}$ depending on the muon beam intensity;
    \item [$A_{\mathrm{Michel}}$:] geometrical acceptance of the spectrometer for Michel e$^+$s;
    \item [$\epsilon_{\mathrm{Michel}}$:] e$^+$ efficiency for Michel events within the geometrical acceptance of the spectrometer.
\end{description}

The number of MEx2G events is given by
\bea
N_\mathrm{MEx2G}
&=& N_{\muup^+}
\cdot \frac{\mathcal{B}_{\mathrm{MEx2G}} }{p_{\mathrm{MEG}}}
\cdot A_{\mathrm{e^+}}
\cdot \epsilon_{\mathrm{e^+}}
\cdot \epsilon_{2\gammaup}
\cdot \epsilon_\mathrm{DM}
\cdot \epsilon_\mathrm{selection},
\label{eq:nMEx2G}
\eea
where
\begin{description}
%    \item [$N_{2\gammaup}$:] the number of observed signal MEx2G events;
%    \item [$\mathcal{B}_{\mathrm{MEx2G}}$:] branching ratio of the MEx2G decay (unknown);
    \item [$p_{\mathrm{MEG}}$:] prescaling factor of the \meg trigger (=1);
    \item [$A_{\mathrm{e^+}}$:] geometrical acceptance of the spectrometer for MEx2G e$^+$s;
    \item [$\epsilon_{\mathrm{e^+}}$:] e$^+$ efficiency for MEx2G events conditional to e$^+$s in the geometrical acceptance of the spectrometer;
    \item [$\epsilon_{2\gammaup}$:] the product of 2$\gammaup$ geometrical acceptance and 2$\gammaup$ trigger, detection, and reconstruction efficiency, conditional to the e$^+$ detection;
    \item [$\epsilon_\mathrm{DM}$:] the trigger direction match efficiency conditional to the e$^+$ and $2\gammaup$ detection (\fref{fig:DMEfficiency});
    \item [$\epsilon_\mathrm{selection}$:] the signal selection efficiency.
\end{description}

\begin{figure}[tb]
    \centering
    \includegraphics[width=\columnwidth]{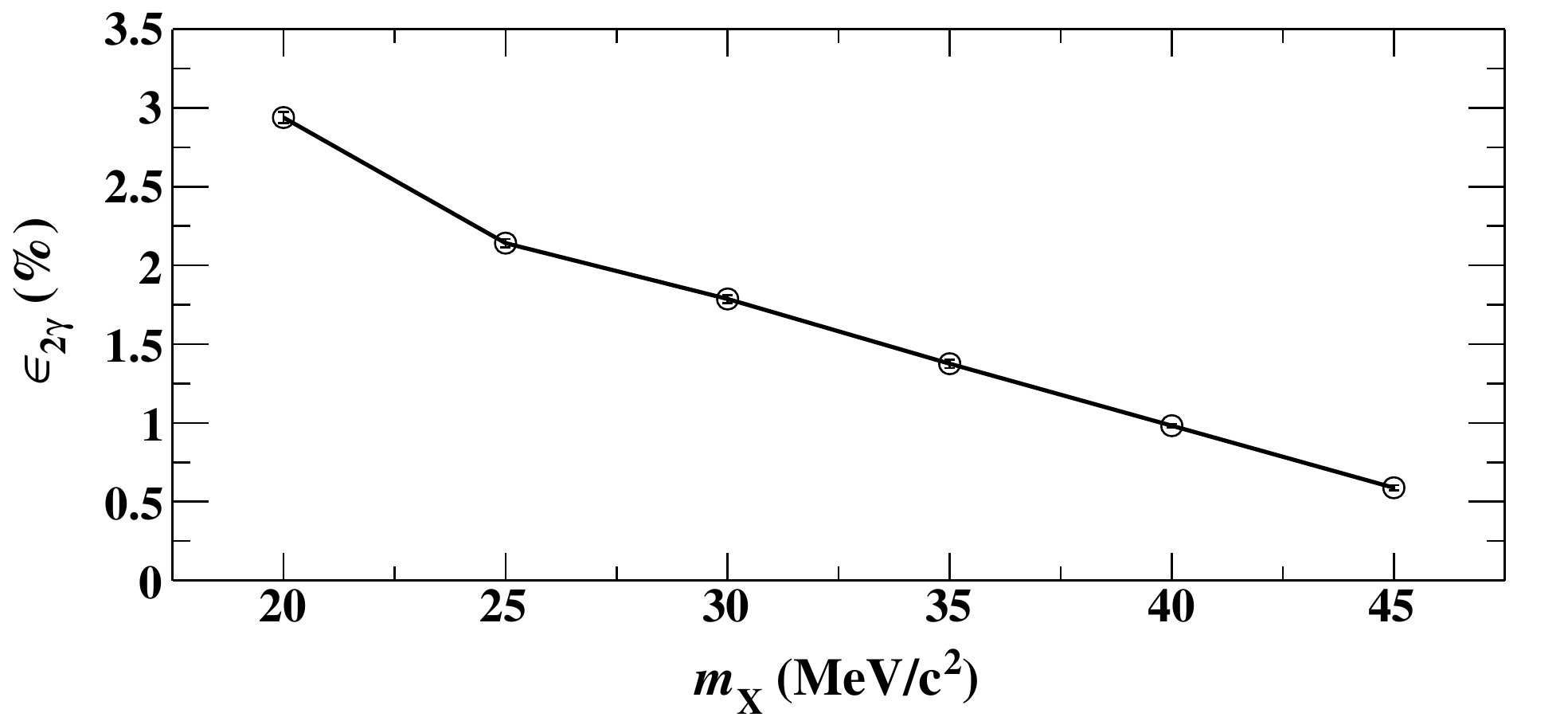}
    \caption{$\epsilon_{2\gammaup}$ (see the text for the definition) versus $m_\mathrm{X}$ for $\tau_\mathrm{X}= 20$~ps.}
    \label{fig:Efficiency}
\end{figure}

Using \erefs{eq:bMichel}--(\ref{eq:nMEx2G}), an estimate of the SES ($s_0$) is given by
\bea
s_0^{-1}&=& N_\mathrm{Michel}
  \frac{1}{\mathcal{B}_{\mathrm{Michel}}\cdot f_{\mathrm{Michel}}}
\cdot \frac{p_{\mathrm{Michel}}\cdot p_{\mathrm{correction}}}{p_{\mathrm{MEG}}}\nn
&\ & \cdot \frac{A_{\mathrm{e^+}}}{A_{\mathrm{Michel}}}
\cdot \frac{\epsilon_{\mathrm{e^+}}}{\epsilon_{\mathrm{Michel}}}
\cdot \epsilon_{2\gammaup}
\cdot \epsilon_\mathrm{DM}
\cdot \epsilon_\mathrm{selection}.
\label{eq:norm1}
\eea
The geometrical acceptance of the spectrometer is common, hence $A_{\mathrm{e^+}}/A_{\mathrm{Michel}}= 1$; the estimate of the relative e$^+$ efficiency is
$\epsilon_{\mathrm{e^+}}/\epsilon_{\mathrm{Michel}}=89\%\ (m_\mathrm{X} =45$ MeV/c$^2)\ -\ 97\%\ (m_\mathrm{X} =20$ MeV/c$^2)$ increasing monotonically with $m_\mathrm{X}$.
The estimate of $\epsilon_{2\gammaup}$ is shown in \fref{fig:Efficiency}; $\epsilon_{2\gammaup} = 0.6\%$ ($m_\mathrm{X} =45$ MeV/c$^2$) -- 2.9\% ($m_\mathrm{X} =20$ MeV/c$^2$), decreasing monotonically with $m_{\mathrm{X}}$.
This dependence comes mainly from the 2$\gammaup$ acceptance: %and the direction match trigger efficiency.
for increasing $m_{\mathrm{X}}$, the opening angle between the 2$\gammaup$s becomes larger, resulting in a decreasing efficiency.

%%%% systematic uncertainties %%%%
The systematic uncertainties are summarised in \tref{tab:sys_signal_eff}.
The uncertainty in the 2$\gammaup$ detection efficiency and that in the MC smearing parameters are the dominant components. 

\begin{table*}[bt]
\caption{Systematic uncertainties in the single event sensitivity ($\tau_\mathrm{X}=20$~ps).}
\label{tab:sys_signal_eff}
\centering
\begin{tabular}{lrrrrrr}
\hline
$m_\mathrm{X}$ (MeV/c$^2$)           & 20     & 25     & 30     & 35     & 40     & 45 \\ \hline
Michel e$^+$ counting                 & 0.99\%    & 1.1\%   & 1.1\%  &  0.93\%  & 1.6\% & 2.8\% \\
Relative e$^+$ efficiency               & 1.4\%     & 0.18\% & 0.31\%  & 0.55\% & 0.89\%  & 1.3\% \\ 
2$\gammaup$ acceptance            & 1.3\%      & 2.0\%      & 3.4\%      & 2.9\%      & 5.4\%      & 1.9\%  \\
$\gammaup$ trigger efficiency    & 0.98\%      & 0.32\%      & 0.26\%      & 0.52\%      & 1.4\%      & 3.2\%  \\
2$\gammaup$ detection efficiency& 7.9\%      & 7.9\%      & 7.9\%      & 7.9\%      & 7.9\%      & 7.9\%  \\
%2$\gammaup$ detection efficiency& 7.4\%      & 7.4\%      & 7.4\%      & 7.4\%      & 7.4\%      & 7.4\%  \\
%2$\gammaup$ pileup inefficiency  & 2.8\%      & 2.8\%      & 2.8\%      & 2.8\%      & 2.8\%      & 2.8\%  \\
MC statistics                             &  1.8\%  & 1.9\%  & 2.2\% & 3.1\% & 1.9\% & 4.7\% \\
MC smearing                            & 4.8\% & 3.4\% & 3.8\% & 5.3\% & 3.3\% & 14\% \\ \hline
Total                                         & 9.7\%  &  9.1\% & 9.7\%  & 11\%   & 10\%  & 17\% \\ \hline
\end{tabular}
\end{table*}

%%%% SES %%%%
The estimated value of SES is $s_0 = (2.9 \pm 0.3)\times 10^{-12}$ (20~MeV/c$^2$) -- $(6.3 \pm 1.1)\times 10^{-10}$ (45~MeV/c$^2$) for $\tau_\mathrm{X} = 20$~ps increasing monotonically with $m_\mathrm{X}$.
The e$^+$ efficiency is $\epsilon_{\mathrm{e^+}} = 1\%$ (45 MeV/c$^2$) -- 36\% (20~MeV/c$^2$) decreasing monotonically with $m_{\mathrm{X}}$, estimated with the MC,
although this quantity is not necessary for the normalisation.
The overall efficiency for the MEx2G events conditional to the e$^+$ in the geometrical acceptance of the spectrometer is therefore $\epsilon_{\mathrm{MEx2G}} = 2.0\times 10^{-5}$ (45~MeV/c$^2$) -- $4.7\times 10^{-3}$ (20~MeV/c$^2$) decreasing monotonically with $m_{\mathrm{X}}$.

\section{Statistical treatment of background and signal}
%\section{Background estimation}
\label{sec:BG_estimation}

\begin{figure}[tb]
	\centering
	\includegraphics[width=\columnwidth]{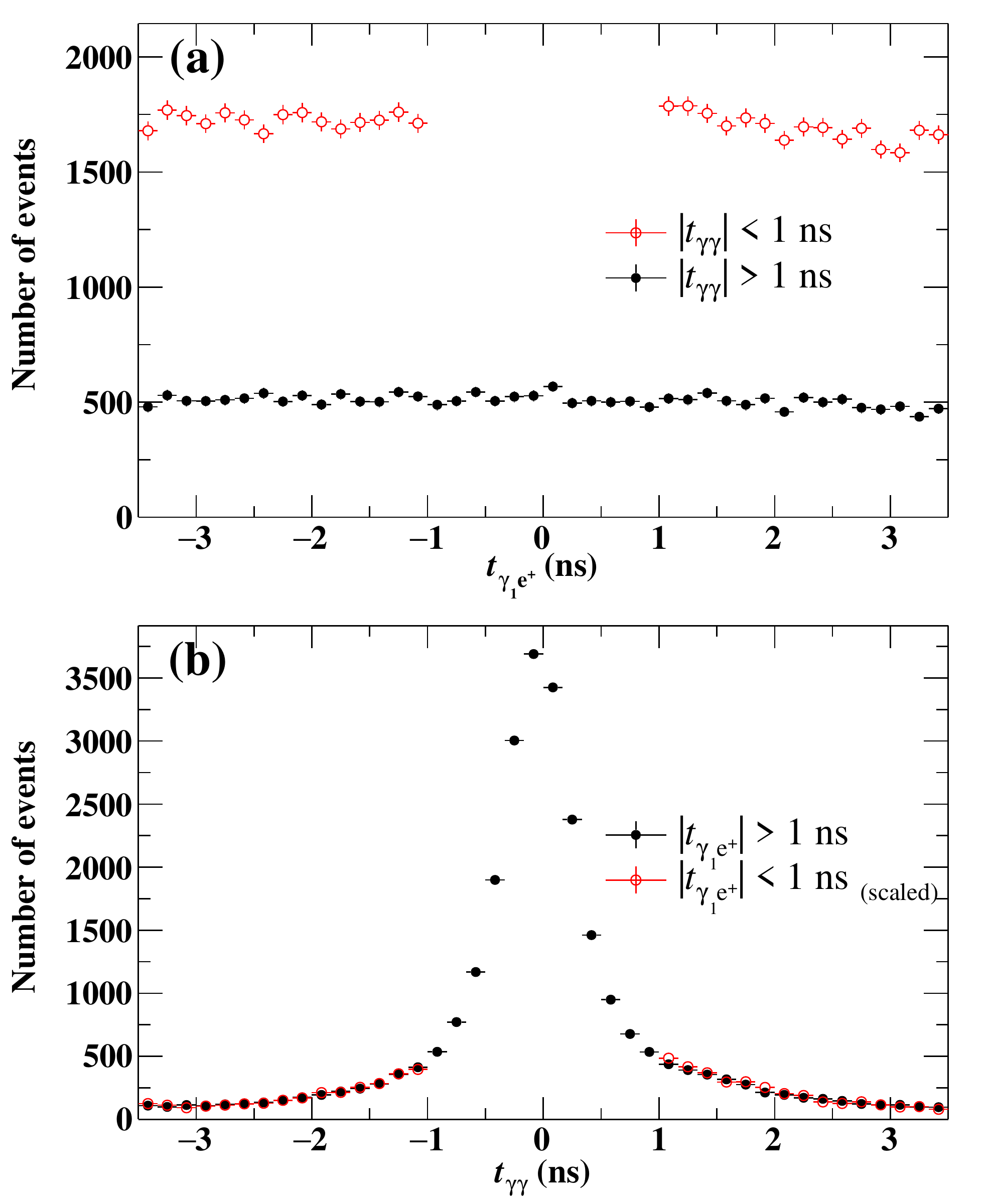}
	\caption{Time distributions in the sideband regions for (20~MeV/c$^2$, 20~ps). (a) $\teg$ distributions for $|\tgg|<1$~ns (red open circles) and for $1<|\tgg|<3.5$~ns (black closed circles).
	 (b) $\tgg$ distributions for $|\teg|<1$~ns (red open circles) and for $1<|\teg|<3.5$~ns scaled by the ratio of the time ranges (black closed circles).
	 A loose cut is applied: $|E_\mathrm{e^+}-E_\mathrm{e^+}^{m_\mathrm{X}}|<1~\mathrm{MeV} \land E_\mathrm{sum}<115~\mathrm{MeV} \land |\vector{P}_\mathrm{sum}|<30~\mathrm{MeV/c} \land d_{uv}<90~\mathrm{cm} \land \chi^2_\mathrm{vtx}<80$.}
	\label{fig:MEx2GTimeDistribution}
\end{figure}

In the following, we describe how we estimate the expected number of background events in the signal region  ($N_{\mathrm{BG}}$) from the numbers of events observed in sidebands A, B, and C ($N_\mathrm{A}^\mathrm{obs}$, $N_\mathrm{B}^\mathrm{obs}$, and $N_\mathrm{C}^\mathrm{obs}$). %\footnote{For instance, $N_\mathrm{A}^\mathrm{obs}$ is the sum of the number of events in A$_1$ and A$_2$.}.
%There are three time sideband regions: A (A$_1$ and A$_2$), B (B$_1$ and B$_2$), and C (C$_1$, C$_2$, C$_3$, and C$_4$) as shown in \fref{fig:time2D}.
%Different alphabets indicate different off-timing sidebands.

There are three types of accidental background events defined in \sref{sec:strategy}.
The expected number of background events in the signal region is given by
\be
N_{\mathrm{BG}} = N_1 + N_2 + N_3,
\ee
where $N_1, N_2, N_3$ are the expected numbers of background events in the signal region from the types 1,  2, and 3, respectively. 
%There are three types of accidental background events:
%\begin{itemize}
%    \item \textbf{Type 1}: e$^+$ and one of $\gammaup$s are the same origin, and the other $\gammaup$ is accidental.
%    \item \textbf{Type 2}: 2$\gammaup$s are the same origin, and e$^+$ is accidental.
%    \item \textbf{Type 3}: All the particles are accidental.
%\end{itemize}
%The numbers right-hand side of the each box in \fref{fig:timesideband2} show which type contributes to each sideband; 
Sideband A has the contributions from types 2 and 3, 
B has the contributions from types 1 and 3, and
C has the contribution from type 3.

\Fref{fig:MEx2GTimeDistribution} shows the time distributions in the sideband regions. A peak of type 2 on a flat component of type~3 is observed in the $\tgg$ distribution, while a peak of type 1 is not clearly visible in the $\teg$ distribution.
The uniformity of the accidental backgrounds is examined using these distributions; the number of events in ($|\teg|<1~\mathrm{ns} \land 1<|\tgg|<3.5 ~\mathrm{ns}$) is compared to
the the number of events interpolated from the region ($1<|\teg|<3.5~\mathrm{ns} \land 1<|\tgg|<3.5~\mathrm{ns}$) scaled by the ratio of the widths of the time ranges (2~ns/5~ns).
They agree within 1.7\% (the central part, including type 1, is 1.7\% larger than the interpolation).
In \fref{fig:MEx2GTimeDistribution}b, $\tgg$ the distribution for $1<|\teg|<3.5$~ns is superimposed on that for $|\teg|<1$~ns after scaling by the time range ratio. The tail component of type 2 is consistent in these regions. %, showing that the type 2 component is distributed uniformly in $\teg$.
The errors on the background estimations by the interpolation are thus negligibly small compared with the statistical uncertainties in  $N_\mathrm{A}^\mathrm{obs}, N_\mathrm{B}^\mathrm{obs}, N_\mathrm{C}^\mathrm{obs}$.

Using  $N_1, N_2, N_3$, the expected numbers of events in sidebands A, B, C can be calculated as follows:
%assuming the background distribution is linear: %\footnote{We confirmed that deviation from this assumption is negligible compared with statistical uncertainty of $N_\mathrm{A}^\mathrm{obs}, N_\mathrm{B}^\mathrm{obs}, N_\mathrm{C}^\mathrm{obs}$.}:
\bea
N_\mathrm{A}^\mathrm{exp} &=& N_2 \times \frac{2y_{\mathrm{C}}}{y_{\mathrm{B}}} + N_3 \times \frac{2y_{\mathrm{C}}}{y_{\mathrm{B}}},\\
\label{eq:nAexp}
N_\mathrm{B}^\mathrm{exp} &=& N_1 \times \frac{2x_{\mathrm{C}}}{x_{\mathrm{A}}} + N_3 \times \frac{2x_{\mathrm{C}}}{x_{\mathrm{A}}} + N_2 \times f_{\mathrm{escape}},\\
\label{eq:nBexp}
N_\mathrm{C}^\mathrm{exp} &=& N_3 \times \frac{2y_{\mathrm{C}}}{y_{\mathrm{B}}} \times  \frac{2x_{\mathrm{C}}}{x_{\mathrm{A}}} 
 + N_2 \times f_{\mathrm{escape}}\times\frac{2y_{\mathrm{C}}}{y_{\mathrm{B}}},
\label{eq:nCexp}
\eea
where $x_\mathrm{A(C)}$ and $y_\mathrm{B(C)}$ are the sizes of the signal regions (sideband regions) in $\tgg$ and $\teg$, respectively, as defined in \fref{fig:time2D},
and $f_{\mathrm{escape}}=0.171 \pm 0.003$ is the fraction of type~2 events in $|\tgg| > 1\,\mathrm{ns}$. %, the origin of which is $\tgg$ peak.
%This factor is estimated to be 0.171. Its uncertainty is negligible.

The likelihood function for $N_{\mathrm{BG}}$ is given from the Poisson statistics as,
\bea
&\mathcal{L}&(N_\mathrm{BG} \mid N_\mathrm{A}^\mathrm{obs}, N_\mathrm{B}^\mathrm{obs}, N_\mathrm{C}^\mathrm{obs}) \nn
&=& P_\mathrm{Poi}(N_\mathrm{A}^\mathrm{obs} \mid N_\mathrm{A}^\mathrm{exp})P_\mathrm{Poi}(N_\mathrm{B}^\mathrm{obs} \mid N_\mathrm{B}^\mathrm{exp})P_\mathrm{Poi}(N_\mathrm{C}^\mathrm{obs} \mid N_\mathrm{C}^\mathrm{exp}).
\label{eq:BG_likehood}
\eea
The best estimate of $N_\mathrm{BG}$ can be obtained by maximising \eref{eq:BG_likehood} (listed in \tref{tab:number_of_BG2}).
%However, as we discuss in the following, $N_\mathrm{BG}$ is not inferred separately from the signal but inferred together with the signal.
However, we do not use this estimated $N_\mathrm{BG}$ in the inference of the signal but  use $(N_\mathrm{A}^\mathrm{obs}, N_\mathrm{B}^\mathrm{obs}, N_\mathrm{C}^\mathrm{obs})$ as discussed in the following.

%\section{Signal estimation}
\label{sec:signal_likelihood}
Our goal is to estimate the branching ratio of the MEx2G decay ($\mathcal{B}_{\mathrm{MEx2G}}$).
%For this purpose, we use the following items:
%\begin{itemize}
%    \item the number of events observed in the time sidebands,
%    \item the normalisation factor and its uncertainty,
%    \item the number of events observed in the signal region.
%\end{itemize}
The likelihood function \eref{eq:BG_likehood} is extended to include $\mathcal{B}_{\mathrm{MEx2G}}$ as a parameter and the number of events in the signal region ($N_\mathrm{S}^\mathrm{obs}$) as an observable.
In addition, to incorporate  the uncertainty in the SES into the $\mathcal{B}_{\mathrm{MEx2G}}$ estimation, the estimated SES ($s_0$) and the true value ($s$) are included into the likelihood function:
\bea
\mathcal{L}(\mathcal{B}_{\mathrm{MEx2G}}, N_\mathrm{BG}, s \mid N_\mathrm{S}^\mathrm{obs}, N_\mathrm{A}^\mathrm{obs}, N_\mathrm{B}^\mathrm{obs}, N_\mathrm{C}^\mathrm{obs}, s_0).
\label{eq:signal_likelihood}
\eea
%
%The expected number of events in the signal region ($N^{\mathrm{exp}}_S$) is written as, %with which $\mathcal{B}_{\mathrm{MEx2G}}=N_{\mathrm{S}}/k$:
%\bea
%N^{\mathrm{exp}}_\mathrm{S} = N_1 + N_2 + N_3 + k\mathcal{B}_{\mathrm{MEx2G}}.
%\label{eq:likelihood_start}
%\eea
%
Using $N_1, N_2, N_3$ and  a Gaussian PDF for the inverse of SES, it can be written as,
%\begin{strip}
\bea
&\mathcal{L}&(\mathrm{\mathcal{B}_{\mathrm{MEx2G}}}, N_1, N_2, N_3, s \mid N_\mathrm{S}^\mathrm{obs}, N_\mathrm{A}^\mathrm{obs}, N_\mathrm{B}^\mathrm{obs}, N_\mathrm{C}^\mathrm{obs}, s_0)\nn
&=& P_\mathrm{Poi}(N_\mathrm{S}^\mathrm{obs} \mid N^{\mathrm{exp}}_\mathrm{S})P_\mathrm{Poi}(N_\mathrm{A}^\mathrm{obs} \mid N_\mathrm{A}^\mathrm{exp})P_\mathrm{Poi}(N_\mathrm{B}^\mathrm{obs} \mid N_\mathrm{B}^\mathrm{exp})\nn
%&&\timesP_\mathrm{Poi}(N_\mathrm{C}^\mathrm{obs} \mid N_\mathrm{C}^\mathrm{exp})P_\mathrm{Gaus}(k_0 \mid k) \nn
%&=& P_\mathrm{Poi}(N_\mathrm{S}^\mathrm{obs} \mid k\mathrm{\mathcal{B}_{\mathrm{MEx2G}}} + N_1 + N_2 + N_3) 
%P_\mathrm{Poi}(N_\mathrm{A}^\mathrm{obs} \mid N_\mathrm{A}^\mathrm{exp})\nn
%&&\timesP_\mathrm{Poi}(N_\mathrm{B}^\mathrm{obs} \mid N_\mathrm{B}^\mathrm{exp})
&& \times P_\mathrm{Poi}(N_\mathrm{C}^\mathrm{obs} \mid N_\mathrm{C}^\mathrm{exp})P_\mathrm{Gaus}(s_0^{-1} \mid s^{-1}),
\label{eq:likehood_modified}
\eea
where $N^{\mathrm{exp}}_\mathrm{S}=N_1+N_2+N_3+\mathcal{B}_{\mathrm{MEx2G}}/s$ is the expected number of events in the signal region.
%%\end{strip}
%Once $\mathcal{B}_{\mathrm{MEx2G}}$ is determined, the number of signal can be changed by estimation of normalisation factor.
%We define $s$ to be the signal number assuming normalisation factor to be $k_0$ and satisfies $\mathcal{B}_{\mathrm{MEx2G}} = s/k_0$ and $N_{\mathrm{S}}=rs$.
%Dividing by a factor $k_0$ (a constant value) does not change the form of the likelihood function, which gives:
%\begin{strip}
%\bea
%&&\mathcal{L}(s, N_1, N_2, N_3, r \mid N_\mathrm{S}^\mathrm{obs}, N_\mathrm{A}^\mathrm{obs}, N_\mathrm{B}^\mathrm{obs}, N_\mathrm{C}^\mathrm{obs}, r_0) \nn
%&=& P_\mathrm{Poi}(N_\mathrm{S}^\mathrm{obs} \mid rs + N_1 + N_2 + N_3)P_\mathrm{Poi}(N_\mathrm{A}^\mathrm{obs} \mid N_\mathrm{A}^\mathrm{exp})P_\mathrm{Poi}(N_\mathrm{B}^\mathrm{obs} \mid N_\mathrm{B}^\mathrm{exp})\nn
%&&\timesP_\mathrm{Poi}(N_\mathrm{C}^\mathrm{obs} \mid N_\mathrm{C}^\mathrm{exp})P_\mathrm{Gaus}(r_0 \mid r),
%\label{eq:likehood_modified3}
%\eea
%%\end{strip}
%where  $r, r_0$ are defined by $r = k/k_0, r_0 = k_0/k_0(=1)$, respectively.
%We assume a Gaussian for the PDF of the normalisation.
%\bea
%P_\mathrm{Gaus}(r_0 \mid r)\sim\mathcal{N}(1 \mid r, \sigma_r), 
%\eea
%where $\sigma_r$ is the relative uncertainty of $k$.

The best estimated values of the parameter set $\{\mathcal{B}_{\mathrm{MEx2G}}$, $N_1$, $N_2$, $N_3$, $s\}$ are obtained by maximising \eref{eq:likehood_modified}. 
Among them, only $\mathcal{B}_{\mathrm{MEx2G}}$ is the interesting parameter, while the others are regarded as nuisance parameters $\vector{\nu}=(N_1, N_2, N_3, s)$.
%Once $s$ is obtained, $\mathcal{B}_{\mathrm{MEx2G}}$ is calculated by $\mathrm{\mathcal{B}_{\mathrm{MEx2G}}}=s/k_0$.

A frequentist test of the null (background-only) hypothesis is performed with the following profile likelihood ratio $\lambda_p$  as the test statistic\,\cite{PDG2018}:
\bea
\lambda_p(\mathcal{B}_{\mathrm{MEx2G}})=\frac{\mathcal{L}(\mathcal{B}_{\mathrm{MEx2G}}, \hat{\hat{\vector{\nu}}})}{\mathcal{L}(\hat{\mathcal{B}}_{\mathrm{MEx2G}}, \hat{\vector{\nu}})},
\label{eq:test_statistic}
\eea
where $\hat{\mathcal{B}}_{\mathrm{MEx2G}}$ and $\hat{\vector{\nu}}$ are the best-estimated values, and $\hat{\hat{\vector{\nu}}}$ is the value of $\vector{\nu}$ that maximises the likelihood at the fixed $\mathcal{B}_{\mathrm{MEx2G}}$.
The systematic uncertainties of  the background estimation and the SES are incorporated into the test by profiling the likelihood about $\vector{\nu}$.
%The p-value is calculated from the distribution of $\lambda_p$ and the value observed with the data. 
The local\footnote{Assuming that the signal is at the assumed $m_\mathrm{X}$.} significance is quantified by the p-value $p_\mathrm{local}$, defined as the probability to find $\lambda_p$ that is equally or less compatible with the null hypothesis than that observed with the data when the signal does not exist.

Since $m_\mathrm{X}$ is unknown, we need to take the look-elsewhere effect \cite{PDG2018} into account to calculate the global significance.
We estimate this effect following the approaches  in~\cite{Gross2010,Davies1987}, in which the trial factor of the search is estimated using an asymptotic property of $\lambda_p$, obeying the chi-square distribution. The smallest $p_\mathrm{local}$ in the $m_\mathrm{X}$ scan is converted into the global p-value $p_\mathrm{global}$ assuming that the signal can appear only at one $m_\mathrm{X}$.

The range of $\mathcal{B}_{\mathrm{MEx2G}}$ at 90\% C.L.\ is constructed based on the Feldman--Cousins unified approach~\cite{Cousins1998}
extended to use the profile-likelihood ratio as the ordering statistic 
in order to incorporate the systematic uncertainties~\cite{Cranmer2005}.

\begin{table}[tb]
\centering
\caption[The number of observed events in signal region]{The number of observed events in the sideband regions and the signal region and the expected number of background events in the signal region.}
\label{tab:number_of_BG2}
\begin{tabular}{lccccc}
\hline
$m_\mathrm{X}$ (MeV/c$^2$) & $N_\mathrm{A}^\mathrm{obs}$ & $N_\mathrm{B}^\mathrm{obs}$ & $N_\mathrm{C}^\mathrm{obs}$ & $N_{\mathrm{BG}}$ & $N_\mathrm{S}^\mathrm{obs}$ \\ \hline             \vspace{1mm}
20                          & 0        & 0       & 1       & $0.048_{-0.046}^{+0.202}$ & 1                   \\ \vspace{1mm}
21                          & 0        & 0       & 3       & $0.146_{-0.084}^{+0.198}$ & 0                   \\ \vspace{1mm}
22                          & 1        & 0       & 5       & $0.292_{-0.140}^{+0.211}$ & 0                   \\ \vspace{1mm}
23                          & 3        & 0       & 3       & $0.622_{-0.330}^{+0.425}$ & 0                   \\ \vspace{1mm}
24                          & 2        & 0       & 1       & $0.414_{-0.260}^{+0.346}$ & 1                   \\ \vspace{1mm}
25                          & 2        & 0       & 3       & $0.414_{-0.261}^{+0.346}$ & 0                   \\ \hline \vspace{1mm}
26                          & 0        & 0       & 3       & $0.150_{-0.091}^{+0.189}$ & 0                   \\ \vspace{1mm}
27                          & 0        & 0       & 1       & $0.050_{-0.049}^{+0.200}$ & 0                   \\ \vspace{1mm}
28                          & 0        & 0       & 1       & $0.048_{-0.046}^{+0.202}$ & 0                   \\ \vspace{1mm}
29                          & 0        & 0       & 1       & $0.048_{-0.046}^{+0.202}$ & 1                   \\ \vspace{1mm}
30                          & 0        & 0       & 0       & $0.000_{-0.000}^{+0.170}$ & 0                   \\ \hline \vspace{1mm}
31                          & 0        & 0       & 1       & $0.048_{-0.046}^{+0.202}$ & 0                   \\ \vspace{1mm}
32                          & 0        & 0       & 0       & $0.000_{-0.000}^{+0.170}$ & 0                   \\ \vspace{1mm}
33                          & 0        & 0       & 0       & $0.000_{-0.000}^{+0.210}$ & 0                   \\ \vspace{1mm}
34                          & 0        & 0       & 0       & $0.000_{-0.000}^{+0.210}$ & 1                   \\ \vspace{1mm}
35                          & 0        & 0       & 0       & $0.000_{-0.000}^{+0.210}$ & 2                   \\ \hline \vspace{1mm}
36                          & 0        & 0       & 0       & $0.000_{-0.000}^{+0.210}$ & 2                   \\ \vspace{1mm}
37                          & 1        & 0       & 0       & $0.400_{-0.301}^{+0.517}$ & 1                   \\ \vspace{1mm}
38                          & 0        & 0       & 2       & $0.168_{-0.105}^{+0.183}$ & 0                   \\ \vspace{1mm} 
39                          & 0        & 0       & 1       & $0.084_{-0.084}^{+0.201}$ & 0                   \\ \vspace{1mm}
40                          & 0        & 0       & 0       & $0.000_{-0.000}^{+0.210}$ & 0                   \\\hline \vspace{1mm}
41                          & 0        & 0       & 0       & $0.000_{-0.000}^{+0.210}$ & 0                   \\ \vspace{1mm}
42                          & 0        & 0       & 0       & $0.000_{-0.000}^{+0.210}$ & 0                   \\ \vspace{1mm}
43                          & 0        & 0       & 1       & $0.084_{-0.084}^{+0.201}$ & 0                   \\ \vspace{1mm}
44                          & 0        & 0       & 0       & $0.000_{-0.000}^{+0.210}$ & 0                   \\ \vspace{1mm}
45                          & 0        & 0       & 0       & $0.000_{-0.000}^{+0.210}$  & 0                  \\ \hline
\end{tabular}
\end{table}
\section{Results and discussion}
\label{sec:results}
%\subsubsection{Observed events in the signal region}
%\label{sec:unblinding}
\tref{tab:number_of_BG2} summarises the numbers of  events in the signal region and the sidebands as well as the expected number of background events in the signal region.
We observe non-zero events in the signal region for some masses.
Note that the adjacent $m_\mathrm{X}$ bins are not statistically independent.
Summing up the observed events gives nine events but five of them are unique events.
One event appears in four bins ($m_\mathrm{X}=$ 34, 35, 36, 37~MeV/c$^2$) and another event appears in two bins ($m_\mathrm{X}=$ 35, 36~MeV/c$^2$).
%\subsubsection{Branching ratio limits}

We discuss the results for $\tau_\mathrm{X}=20$~ps below. The results for other $\tau_\mathrm{X}$ are similar, with small changes in the efficiency.
The results are presented in detail in Appendix~\ref{sec:detailed_results}.

\Fref{fig:MEx2GFinal_summary_comp_20} shows 90\% confidence intervals on $\mathcal{B}_\mathrm{MEx2G}$ obtained from this analysis together with the sensitivities and the previous upper limits due to Crystal Box.
The sensitivities %\footnote{%Assuming a null signal hypothesis, the expected number of observed events ($N_{\mathrm{exp}, \mathrm{obs}}$) is 0 or 1.
%We define the sensitivity as the mean upper limit for a null signal hypothesis.} %as weighted average of probabilities of these two cases;
%The sensitivity is given by internal division points between expected branching ratio limits for $N_{\mathrm{exp}, \mathrm{obs}}=0$ and $N_{\mathrm{exp}, \mathrm{obs}}=1$ with their Poisson probabilities.} 
are evaluated by the mean of the branching ratio limits at 90\% C.L.\ under the null hypothesis.
%The upper bound of the blue region corresponds to the upper limits of the branching ratio of \mextwog.
Note that since we adopt the Feldman--Cousins unified approach, a one-sided or two-sided interval is automatically determined according to the data. Therefore, lower limits can be set in $m_\mathrm{X}$ regions where non-zero events are observed with small $N_\mathrm{BG}$.
%Since we observed excess of events and uses Feldman--Cousins to calculate the confidence interval of the signal number, some masses also have lower limits.
\begin{figure}[tb]
	\centering
	\includegraphics[width=\columnwidth]{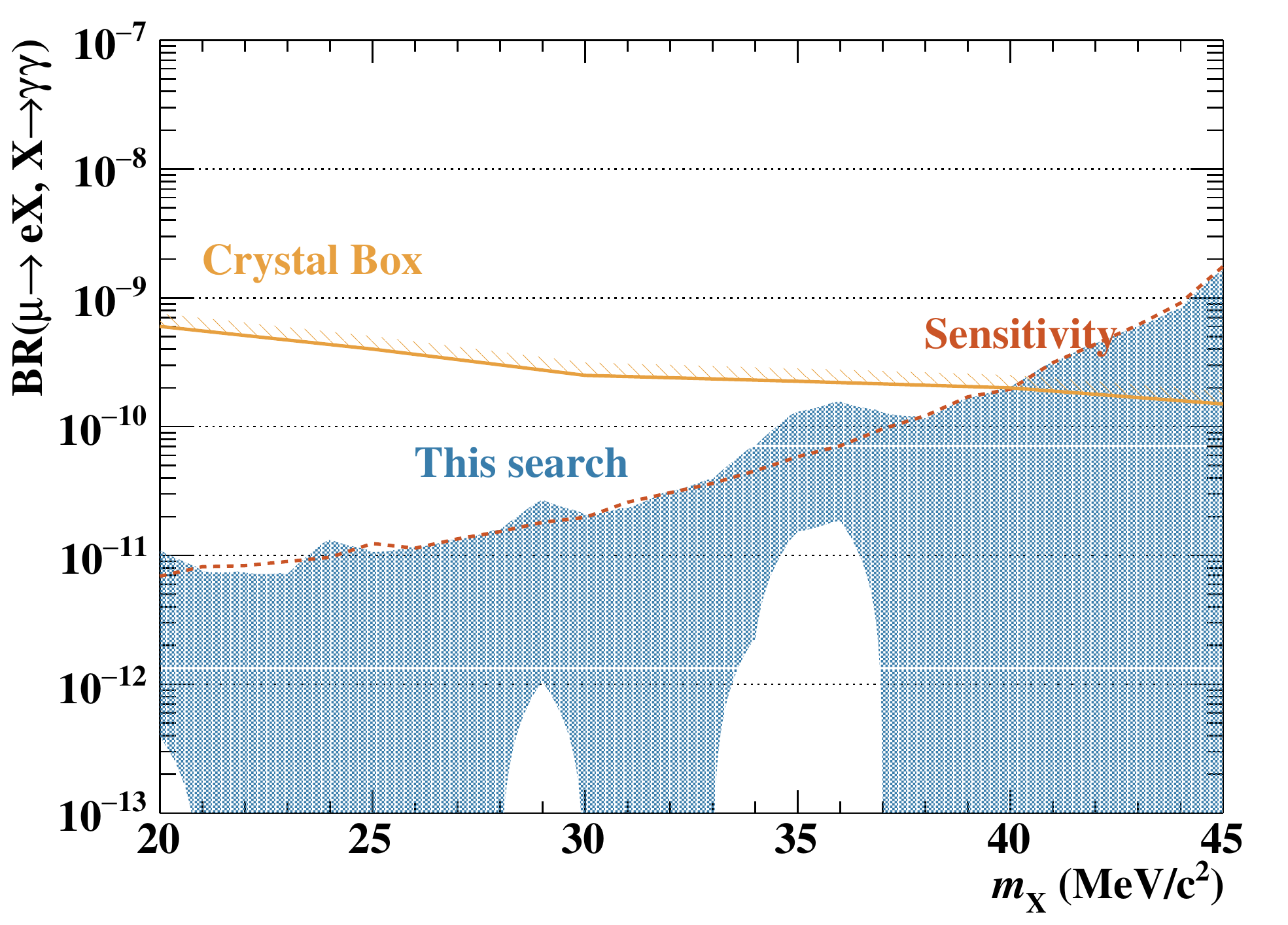}
	\caption[Branching ratio limits]{Confidence intervals (90\% C.L.) on $\mathcal{B}_\mathrm{MEx2G}$ (blue band) for $\tau_\mathrm{X}=20$~ps. The red broken line shows the expected upper limits under the null hypothesis and the yellow line shows the limits extracted by Crystal Box analysis.}
	\label{fig:MEx2GFinal_summary_comp_20}
\end{figure}

%\subsubsection{p-value and significance}
%\paragraph{p-value}
\begin{figure}[t!]
	\centering
	\includegraphics[width=\columnwidth]{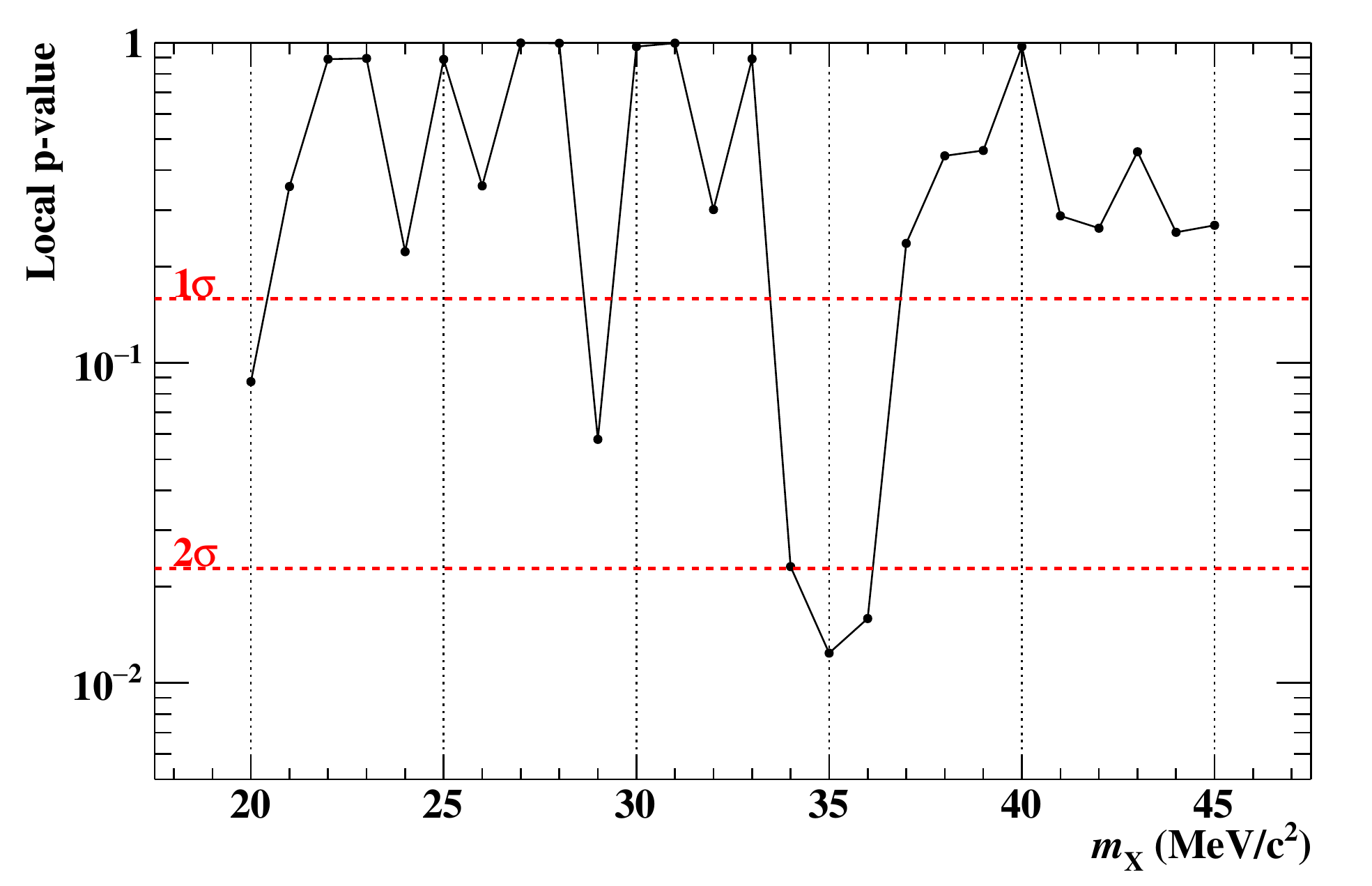}
	\caption{Local p-value under null hypothesis as a function of assumed $m_\mathrm{X}$.}
	\label{fig:summary_pvalue}
\end{figure}

The statistical significance of the excesses is tested against the null hypothesis.
\Fref{fig:summary_pvalue} shows  $p_\mathrm{local}$ versus $m_\mathrm{X}$.
We observe the lowest $p_\mathrm{local} = 0.012$ at $m_\mathrm{X}=35$~MeV/c$^2$, which corresponds to 2.2$\sigma$ significance.
%\paragraph*{Look elsewhere effect}\,
The global p-value is calculated to be $p_\mathrm{global} \approx 0.10$
%\bea
%p_{\mathrm {global }} \approx p_{\mathrm {local }}+\mathrm{correction}\approx0.10,
%\eea
by taking the look-elsewhere effect into account.
This corresponds to 1.3$\sigma$, that is not statistically significant.
%\section{Discussion}

%We observed excess events in some mass regions.
%The lowest p-value of 0.012 was observed at 35 MeV, which corresponds to 2.2 $\sigma$ (local) and 1.3 $\sigma$ (global) significance.
%Therefore, this excess is not statistically significant.
%
Owing to the large statistics of the MEG dataset, the branching ratio upper limits have been reduced to the level of $\mathcal{O}(10^{-11})$.
Our results improves the upper limits from the Crystal Box experiment for $m_{\mathrm{X}}< 40\,\mathrm{MeV/c^2}$, by a factor of 60 at most. 
%These limits are improved by a factor of 4.4--13.0 depending on $m_{\mathrm{X}}$ from the preliminary results obtained using the first two years of the MEG data\,\cite{Natori2012}.
%The improvement is higher for larger $m_{\mathrm{X}}$.
%This improvement with respect to the earlier result is due to both increased statistics and improvement in the analysis, especially to a better background cut optimisation.
%%This values include the statistical fluctuation of the observed numbers of events.
%%
%%Single event sensitivity (SES) is improved by a factor of 5.6--13 depending on $m_{\mathrm{X}}$.
%%Statistics and positron analysis updates contribute at most $\sim$ 5 times smaller SES compared with the previous analysis.
%%$m_{\mathrm{X}}$ dependence of the improvements mainly come from the optimisation of selection conditions.
%%We set the criteria to optimise selection conditions and $m_{\mathrm{X}}$ dependent thresholds are determined in Sect.~\ref{sec:cut_condition} while selection conditions were not optimised well in the previous analysis.
%%Selection efficiency is improved by at most $\sim$3 in 45 MeV.

%Since we used the full dataset of the MEG experiment, we need new experiments for additional explorations of the decay, e.g.\ to test whether the small excess observed in this search grows.
This publication reports results from the full MEG dataset. Hence, new experiments will be needed for further exploration of this decay, e.g.\ to test whether the small excess observed in this search grows.
An upgraded experiment, MEG II, is currently being prepared\,\cite{baldini_2018}.
A brief prospect for improved sensitivity to MEx2G in MEG II  is discussed below.
%The beam intensity was forced to be reduced in order to suppress the accidental background and operate the detector stably.
%We plan to upgrade all the detectors to make the maximum use of the muon beam at the world's highest intensity ($7\times10^7\muup/$s) at Paul Scherrer Institut in Switzerland.
%The experimental sensitivity of the MEx2G search is expected to be improved by one order of magnitude using the MEG II datasets in all mass regions.
%statistics, positron efficiency
In this analysis the sensitivity worsens with increasing $m_{\mathrm{X}}$, mainly due to the 2$\gammaup$ acceptance and direction match efficiencies.
The acceptance is determined by the geometry of the LXe photon detector and is not changed by the upgrade.
The direction match efficiency can even worsen if we  only  consider the \meg search;
the $\gammaup$ position resolution is expected to improve by a factor two, which enables tightening the direction match trigger condition.
%This means that non back-to-back events are more unlikely to be triggered, resulting in a worse trigger efficiency for \mextwog.
However, the MEG II trigger development is underway and the trigger efficiency for high mass can be improved up to a factor $\sim 2$ if a dedicated trigger is prepared.
%This will be a matter of discussion among the MEG II collaboration.
Basically, MEG II will collect ten times more $\muup^+$ decays and the resolutions of each kinematic variable will improve by roughly a factor two, leading to
higher efficiency while maintaining low background. It is therefore possible to improve the sensitivity by one order of magnitude.

%All the resolutions will be improved by factor two, which will also improve the signal selection efficiency to some extent.
%%The improvements of $\gammaup$ position resolution makes separation of 2$\gammaup$ better.
%There is no dedicated time calibration source for 2$\gammaup$ events in this analysis.
%Developments of dedicated analysis tools for 2$\gammaup$ from annihilation-in-flight positrons provide us with the new time calibration source.
%We may set tighter selection conditions for timing distributions, resulting in better selection efficiency.

% --------------------------- 4_conclusion ---------------------------
\section{Conclusions}
\label{sec:conclusions}

We have searched for a lepton-flavour-violating muon decay mediated by a new light particle,
%The charged lepton flavor violation is one of the powerful tools to search for new physics beyond the standard model.
%On the other hand, light new physics has attracted a great deal of attention.
%In the analysis performed in this thesis, we combined these two different directions and have searched for the 
\mextwog decay, for the first time using the full dataset (2009--2013) of the MEG experiment.
No significant excess was found in the mass range $m_\mathrm{X} = 20$--45~MeV/c$^2$ and $\tau_\mathrm{X}< 40$~ps,
and we set new branching ratio upper limits in the mass range $m_\mathrm{X} = 20$--40~MeV/c$^2$.
In particular, the upper limits are lowered to the level of $\mathcal{O}(10^{-11})$ for $m_\mathrm{X} = 20$--30~MeV/c$^2$.
The result is up to 60 times more stringent than the bound converted from the previous experiment, Crystal Box.
\begin{appendices}
\section{
Detailed results for different lifetimes}
\label{sec:detailed_results}
The detailed results for different lifetimes $\tau_\mathrm{X}$ are summarised in Tables~\ref{tab:results_5ps}--\ref{tab:results_40ps}.

\begin{table}[tb]
\centering
\caption{Results for $\tau_\mathrm{X}=5$~ps. LL and UL denote the lower limit and upper limit of the 90\% confidence interval.}
\label{tab:results_5ps}
\begin{tabular}{lccc}
\hline
$m_\mathrm{X}$ (MeV/c$^2$) & $s_0$ & LL & UL \\ \hline
20          & $(3.29 \pm 0.31)\times10^{-12}$     & $4.60\times 10^{-13}$  &  $1.28 \times 10^{-11}$    \\
21         & $(3.56 \pm 0.44)\times10^{-12}$      & --       & $7.78 \times 10^{-12}$    \\
22         & $(3.73 \pm 0.44)\times10^{-12}$      & --       & $8.30 \times 10^{-12}$     \\
23         & $(3.96 \pm 0.45)\times10^{-12}$      & --       & $7.72 \times 10^{-12}$     \\
24         & $(4.26 \pm 0.49)\times10^{-12}$      & --       & $1.43 \times 10^{-11}$        \\
25         & $(5.41 \pm 0.55)\times10^{-12}$      & --       & $1.08 \times 10^{-11}$ \\ \hline
26          & $(5.16 \pm 0.61)\times10^{-12}$     & --       & $1.13 \times 10^{-11}$       \\
27          & $(5.81 \pm 0.70)\times10^{-12}$     & --       & $1.39 \times 10^{-11}$       \\
28          & $(6.62 \pm 0.81)\times10^{-12}$     & --       & $1.58 \times 10^{-11}$       \\
29          & $(7.67 \pm 0.95)\times10^{-12}$     & $8.53 \times 10^{-13}$  & $2.93 \times 10^{-11}$     \\
30          & $(8.62 \pm 0.85)\times10^{-12}$     & --       & $2.26 \times 10^{-11}$       \\ \hline
31          & $(1.07 \pm 0.14)\times10^{-11}$     & --       & $2.54 \times 10^{-11}$       \\
32          & $(1.29 \pm 0.17)\times10^{-11}$     & --       & $3.20 \times 10^{-11}$       \\
33          & $(1.59 \pm 0.21)\times10^{-11}$     & --       & $4.00 \times 10^{-11}$        \\
34          & $(1.97 \pm 0.26)\times10^{-11}$     & $1.97 \times 10^{-12}$    & $7.66 \times 10^{-11}$  \\
35          & $(2.60 \pm 0.31)\times10^{-11}$     & $1.65 \times 10^{-11}$   & $1.40 \times 10^{-10}$   \\ \hline
36          & $(3.18 \pm 0.42)\times10^{-11}$     & $1.98 \times 10^{-11}$     & $1.72 \times 10^{-10}$      \\
37          & $(4.12 \pm 0.53)\times10^{-11}$     & --      & $1.43 \times 10^{-10}$      \\
38          & $(5.43 \pm 0.70)\times10^{-11}$     & --      & $1.27 \times 10^{-10}$       \\
39          & $(7.25 \pm 0.93)\times10^{-11}$     & --      & $1.79 \times 10^{-10}$      \\
40          & $(9.01 \pm 0.93)\times10^{-11}$     & --      & $2.20 \times 10^{-10}$       \\ \hline
41          & $(1.35 \pm 0.19)\times10^{-10}$     & --      & $3.37 \times 10^{-10}$        \\
42          & $(1.88 \pm 0.27)\times10^{-10}$     & --      & $4.72 \times 10^{-10}$        \\
43          & $(2.66 \pm 0.39)\times10^{-10}$     & --      & $6.47 \times 10^{-10}$        \\
44          & $(3.81 \pm 0.57)\times10^{-10}$     & --      & $1.01 \times 10^{-9}$      \\
45          & $(6.25 \pm 0.88)\times10^{-10}$     & --      & $1.59 \times 10^{-9}$      \\ \hline
\end{tabular}
\end{table}

\begin{table}[tb]
\centering
\caption{Results for $\tau_\mathrm{X}=20$~ps. LL and UL denote the lower limit and upper limit of the 90\% confidence interval.}
\label{tab:results_20ps}
\begin{tabular}{lccc}
\hline
$m_\mathrm{X}$ (MeV/c$^2$) & $s_0$ & LL & UL \\ \hline
20         & $(2.92 \pm 0.28)\times10^{-12}$     & $3.94\times 10^{-13}$  &  $1.10 \times 10^{-11}$    \\
21         & $(3.18 \pm 0.39)\times10^{-12}$     & --       & $7.54 \times 10^{-12}$    \\
22         & $(3.35 \pm 0.39)\times10^{-12}$     & --       & $7.36 \times 10^{-12}$     \\
23         & $(3.57 \pm 0.40)\times10^{-12}$     & --       & $7.14 \times 10^{-12}$     \\
24         & $(3.86 \pm 0.43)\times10^{-12}$     & --       & $1.33 \times 10^{-11}$        \\
25         & $(4.74 \pm 0.43)\times10^{-12}$     & --       & $1.06 \times 10^{-11}$ \\ \hline
26         & $(4.71 \pm 0.53)\times10^{-12}$     & --       & $1.15 \times 10^{-11}$       \\
27         & $(5.31 \pm 0.62)\times10^{-12}$     & --       & $1.33 \times 10^{-11}$       \\
28         & $(6.07 \pm 0.72)\times10^{-12}$     & --       & $1.58 \times 10^{-11}$       \\
29         & $(7.04 \pm 0.85)\times10^{-12}$     & $1.04 \times 10^{-12}$  & $2.70 \times 10^{-11}$     \\
30         & $(7.94 \pm 0.78)\times10^{-12}$     & --       & $2.07 \times 10^{-11}$       \\ \hline
31         & $(9.86 \pm 1.26)\times10^{-12}$     & --       & $2.33 \times 10^{-11}$       \\
32         & $(1.19 \pm 0.15)\times10^{-11}$     & --       & $3.14 \times 10^{-11}$       \\
33         & $(1.46 \pm 0.19)\times10^{-11}$     & --       & $3.92 \times 10^{-11}$        \\
34         & $(1.82 \pm 0.23)\times10^{-11}$     & $2.25 \times 10^{-12}$    & $7.10 \times 10^{-11}$  \\
35         & $(2.38 \pm 0.25)\times10^{-11}$     & $1.53 \times 10^{-11}$   & $1.31 \times 10^{-10}$   \\ \hline
36         & $(2.93 \pm 0.37)\times10^{-11}$     & $1.85 \times 10^{-11}$     & $1.56 \times 10^{-10}$      \\
37         & $(3.79 \pm 0.47)\times10^{-11}$     & --      & $1.29 \times 10^{-10}$      \\
38         & $(4.99 \pm 0.62)\times10^{-11}$     & --      & $1.16 \times 10^{-10}$       \\
39         & $(6.65 \pm 0.83)\times10^{-11}$     & --      & $1.66 \times 10^{-10}$      \\
40         & $(8.20 \pm 0.87)\times10^{-11}$     & --      & $2.04 \times 10^{-10}$       \\ \hline
41         & $(1.23 \pm 0.20)\times10^{-10}$     & --      & $3.15 \times 10^{-10}$        \\
42         & $(1.71 \pm 0.28)\times10^{-10}$     & --      & $4.44 \times 10^{-10}$        \\
43         & $(2.41 \pm 0.41)\times10^{-10}$     & --      & $5.98 \times 10^{-10}$        \\
44         & $(3.44 \pm 0.61)\times10^{-10}$     & --      & $8.25 \times 10^{-10}$      \\
45         & $(6.29 \pm 1.12)\times10^{-10}$     & --      & $1.63 \times 10^{-9}$      \\ \hline
\end{tabular}
\end{table}

\begin{table}[tb]
\centering
\caption{Results for $\tau_\mathrm{X}=40$~ps. LL and UL denote the lower limit and upper limit of the 90\% confidence interval.}
\label{tab:results_40ps}
\begin{tabular}{lccc}
\hline
$m_\mathrm{X}$ (MeV/c$^2$) & $s_0$ & LL & UL \\ \hline
20         & $(3.02 \pm 0.29)\times10^{-12}$     & $3.78\times 10^{-13}$  &  $1.19 \times 10^{-11}$    \\
21         & $(3.28 \pm 0.39)\times10^{-12}$     & --       & $7.44 \times 10^{-12}$    \\
22         & $(3.44 \pm 0.39)\times10^{-12}$     & --       & $7.80 \times 10^{-12}$     \\
23         & $(3.66 \pm 0.40)\times10^{-12}$     & --       & $7.41 \times 10^{-12}$     \\
24         & $(3.94 \pm 0.43)\times10^{-12}$     & --       & $1.43 \times 10^{-11}$        \\
25         & $(4.83 \pm 0.43)\times10^{-12}$     & --       & $1.04 \times 10^{-11}$ \\ \hline
26         & $(4.76 \pm 0.54)\times10^{-12}$     & --       & $1.09 \times 10^{-11}$       \\
27         & $(5.35 \pm 0.61)\times10^{-12}$     & --       & $1.33 \times 10^{-11}$       \\
28         & $(6.09 \pm 0.71)\times10^{-12}$     & --       & $1.51 \times 10^{-11}$       \\
29         & $(7.03 \pm 0.83)\times10^{-12}$     & $9.22 \times 10^{-13}$  & $2.71 \times 10^{-11}$     \\
30         & $(7.87 \pm 0.78)\times10^{-12}$     & --       & $1.88 \times 10^{-11}$       \\ \hline
31         & $(9.78 \pm 1.21)\times10^{-12}$     & --       & $2.33 \times 10^{-11}$       \\
32         & $(1.18 \pm 0.15)\times10^{-11}$     & --       & $2.92 \times 10^{-11}$       \\
33         & $(1.44 \pm 0.18)\times10^{-11}$     & --       & $3.57 \times 10^{-11}$        \\
34         & $(1.78 \pm 0.22)\times10^{-11}$     & $1.97 \times 10^{-12}$    & $6.81 \times 10^{-11}$  \\
35         & $(2.30 \pm 0.23)\times10^{-11}$     & $1.47 \times 10^{-11}$   & $1.26 \times 10^{-10}$   \\ \hline
36         & $(2.84 \pm 0.34)\times10^{-11}$     & $1.78 \times 10^{-11}$     & $1.54 \times 10^{-10}$      \\
37         & $(3.67 \pm 0.43)\times10^{-11}$     & --      & $1.28 \times 10^{-10}$      \\
38         & $(4.80 \pm 0.56)\times10^{-11}$     & --      & $1.12 \times 10^{-10}$       \\
39         & $(6.37 \pm 0.75)\times10^{-11}$     & --      & $1.51 \times 10^{-10}$      \\
40         & $(7.94 \pm 0.78)\times10^{-11}$     & --      & $1.90 \times 10^{-10}$       \\ \hline
41         & $(1.17 \pm 0.18)\times10^{-10}$     & --      & $2.84 \times 10^{-10}$        \\
42         & $(1.62 \pm 0.25)\times10^{-10}$     & --      & $4.28 \times 10^{-10}$        \\
43         & $(2.27 \pm 0.37)\times10^{-10}$     & --      & $5.94 \times 10^{-10}$        \\
44         & $(3.23 \pm 0.54)\times10^{-10}$     & --      & $8.31 \times 10^{-10}$      \\
45         & $(5.65 \pm 0.97)\times10^{-10}$     & --      & $1.53 \times 10^{-9}$      \\ \hline
\end{tabular}
\end{table}
\end{appendices}

\section*{Acknowledgments}

We are grateful for the support and co-operation provided 
by PSI as the host laboratory and to the technical and 
engineering staff of our institutes. This work is
supported by DOE DEFG02-91ER40679 (USA); INFN
(Italy); MIUR Montalcini D.M. 2014 n. 975 (Italy); JSPS KAKENHI numbers JP22000004, JP26000004, JP17J04114, and JSPS Core-to-Core Program, A. Advanced Research Networks JPJSCCA20180004 (Japan);
Schweizerischer Nationalfonds (SNF) Grant 200021\_137738 and Grant 200020\_172706; the Russian Federation Ministry of Education and Science, and Russian Fund
for Basic Research grant RFBR-14-22-03071.

\bibliographystyle{my}
\bibliography{MEG}

\end{document}

%% file: author-meg-epjc.tex
\input{institute}

\date{Received: date / Accepted: date}

\author{
The MEG collaboration\\\\
        A.~M.~Baldini~\thanksref{addr4}$^a$ \and
%        Y.~Bao~\thanksref{addr1} \and      
%        E.~Baracchini~\thanksref{addr3,addr17} \and
%       C.~Bemporad~\thanksref{addr4}$^{ab}$ \and
        F.~Berg~\thanksref{addr1,addr2} \and
        M.~Biasotti~\thanksref{addr5}$^{ab}$ \and
        G.~Boca~\thanksref{addr7}$^{ab}$ \and
        P.~W.~Cattaneo~\thanksref{addr7}$^{a}$  \and
        G.~Cavoto~\thanksref{addr8}$^{ab}$ \and
        F.~Cei~\thanksref{addr4}$^{ab}$ \and
%        C.~Cerri~\thanksref{addr4}$^{a}$ \and
        M.~Chiappini~\thanksref{addr4}$^{ab}$ \and
        G.~Chiarello~\thanksref{addr8}$^{ab}$ \and
        C.~Chiri~\thanksref{addr6}$^{ab}$ \and
        A.~Corvaglia~\thanksref{addr6}$^{ab}$ \and
        A.~de~Bari~\thanksref{addr7}$^{ab}$ \and
        M.~De~Gerone~\thanksref{addr5}$^{a}$ \and
%        T.~Doke~\thanksref{e2,addr10} \and
%        A.~D'Onofrio~\thanksref{addr4}$^{ab}$ \and
%        S.~Dussoni~\thanksref{addr4}$^{a}$\and
%        J.~Egger~\thanksref{addr1} \and
        M.~Francesconi~\thanksref{addr4}$^{a}$ \and
%        Y.~Fujii~\thanksref{addr3}  \and
        L.~Galli~\thanksref{addr4}$^{a}$ \and
        F.~Gatti~\thanksref{addr5}$^{ab}$ \and
        F.~Grancagnolo~\thanksref{addr6}$^{a}$ \and
        M.~Grassi~\thanksref{addr4}$^{a}$ \and
%        A.~Graziosi~\thanksref{addr8}$^{ab}$ \and
        D.~N.~Grigoriev~\thanksref{addr12,addr14,addr15} \and
%        T.~Haruyama~\thanksref{addr9} \and
        M.~Hildebrandt~\thanksref{addr1} \and
        Z.~Hodge~\thanksref{addr1,addr2} \and
        K.~Ieki~\thanksref{addr3}  \and
        F.~Ignatov~\thanksref{addr12,addr15} \and
        R.~Iwai~\thanksref{addr3}  \and
        T.~Iwamoto~\thanksref{addr3}  \and
%        D.~Kaneko~\thanksref{addr3}  \and
        S.~Kobayashi~\thanksref{addr3}  \and
%        T.~I.~Kang~\thanksref{addr11}  \and
        P.-R.~Kettle~\thanksref{addr1} \and
%        B.~I.~Khazin~\thanksref{e2,addr12,addr15} \and
        W.~Kyle~\thanksref{addr11} \and
        N.~Khomutov~\thanksref{addr13} \and
%        A.~Korenchenko~\thanksref{e2,addr13}  \and
        A.~Kolesnikov~\thanksref{addr13}  \and
        N.~Kravchuk~\thanksref{addr13}  \and
%        V.~Krylov~\thanksref{addr13}  \and
        N.~Kuchinskiy~\thanksref{addr13}  \and
        T.~Libeiro~\thanksref{addr11} \and
        G.~M.~A.~Lim~\thanksref{addr11} \and
%        A.~Maki~\thanksref{addr9}  \and
        V.~Malyshev~\thanksref{addr13}  \and
        N.~Matsuzawa~\thanksref{addr3}  \and
        M.~Meucci~\thanksref{addr8}$^{ab}$ \and
        S.~Mihara~\thanksref{addr9}  \and
        W.~Molzon~\thanksref{addr11} \and
        Toshinori~Mori~\thanksref{addr3}  \and
%        F.~Morsani~\thanksref{addr4}$^{a}$ \and
        A.~Mtchedilishvili~\thanksref{addr1}  \and
%        D.~Mzavia~\thanksref{e2,addr13}  \and 
        M.~Nakao~\thanksref{e1, addr3}  \and 
%        S.~Nakaura~\thanksref{addr3}  \and
        H.~Natori~\thanksref{addr3} \and 
%        R.~Nard\`o~\thanksref{addr7}$^{ab}$ \and
        D.~Nicol\`o~\thanksref{addr4}$^{ab}$ \and
        H.~Nishiguchi~\thanksref{addr9}  \and
        M.~Nishimura~\thanksref{addr3}  \and 
        S.~Ogawa~\thanksref{addr3}  \and
        R.~Onda~\thanksref{addr3}  \and
        W.~Ootani~\thanksref{addr3}  \and
%        S.~Orito~\thanksref{e2,addr3}  \and
        A.~Oya~\thanksref{addr3}  \and
        D.~Palo~\thanksref{addr11} \and
        M.~Panareo~\thanksref{addr6}$^{ab}$ \and
        A.~Papa~\thanksref{addr1,addr4}$^{ab}$ \and
%        R.~Pazzi~\thanksref{e2,addr4} \and
%        A.~Pepino~\thanksref{addr6}$^{ab}$ \and
        V.~Pettinacci~\thanksref{addr8}$^{a}$ \and
%        G.~Piredda~\thanksref{addr8}$^{a}$ \and
        G.~Pizzigoni~\thanksref{addr5}$^{ab}$ \and
        A.~Popov~\thanksref{addr12,addr15} \and
%        F.~Raffaelli~\thanksref{addr4}$^{a}$ \and
        F.~Renga~\thanksref{addr8}$^{a}$ \and
%        E.~Ripiccini~\thanksref{addr8}$^{ab}$ \and
        S.~Ritt~\thanksref{addr1} \and
        A.~Rozhdestvensky~\thanksref{addr13}  \and
        M.~Rossella~\thanksref{addr7}$^{a}$ \and
%        G.~Rutar~\thanksref{addr1,addr2} \and
        R.~Sawada~\thanksref{addr3} \and
        P.~Schwendimann~\thanksref{addr1} \and
%        F.~Sergiampietri~\thanksref{addr4}$^{a}$ \and
        G.~Signorelli~\thanksref{addr4}$^{a}$ \and
%        M.~Simonetta~\thanksref{addr7}$^{ab}$  \and
        A.~Stoykov~\thanksref{addr1} \and
        G.~F.~Tassielli~\thanksref{addr6}$^{a}$ \and
%        F.~Tenchini~\thanksref{addr4}$^{ab}$ \and
        K.~Toyoda~\thanksref{addr3} \and
        Y.~Uchiyama~\thanksref{addr3} \and
        M.~Usami~\thanksref{addr3} \and
%        M.~Venturini~\thanksref{addr4}$^{a,}$~\thanksref{addr16} \and
        C.~Voena~\thanksref{addr8}$^{a}$ \and
%        A.~Yamamoto~\thanksref{addr9} \and
        K.~Yanai~\thanksref{addr3} \and
%        K.~Yoshida~\thanksref{addr3} \and
%        Z.~You~\thanksref{addr11} \and
        Yu.V.~Yudin~\thanksref{addr12,addr15} %\and
%        D.~Zanello~\thanksref{addr8}
}

\institute{\INFNPi \label{addr4}
           \and
             \PSI \label{addr1} 
           \and
              \ETHZ \label{addr2}
%             \email{fauthor@example.com}           %  \\
           \and
             \INFNGe \label{addr5}
           \and
             \INFNPv \label{addr7}
           \and
             \INFNRm \label{addr8}
           \and
             \INFNLe \label{addr6} 
           \and
             \BINP   \label{addr12}
           \and
             \NOVST  \label{addr14}
           \and
             \NOVS   \label{addr15}
           \and
              \ICEPP \label{addr3}
           \and
             \UCI    \label{addr11}
           \and
             \JINR   \label{addr13}
%             \Waseda \label{addr10}
           \and
             \KEK    \label{addr9}
%             \ScuolaPi \label{addr16}   
%             \INFNLNF \label{addr17}   
}

%% file: institute.tex
\newcommand*{\INFNPi}{INFN Sezione di Pisa$^{a}$; Dipartimento di Fisica$^{b}$ dell'Universit\`a, Largo B.~Pontecorvo~3, 56127 Pisa, Italy}
\newcommand*{\INFNGe}{INFN Sezione di Genova$^{a}$; Dipartimento di Fisica$^{b}$ dell'Universit\`a, Via Dodecaneso 33, 16146 Genova, Italy}
\newcommand*{\INFNPv}{INFN Sezione di Pavia$^{a}$; Dipartimento di Fisica$^{b}$ dell'Universit\`a, Via Bassi 6, 27100 Pavia, Italy}
\newcommand*{\INFNRm}{INFN Sezione di Roma$^{a}$; Dipartimento di Fisica$^{b}$ dell'Universit\`a ``Sapienza'', Piazzale A.~Moro, 00185 Roma, Italy}
\newcommand*{\INFNLe}{INFN Sezione di Lecce$^{a}$; Dipartimento di Matematica e Fisica$^{b}$ dell'Universit\`a del Salento, Via per Arnesano, 73100 Lecce, Italy}
\newcommand*{\ICEPP} {ICEPP, The University of Tokyo, 7-3-1 Hongo, Bunkyo-ku, Tokyo 113-0033, Japan }
\newcommand*{\UCI}   {University of California, Irvine, CA 92697, USA}
\newcommand*{\KEK}   {KEK, High Energy Accelerator Research Organization, 1-1 Oho, Tsukuba, Ibaraki 305-0801, Japan}
\newcommand*{\PSI}   {Paul Scherrer Institut PSI, 5232 Villigen, Switzerland}
\newcommand*{\Waseda}{Research Institute for Science and Engineering, Waseda~University, 3-4-1 Okubo, Shinjuku-ku, Tokyo 169-8555, Japan}
\newcommand*{\BINP}  {Budker Institute of Nuclear Physics of Siberian Branch of Russian Academy of Sciences, 630090 Novosibirsk, Russia}
\newcommand*{\JINR}  {Joint Institute for Nuclear Research, 141980 Dubna, Russia}
\newcommand*{\ETHZ}  {Swiss Federal Institute of Technology ETH, 8093 Z\" urich, Switzerland}
\newcommand*{\NOVS}  {Novosibirsk State University, 630090 Novosibirsk, Russia}
\newcommand*{\NOVST} {Novosibirsk State Technical University, 630092 Novosibirsk, Russia}
\newcommand*{\ScuolaPi}{Scuola Normale Superiore, Piazza dei Cavalieri 7, 56126 Pisa, Italy}
\newcommand*{\INFNLNF}{\textit{Present Address}: INFN, Laboratori Nazionali di Frascati, Via 
E. Fermi, 40-00044 Frascati, Rome, Italy}